\documentclass[twocolumn,final]{svjour3b}

\RequirePackage{fix-cm}

\usepackage{mathtools}
\usepackage{MnSymbol}
\usepackage{graphicx}
\usepackage{xcolor}
\usepackage{tikz}
\usepackage[labelfont=bf]{caption}
\usepackage{subcaption}
\usepackage[edges]{forest}
\usepackage{pgfplots}
\usepackage{pgfplotstable}
\usetikzlibrary{patterns}
\usetikzlibrary{pgfplots.groupplots}
\usetikzlibrary{backgrounds}
\usepackage{enumitem}
\usepackage{xfrac}
\usepackage[algo2e,ruled,vlined,linesnumbered]{algorithm2e}
\usepackage{listings}
\usepackage{pdflscape}
\usepackage{pdfpages}
\usepackage{hhline}
\usepackage{pdfpages}
\usepackage{siunitx}
\usepackage{balance}
\usepackage{colortbl}
\usepackage[hidelinks]{hyperref}
\usepackage[all]{nowidow}
\usepackage{times}
\usepackage[scaled=0.86]{helvet}
\usepackage{titlesec}
\usepackage{bbold}

\titlespacing*{\section}{0pt}{3ex}{2ex}
\titlespacing*{\subsection}{0pt}{3ex}{2ex}
\titlespacing*{\subsubsection}{0pt}{3ex}{2ex}

\DeclareMathOperator{\EX}{\mathbb{E}}

\newcommand{\nil}{\textnormal{\scshape{Nil}}\xspace}

\newcommand\Tstrut{\rule{0pt}{2.6ex}}       

\DeclarePairedDelimiter{\ceil}{\lceil}{\rceil}

\newcommand{\pstr}[1]{\textcolor{blue}{#1}}

\newcommand{\ptstr}[1]{\texttt{\textcolor{blue}{#1}}}
\newcommand{\vtstr}[1]{\texttt{\textit{\textcolor{red}{#1}}}}
\newcommand{\vtistr}[1]{\texttt{\textit{#1}}}
\newcommand{\ptbstr}[1]{\textbf{\texttt{\textcolor{blue}{#1}}}}
\newcommand{\vtbstr}[1]{\textbf{\texttt{\textit{\textcolor{red}{#1}}}}}

\definecolor{colx1}{RGB}{252, 26,135}  
\definecolor{colx2}{RGB}{ 11, 36,251}  
\definecolor{colx3}{RGB}{ 17,128,127}  
\definecolor{colx4}{RGB}{183,108, 50}  
\definecolor{colx5}{RGB}{253,128, 35}  
\definecolor{colx6}{RGB}{127, 15,126}  
\definecolor{colx7}{RGB}{200,  0, 20}  
\definecolor{colx8}{RGB}{191,191,191}  
\definecolor{colx9}{RGB}{100,  0,  4}  
\definecolor{colxa}{RGB}{  0,139, 60}  
\definecolor{colxb}{RGB}{  0, 98,139}  
\definecolor{colxc}{RGB}{242,112, 36}  
\definecolor{colxd}{RGB}{222, 23, 23}  
\definecolor{colxe}{RGB}{211,211,211}  
\definecolor{colxe}{RGB}{211,211,211}  
\definecolor{colxee}{RGB}{190,190,190}  
\definecolor{colxf}{RGB}{236,217,198}  
\definecolor{colxff}{RGB}{216,200,162}  
\definecolor{colxg}{RGB}{ 70,192,111}  
\definecolor{colxh}{RGB}{  0,185,242}  
\definecolor{colxi}{RGB}{225,222,236}  
\definecolor{coltbd}{RGB}{255,244,235}  
\definecolor{colptb}{RGB}{238,255,230}  
\definecolor{colxrl}{RGB}{253,179,180} 
\definecolor{colxrd}{RGB}{252, 13, 27} 
\definecolor{colxbl}{RGB}{179,180,253} 
\definecolor{colxbd}{RGB}{ 11, 36,251} 
\definecolor{colxy}{RGB}{227,236,198}  

\newcommand*{\tikzmk}[1]{
  \tikz[remember picture,overlay,]\node(#1){};\ignorespaces}

\newcommand{\inRegularPaper}[2]{%
\ifdefined\IsRegularPaper%
#1%
\else%
#2%
\fi}

\tikzset{
  rscasnode/.style={
    anchor=north,
    font={\footnotesize},
    align=center,
  },
  cmbrace/.style={
    decorate,
    decoration={mirror,brace,amplitude=4pt},
  },
  cbrace/.style={
    decorate,
    decoration={brace,amplitude=4pt},
  },
  partition/.style={
    inner sep=0,
    outer sep=0,
    anchor=north west,
    font={\tiny},
    align=left,
  },
  littlemarker/.style={
    fill=black,
    outer sep=0,
    inner sep=0,
    minimum width=2pt,
    minimum height=2pt,
  },
  ptable/.style={
    anchor=north,
    minimum width=28mm,
    minimum height=2mm,
    draw,
    fill=colptb,
  },
  ptablenode/.style={
    inner sep=1.5pt,
    anchor=center,
    font={\bf \ttfamily \tiny},
  },
  boxlabel/.style={
    anchor=north east,
    font={\bf},
    fill=white,
    draw,
    outer sep=0,
    inner sep=1.5pt,
  },
}

\forestset{
  casindex/.style={
    for tree={
      font={\ttfamily \small},
      align=center,
      parent anchor=south,
      child anchor=north,
      s sep-=2mm,
    },
  },
  declare toks={elo}{
    font={\ttfamily \small}
  },
  dot/.style={
    tikz+={
      \fill[#1] (.child anchor) circle[radius=2.5pt];
    }
  },
  etype/.style n args={3}{
    edge path={
      \noexpand \path[\forestoption{edge},#3]
        (!u.parent anchor) -- (.child anchor)
        \forestoption{edge label};
      \noexpand \path[#1] (.child anchor) circle[radius=2pt];
      \noexpand \path[#2] (!u.parent anchor) circle[radius=2pt];
    },
  },
  ppnode/.style={
    etype={fill=black}{draw=none}{black},
  },
  pvnode/.style={
    etype={fill=black}{draw=none}{black},
  },
  vpnode/.style={
    etype={fill=black}{draw=none}{black},
  },
  vvnode/.style={
    etype={fill=black}{draw=none}{black},
  },
  lpnode/.style={
    etype={fill=black}{draw=none}{black},
  },
  lvnode/.style={
    etype={fill=black}{draw=none}{black},
  },
  bbnode/.style={
    etype={fill=black}{draw=none}{black},
  },
  lbnode/.style={
    etype={fill=black}{draw=none}{black},
  },
  elabel/.style n args=2{
    edge label/.expanded={
      node[midway,anchor=#1,\forestoption{elo}]{\strut\unexpanded{#2}}
    }
  },
  searchindex/.style={
    casindex,
    for tree={
      parent anchor=north,
      l=0mm,
      s sep+=5pt,
      bbnode,
    },
  },
}

\pgfplotsset{
  idy/.style={
    red,
    fill=white,
    postaction={
      pattern=crosshatch dots,
      pattern color=red,
    },
  },
  ipv/.style={
    blue,
    fill=white,
    postaction={
      pattern=north east lines,
      pattern color=blue,
    },
  },
  ivp/.style={
    blue,
    fill=white,
    postaction={
      pattern=north west lines,
      pattern color=blue,
    },
  },
  ione/.style={
    brown!60!black,
    fill=white,
    postaction={
      pattern=horizontal lines,
      pattern color=brown!60!black,
    },
  },
  itwo/.style={
    brown!60!black,
    fill=white,
    postaction={
      pattern=crosshatch,
      pattern color=brown!60!black,
    },
  },
}

\pgfplotsset {
  col2/.style={
    height=33mm,
    width=50mm,
    ticklabel style={font=\footnotesize},
    xlabel near ticks,
    ylabel near ticks,
    ylabel/.append style={align=center},
    ylabel/.append style={font=\footnotesize},
    xlabel/.append style={font=\footnotesize},
  },
}

\pgfplotsset{
  rscas/.style={
    color=colxb,
    mark=*,
  },
  lucene/.style={
    color=colx6,
    mark=triangle*,
  },
  lucenesorted/.style={
    color=colx6,
    mark=triangle,
  },
  barRCAS/.style={
    colxb,
    mark=none,
    fill=white,
    postaction={
      pattern=crosshatch dots,
      pattern color=colxb,
    },
  },
  barRCASsolid/.style={
    colxb,
    mark=none,
    fill=colxb!60!white,
  },
  barLucene/.style={
    colxc,
    mark=none,
    fill=white,
    postaction={
      pattern=crosshatch,
      pattern color=colxc,
    },
  },
  barLuceneSorted/.style={
    colxc,
    mark=none,
    fill=white,
    postaction={
      pattern=north west lines,
      pattern color=colxc,
    },
  },
  barPV/.style={
    colx6,
    fill=white,
    mark=none,
    postaction={
      pattern=north west lines,
      pattern color=colx6,
    },
  },
  barVP/.style={
    colxa,
    fill=white,
    mark=none,
    postaction={
      pattern=north east lines,
      pattern color=colxa,
    },
  },
  barStructure/.style={
    colxg!30!black,
    fill=colxg,
    mark=none,
  },
  fulldynhatch/.style={
    colx6,
    mark=none,
    fill=white,
    postaction={
      pattern=north east lines,
      pattern color=colx6,
    },
  },
  postgrespv/.style={
    colxa,
    mark=diamond*,
  },
  postgresvp/.style={
    colxc,
    mark=square*,
  },
  gnusort/.style={
    color=colx6,
    mark=diamond*,
  },
  postgres/.style={
    color=colxa,
    mark=square*,
  },
  twopass/.style={
    color=colx4,
    mark=square*,
  },
  bytebybyte/.style={
    color=colxd,
    mark=triangle*,
  },
  allornothing/.style={
    color=orange,
    mark=triangle*,
  },
  uniformatrandom/.style={
    color=red,
    mark=diamond*,
  },
  costmodel/.style={
    color=colxg,
    mark=triangle*,
  },
  rcasplusdirectio/.style={
    color=colxb,
    mark=o,
  },
  twopassdirectio/.style={
    color=colx4,
    mark=square,
  },
  bytebybytedirectio/.style={
    color=colxd,
    mark=triangle,
  },
  allornothingdirectio/.style={
    color=orange,
    mark=triangle,
  },
}

\pgfplotsset{compat=1.14,
  /pgfplots/ybar legend/.style={
    /pgfplots/legend image code/.code={%
      \draw[##1,/tikz/.cd,yshift=-0.25em]
      (0cm,0cm) rectangle (7pt,1.1em);
    },
  },
}

\definecolor{codegray}{rgb}{0.5,0.5,0.5}
\lstdefinestyle{pseudocode}{
  language=c,
  showspaces=false,
  basicstyle={\ttfamily \scriptsize},
  numberstyle=\tiny\color{codegray},
  keywordstyle=\color{magenta},
  numbers=left,
  numbersep=5pt,
  mathescape=true,
  numberblanklines=false,
  morekeywords={uint8_t,uint16_t,int16_t,in},
}
\lstdefinestyle{pcode}{
  style=pseudocode,
  basicstyle={\ttfamily \normalsize},
}
\lstdefinestyle{pcodesmall}{
  style=pseudocode,
  basicstyle={\ttfamily \small},
}
\newcommand{\pcode}[2][]{\textnormal{\lstinline[style=pcode,#1]|#2|}}
\newcommand{\pcodes}[2][]{\textnormal{\lstinline[style=pseudocode,#1]|#2|}}

\smartqed

\makeatletter
\renewcommand{\tableofcontents}{\scriptsize\@starttoc{toc}}
\makeatother
\usepackage{xcolor,colortbl}
\usepackage{mdframed}

\definecolor{Gray}{gray}{0.98}

\newcommand{\MREV}[1]{{#1}}
\newcommand{\MREVB}[1]{{#1}}

\newcommand{\RCOMMENT}[1]{\medskip\begin{siderules}\small #1 \end{siderules}\vspace{-7pt}}

\newmdenv[
  topline=false,
  bottomline=false,
  rightline=false,
  leftline=true,
  skipabove=\topsep,
  skipbelow=\topsep,
  backgroundcolor=gray!3
]{siderules}

\begin{document}

\edef\oldindent{\the\parindent}
\edef\oldparskip{\the\parskip}
\setlength{\parindent}{0em}
\setlength{\parskip}{2ex}
\pagenumbering{roman}
\newcommand\shorttitle{}
\pagestyle{plain}

\inRegularPaper{
\textbf{Revision report for ``\emph{Robust and Scalable
    Content-and-Structure Indexing}''}

\MREVB{%
  We thank Reviewer \#3 for making suggestions on how to improve the paper
  further.  Again, our changes in the paper are highlighted
  in blue (changes and additions only; typos, minor editing, and
  deletions have been omitted to keep the paper readable). This
  revision report explains in detail how we addressed the comments and
  edited the paper.
}

\rule{\linewidth}{0.4pt}
Reviewer \#3 \\
\rule{\linewidth}{0.4pt}

\RCOMMENT{%
  The revision has addressed most of the concerns from the previous round of
  reviewing. The paper is easy to follow.
}

\MREVB{%
  We are happy to read that there are no issues with the presentation
  and that the paper is readable.  }

\RCOMMENT{%
  There are several papers about LSM tree range-query optimizations
  without using filters: EvenDB (EuroSys'20), REMIX (FAST'21).  }

\MREVB{ Thanks a lot for the pointers.  We integrated the two
  references into the related work section and highlighted the
  differences between CAS queries and range queries.  }

\RCOMMENT{%
  For R4-D6, those are indeed fair points because those challenges
  must be addressed in real systems when you manage data on top of a
  storage abstraction. While the authors prefer to not discuss it in
  the paper, a statement about "those challenges exist and they're out
  of the scope of the paper" is necessary so that readers can know
  whether there is a gap between this paper's ideas and a real
  solution that they can adopt.  }

\MREVB{ We did not want to convey the message that recovery and
  multi-user synchronization are not important.  Clearly, they are
  important.  Section 6.2 was amended to clarify that in order
  to build a full-fledged data management system, these aspects need
  to be taken care of.  }

\cleardoublepage
}

\setlength{\parindent}{\oldindent}%
\setlength{\parskip}{\oldparskip}
\pagenumbering{arabic}


\title{Robust and Scalable Content-and-Structure Indexing}
\subtitle{(Extended Version)}

\author{
  Kevin Wellenzohn \and
  Michael H.~B\"ohlen \and
  Sven Helmer \and
  Antoine Pietri \and
  Stefano Zacchiroli
}


\institute{
  Kevin Wellenzohn \email{wellenzohn@ifi.uzh.ch}
  \and
  Michael H.~B\"ohlen \email{boehlen@ifi.uzh.ch}
  \and
  Sven Helmer \email{helmer@ifi.uzh.ch}
  \at
  Department of Informatics, University of Zurich, Zurich, Switzerland
  \and
  Antoine Pietri \email{antoine.pietri@inria.fr}
  \at
  Inria, Paris, France
  \and
  Stefano Zacchiroli \email{stefano.zacchiroli@telecom-paris.fr}
  \at
  LTCI, T{\'e}l{\'e}com Paris, Institut Polytechnique de Paris, Paris, France
}

\date{}

\maketitle


\begin{abstract}
  Frequent queries on semi-structured hierarchical data are
  Content-and-Structure (CAS) queries that filter data items based on
  their location in the hierarchical structure and their value for
  some attribute.  We propose the Robust and Scalable
  Content-and-Structure (RSCAS) index to efficiently answer CAS
  queries on big semi-structured data.  To get an index that is robust
  against queries with varying selectivities we introduce a novel
  dynamic interleaving that merges the path and value dimensions of
  composite keys in a balanced manner.  We store interleaved keys in
  our trie-based RSCAS index, which efficiently supports a wide range
  of CAS queries, including queries with wildcards and descendant
  axes.  We implement RSCAS as a log-structured merge (LSM) tree to
  scale it to data-intensive applications with a high \MREV{insertion}
  rate.  We illustrate RSCAS's robustness and scalability by indexing
  data from the Software Heritage (SWH) archive, which is the world's
  largest, publicly-available source code archive.

  \keywords{Indexing \and content and structure \and interleaving \and
    hierarchical data \and semi-structured data \and XML \and LSM trees}
\end{abstract}

\section{Introduction}

A lot of the data in business and engineering applications is
semi-structured and inherently hierarchical.  Typical examples are
source code archives \cite{JA18}, bills of materials \cite{RB15},
enterprise asset hierarchies \cite{JF13}, and enterprise resource
planning applications \cite{JF15}.  A common type of queries on such
data are content-and-structure (CAS) queries \cite{CM15}, containing a
\emph{value predicate on the content} of an attribute and a \emph{path
  predicate on the location} of this attribute in the hierarchical
structure.

CAS indexes are being used to support the efficient processing of CAS
queries. There are two important properties that we look for in a CAS
index: robustness and scalability.  \emph{Robustness} means that a CAS
index optimizes the average query runtime over all possible queries.
It ensures that an index can efficiently deal with a wide range of CAS
queries.  Many existing indexes are not robust since the performance
depends on the individual selectivities of its path and value
predicates.  If either the path or value selectivity is high, these
indexes produce large intermediate results even if the combined
selectivity is low.  This happens because existing solutions either
build separate indexes for, respectively, content and structure
\cite{CM15} or prioritize one dimension over the other (i.e., content
over structure or vice versa) \cite{oak,BC01,JL15}.
\emph{Scalability} means that even for large datasets an index can be
efficiently created and updated, and is not constrained by the size of
the available memory.  Existing indexes are often not scalable since
they rely on in-memory data structures that do not scale to large
datasets. For instance, with the memory-based CAS index \cite{KW20} it
is impossible to index datasets larger than 100\,GB on a machine with
400\,GB main memory.

We propose RSCAS, a robust and scalable CAS index. RSCAS's robustness
is rooted in a \emph{well-balanced integration} of the content and
structure of the data in a single index. Its scalability is due to
log-structured merge (LSM) trees \cite{PO96} that combine an in-memory
structure for fast \MREV{insertions} with a series of read-only
disk-based structures for fast sequential reads and writes.

To achieve robustness we propose to interleave the path and value
bytes of composite keys in a balanced manner.  A well-known technique
to interleave composite keys is the $z$-order curve \cite{GM66,JO84},
but applying the $z$-order curve to paths and values is subtle.  Often
the query performance is poor because of long common prefixes, varying
key lengths, different domain sizes, and data skew.  The paths in a
hierarchical structure have, by their very nature, long common
prefixes, but the first byte following a longest common prefix
separates data items.  We call such a byte a \emph{discriminative
  byte} and propose a \emph{dynamic interleaving} that interleaves the
discriminative bytes of paths and values alternatingly.  This leads to
a well-balanced partitioning of the data with a robust query
performance.  We use the dynamic interleaving to define the RSCAS
index for semi-structured hierarchical data.  The RSCAS index is
trie-based and efficiently supports the basic search methods for CAS
queries: \emph{range searches} and \emph{prefix searches}.  Range
searches enable value predicates that are expressed as a value range
and prefix searches support path predicates that contain wildcards and
descendant axes.  Crucially, tries in combination with dynamically
interleaved keys allow us to efficiently evaluate path and value
predicates simultaneously.

To scale the RSCAS index to large datasets and support efficient
insertions, we use LSM trees \cite{PO96} that combine an in-memory
RSCAS trie with a series of disk-resident RSCAS tries whose size is
doubling in each step. \MREV{RSCAS currently supports only insertions
  since our main use case, indexing an append-only archive, does not
  require updates or deletes}.  The in-memory trie is based on the
Adaptive Radix Tree (ART) \cite{VL13}, which is a memory-optimized
trie structure that supports efficient \MREV{insertions}.  Whenever
the in-memory RSCAS trie reaches its maximum capacity, we create a new
disk-based trie.  Since disk-based RSCAS tries are immutable, we store
them compactly on disk and leave no gaps between nodes.  We develop a
partitioning-based bulk-loading algorithm that builds RSCAS on disk
while, at the same time, dynamically interleaving the keys. This
algorithm works well with limited memory but scales nicely with the
amount of memory to reduce the disk I/O during bulk-loading.

\paragraph{Main contributions:}
\begin{itemize}

\item We develop a \emph{dynamic interleaving} to interleave paths and
  values in an alternating way using the concept of
  \emph{discriminative bytes}.  We show how to compute this
  interleaving by a hierarchical partitioning of the data.  We prove
  that our dynamic interleaving is robust against varying
  selectivities (Section~\ref{sec:theory}).

\item We propose the trie-based \emph{Robust and Scalable
    Content-and-Structure (RSCAS) index} for semi-structured
  hierarchical data.  Dynamically interleaved keys give RSCAS its
  robustness. Its scalability is rooted in LSM trees that combine a
  memory-optimized trie for fast in-place \MREV{insertions} with a
  series of disk-optimized tries (Section \ref{sec:rscas}).

\item We propose efficient algorithms for querying, \MREV{inserting},
  bulk-loading, and merging RSCAS tries.  A combination of range and
  prefix searches is used to evaluate CAS queries on the trie-based
  structure of RSCAS. \MREV{Insertions} are performed on the in-memory
  trie using lazy restructuring. Bulk-loading creates large
  disk-optimized tries in the background. Merging is applied when the
  in-memory trie overflows to combine it with a series of
  disk-resident tries (Section \ref{sec:algorithms}).

\item We conduct an experimental evaluation with three real-world and
  one synthetic dataset.  One of the real-world datasets is Software
  Heritage (SWH) \cite{JA18}, the world's largest archive of
  publicly-available source code.  Our experiments show that RSCAS
  delivers robust query performance with up to two orders of magnitude
  improvements over existing approaches, while offering comparable
  bulk-loading and insertion performance (Section
  \ref{sec:experiments}).

\end{itemize}

\section{Application Scenario}
\label{sec:scenario}

As a practical use case we deploy a large-scale CAS index for Software
Heritage (SWH)~\cite{RC17}, the largest public archive of software
source code and its development history.\footnote{As of 2021-11-08 the
  Software Heritage archive contains more than 11 billion source code
  files and 2 billion commits, coming from more than 160 million
  public software projects. The archive can be browsed at:
  \url{https://archive.softwareheritage.org/}.}

At its core, Software Heritage archives version control systems
(VCSs), storing all recorded source code artifacts in a giant,
globally deduplicated Merkle structure~\cite{merkle87} that stores
elements from many different VCSs using cryptographic hashes as keys.
VCSs record the evolution of source code trees over time, an aspect
that is reflected in the data model of Software Heritage
\cite{AP20}. The data model supports the archiving of artifacts, such
as file blobs (byte sequences, corresponding to tree leaves), source
code directories (inner nodes, pointing to sub-directories and files,
giving them local path names), commits (called \emph{revisions} in
this context), releases (commits annotated with memorable names such
as \texttt{"1.0"}), and VCS repository snapshots. Nodes in the data
model are associated with properties that are relevant for querying.
Examples of node properties are: cryptographic node identifiers, as
well as commit and release metadata such as authors, log messages,
timestamps, etc.

Revisions are a key piece of software development workflows. Each of
them, except the very first one in a given repository, is connected to
the previous ``parent'' revision, or possibly multiple parents in case
of merge commits. These connections allow the computation of the popular
\emph{diff} representations of commits that show how and which files have been
changed in any given revision.
\inRegularPaper{}{Computing diffs for all
revisions in the archive makes it possible to look up all revisions
that have changed files of interest.}

Several aspects make Software Heritage a relevant and challenging use
case for CAS indexing. First, the size of the archive is significant:
at the time of writing, the archive consists of about 20 billion nodes
(total file size is about 1 PiB, but we will not index \emph{within}
files, so this measure is less relevant).  Second, the archive grows
constantly by continuously crawling public data sources such as
collaborative development platforms (e.g., GitHub, GitLab), Linux
distributions (e.g., Debian, NixOS), and package manager repositories
(e.g., PyPI, NPM). The archive growth \emph{ratio} is also very
significant: the amount of archived source code artifacts grows
exponentially over time, doubling every 2 to 3 years~\cite{GR20},
which calls for an incremental indexing approach to avoid indexing
lag. For instance, during 2020 alone the archive has ingested about
600 million new revisions and 3 billion new file blobs (i.e., file
contents never seen before).

Last but not least, short of the CAS queries proposed in this paper,
the current querying capabilities for the archive are quite
limited. Entire software repositories can be looked up by full-text
search \emph{on their URLs}, providing entry points into the
archive. From there, users can browse the archive, reaching the
desired revisions (e.g., the most recent revision in the
\texttt{master} branch since the last time a repository was crawled)
and, from there, the corresponding source code trees. It is not
possible to query the ``diff'', i.e., find revisions that modified
certain files in a certain time period, which is limiting for both
user-facing and research-oriented queries (e.g., in the field of
empirical software engineering).

With the approach proposed in this paper we offer functionality
to answer CAS queries like the following:
\begin{quote}
  Find all revisions from June 2021 that modify a C file located in a
  folder whose name begins with \texttt{"ext"}.
\end{quote}
This query consists of two predicates. First, a content predicate on
the revision time, which is a range predicate that matches all
revisions from the first to the last day of June 2021.  Second, a
structure predicate on the paths of the files that were touched by a
revision.  We are only interested in revisions that modify files with
\texttt{.c} extension and that are located in a certain directory.
This path predicate can be expressed as \texttt{/**/ext*/*.c} with
wildcard \texttt{**} to match folders that are nested arbitrarily
deeply in the filesystem of a repository and wildcard \texttt{*} to
match all characters in a directory or file name.

\section{Related Work}
\label{sec:rw}

For related work, two CAS indexing techniques have been investigated:
(a) creating separate indexes for content and structure, and (b)
combining content and structure in one index. We call these two
techniques \emph{separate CAS indexing} and \emph{combined CAS
  indexing}, respectively.

Separate CAS indexing creates dedicated indexes for, respectively, the
content and the structure of the data.  Mathis et al.~\cite{CM15} use
a B+ tree to index the content and a structural summary (i.e., a
DataGuide \cite{RG97}) to index the structure of the data. The
DataGuide maps each unique path to a numeric identifier, called the
path class reference (PCR), and the B+ tree stores the values along
with their PCRs. Thus, the B+ tree stores
$(\texttt{value}, \langle \texttt{nodeId}, \texttt{PCR} \rangle)$
tuples in its leaf nodes, where \texttt{nodeId} points to a node whose
content is \texttt{value} and whose path is given by \texttt{PCR}.  To
answer a CAS query we must look at the path index and the value index
independently.  The subsequent join on the PCR is slow if intermediate
results are large.
\inRegularPaper{}{Mathis et al.\ assume that there are few unique
paths and the path index is small (fewer than 1000 unique paths in
their experiments).}
Kaushik et al.~\cite{RK04} present an approach
that combines a 1-index \cite{TM99} to evaluate path predicates with a
B+ tree to evaluate value predicates, but they do not consider
updates.

A popular system that implements separate indexing is Apache Lucene
\cite{lucene}\inRegularPaper{.}{, which is a scalable and widely
  deployed indexing and search system that underpins Apache Solr and
  Elasticsearch.}  Lucene uses different index types depending on the
type of the indexed attributes.  For CAS indexing, we represent paths
as strings and values as numbers.  Lucene indexes strings with finite
state transducers (FSTs), which are automata that map strings to lists
of sorted document IDs (called postings lists).  Numeric attributes
are indexed in a Bkd-tree \cite{OP03}, which is a disk-optimized
kd-tree.  Lucene answers conjunctive queries, like CAS queries, by
evaluating each predicate on the appropriate index. The indexes return
sorted postings lists that must be intersected to see if a document
matches all predicates of a conjunctive query.  Since the lists are
sorted, the intersection can be performed efficiently.  However, the
independent evaluation of the predicates may yield large intermediate
results, making the approach non-robust.  To scale to large datasets,
Lucene implements techniques that are similar to LSM trees \cite{PO96}
(cf.\ Section~\ref{sec:rscas}).  \inRegularPaper{}{Lucene batches
  insertions in memory before flushing them as read-only segments to
  disk. As the number of segments grows, Lucene continuously compacts
  them by merging small segments into a larger segment.}

The problem with separate CAS-indexing is that it is not robust. If at
least one predicate of a CAS query is not selective, separate indexing
approaches generate large intermediate results.  This is inefficient
if the final result is small.  Since the predicates are evaluated on
different indexes, we cannot use the more selective predicate to prune
the search space.

Combined CAS indexing integrates paths and values in one index.  A
well-known and mature technology are composite indexes, which are
used, e.g., in relational databases to index keys that consist of more
than one attribute.  Composite indexes concatenate the indexed
attributes according to a specified ordering.  In CAS indexing, there
are two possible orderings of the paths and values: the $PV$-ordering
orders the paths before the values, while the $VP$-ordering orders the
values first. The ordering determines what queries a composite index
can evaluate efficiently. Composite indexes are only efficient for
queries that have a small selectivity for the attribute appearing
first.  In our experiments we use the composite B+ tree of
Postgres as the reference point for an efficient and scalable
implementation of composite indexes.

IndexFabric \cite{BC01} is another example of a composite CAS index.
It uses a $PV$-ordering, concatenating the (shortened) paths and values
of composite keys, and storing them in a disk-optimized PATRICIA trie
\cite{DM68}.  IndexFabric shortens the paths to save disk space by
mapping long node labels to short strings (e.g., map label `extension'
to `e'). During query evaluation IndexFabric must first fully evaluate
the path predicate before it can look at the value predicate since it
orders paths before the values in the index. Since it uses shortened
paths, it cannot evaluate wildcards within a node label (e.g.,
\text{ext*} to match extension, exterior, etc.). IndexFabric does not
support bulk-loading.

The problem with composite indexes is that they prioritize the
dimension appearing first.  The selectivity of the predicate in
the first dimension determines the query performance.  If it is high
and the other selectivity is low, the composite index performs badly
because the first predicate must be fully evaluated before the second
predicate can be evaluated. As a result, a composite index is not
robust.

Instead of concatenating dimensions, it is possible to
\emph{interleave} dimensions. The $z$-order curve \cite{GM66,JO84},
for example, is obtained by interleaving the binary representation of
the individual dimensions and is used in UB-trees \cite{FR00} and k-d
tries \cite{BN08,JO84,HS06}.  Unfortunately, the $z$-order curve
deteriorates to the performance of a composite index if the data
contains long common prefixes \cite{KW20}.  This is the case in CAS
indexing where paths have long common prefixes.  \inRegularPaper{}{The
  problem with common prefixes is that they are the same for all data
  items and do not prune the search space during a search.
  Interleaving a common prefix in one dimension with a non-common
  prefix in the other dimension means we prune keys in one dimension
  but not the other \cite{VM99}.}

LSM trees \cite{PO96} are used to create scalable indexing systems
with high write throughput (see, e.g., AsterixDB \cite{SA14}, BigTable
\cite{FC08}, Dynamo \cite{GD07}, etc.).  They turn expensive in-place
updates that cause many random disk I/Os into out-of-place updates
that use sequential writes. To achieve that, LSM trees combine a small
in-memory tree $R^M_0$ with a series of disk-resident trees
$R_0, R_1, \ldots$, each tree being $T$ times larger than the tree in
the previous level. Insertions are performed exclusively in the
main-memory tree $R^M_0$.

Modern LSM tree implementations, see \cite{CL20} for an excellent
recent survey, use sorted string tables (SSTables) or other immutable
data structures at multiple levels. \MREV{Generally, there are two
  different merge policies: leveling and tiering.  With the leveling
  merge policy, each level $i$ contains exactly one structure and when
  the structure at level $i$ grows too big, this structure and the one
  at level $i+1$ are merged. A structure on level $i+1$ is $T$ times
  larger than a structure on level $i$. Tiering maintains multiple
  structures per level. When a level $i$ fills up with $T$ structures,
  they are merged into a structure on level $i+1$. We discuss the
  design decisions regarding LSM-trees and RSCAS in
  Section~\ref{sec:scaling}.}

An LSM tree requires an efficient bulk-loading algorithm to create a
disk-based RSCAS trie when the in-memory trie overflows.
Sort-based algorithms sort the data and
  build an index bottom-up.  Buffer-tree methods
  bulk-load a tree by buffering insertions in nodes
  and flushing them in batches to its children when a buffer
  overflows.  Neither sort- nor buffer-based techniques
\cite{LA03,DA13,JB97} can be used for RSCAS because our dynamic
interleaving must look at \emph{all} keys to correctly interleave
them.  We develop a partitioning-based bulk-loading algorithm for
RSCAS that alternatingly partitions the data in the path and value
dimension to dynamically interleave paths and values.

The combination of the dynamic interleaving with wildcards and range
queries makes it hard to embed RSCAS into an LSM-tree-based key-value
(KV) store. While early, simple KV-stores did not support range
queries at all, more recent KV-stores \MREVB{create Bloom filters for
  a predefined set of fixed prefixes~\cite{myrocks20}, i.e., only
  range queries using these prefixes can be answered efficiently. SuRF
  was one of the first approaches able to handle arbitrary range
  queries by storing minimum-length prefixes in a trie so that all
  keys can be uniquely identified~\cite{surf18}. This was followed by
  Rosetta, which stores all prefixes for each key in a hierarchical
  series of Bloom filters~\cite{rosetta20}.  KV-stores supporting
  ranges queries without filters have also been developed. EvenDB
  optimizes the evaluation of queries exhibiting spatial locality,
  i.e., keys with the same prefixes are kept close together and in
  main memory~\cite{evendb20}.  REMIX offers a globally sorted view of
  all keys with a logical sorting of the data \cite{remix21}.  The
  evaluation of range queries boils down to seeking the first matching
  element in a sorted sequence of keys and scanning to the end of the
  range. CAS queries follow a different pattern. During query
  evaluation, we simultaneously process a range query in the value
  dimension and match strings with wildcards at arbitrary positions in
  the path dimension.  The prefix shared by the matching keys ends at
  the first wildcard, which can occur early in the path.  We prune
  queries with wildcards by regularly switching back to the more
  selective value dimension.  }

\begin{table*}[htb] \centering
\caption{A set $\mathsf{K}^{1..9} = \{ \mathsf{k}_1, \ldots,
\mathsf{k}_9 \}$ of composite keys}
\label{tab:ex2}
\begin{tikzpicture}[
    tick/.style={
      font={\ttfamily \scriptsize},
      anchor=north,
    },
  ]
  \node[inner sep=0, outer sep=0,anchor=south west] (table) at (0,0) {
  \begin{tabular}{llr|c}
    \multicolumn{1}{c}{}
      & \multicolumn{1}{c}{Path Dimension $P$}
      & \multicolumn{1}{c}{Value Dimension $V$ (64 bit unsigned
      integer)}
      & \multicolumn{1}{c}{Revision $R$ (SHA1 hash)} \\
    \cline{2-4}
    $\mathsf{k}_1$
      & \texttt{/Sources/Map.go\$}
      & \textsf{2019-10-17 17:17:46} (\vtistr{00\,00\,00\,00\,5D\,A8\,94\,2A})
      & $\mathsf{r}_1$ (\texttt{A1\,A6\,06\,B0\,B3\,$\ldots$})
      \\
    $\mathsf{k}_2$
      & \texttt{/crypto/ecc.h\$}
      & \textsf{2020-11-24 22:48:36} (\vtistr{00\,00\,00\,00\,5F\,BD\,8D\,C4})
      & $\mathsf{r}_2$ (\texttt{D4\,47\,39\,D8\,F8\,$\ldots$})
      \\
    $\mathsf{k}_3$
      & \texttt{/crypto/ecc.c\$}
      & \textsf{2020-11-24 22:48:36} (\vtistr{00\,00\,00\,00\,5F\,BD\,8D\,C4})
      & $\mathsf{r}_2$ (\texttt{D4\,47\,39\,D8\,F8\,$\ldots$})
      \\
    $\mathsf{k}_4$
      & \texttt{/Sources/Schema.go\$}
      & \textsf{2019-10-17 17:19:24} (\vtistr{00\,00\,00\,00\,5D\,A8\,94\,8C})
      & $\mathsf{r}_3$ (\texttt{41\,D1\,7A\,7B\,4D\,$\ldots$})
      \\
    $\mathsf{k}_5$
      & \texttt{/fs/ext3/inode.c\$}
      & \textsf{2020-06-24 01:20:41} (\vtistr{00\,00\,00\,00\,5E\,F2\,9C\,59})
      & $\mathsf{r}_4$ (\texttt{96\,98\,D9\,F5\,06\,$\ldots$})
      \\
    $\mathsf{k}_6$
      & \texttt{/fs/ext4/inode.h\$}
      & \textsf{2020-05-14 11:56:02} (\vtistr{00\,00\,00\,00\,5E\,BD\,23\,C2})
      & $\mathsf{r}_5$ (\texttt{FF\,CA\,AE\,8F\,57\,$\ldots$})
      \\
    $\mathsf{k}_7$
      & \texttt{/fs/ext4/inode.c\$}
      & \textsf{2020-11-24 17:05:30} (\vtistr{00\,00\,00\,00\,5F\,BD\,3D\,5A})
      & $\mathsf{r}_6$ (\texttt{68\,8D\,97\,3C\,BE\,$\ldots$})
    \\
    $\mathsf{k}_8$
      & \texttt{/Sources/Schedule.go\$}
      & \textsf{2019-10-17 17:32:11} (\vtistr{00\,00\,00\,00\,5D\,A8\,97\,8B})
      & $\mathsf{r}_7$ (\texttt{99\,07\,EE\,0A\,7B\,$\ldots$})
      \\
    $\mathsf{k}_9$
      & \texttt{/Sources/Scheduler.go\$}
      & \textsf{2019-10-17 17:32:11} (\vtistr{00\,00\,00\,00\,5D\,A8\,97\,8B})
      & $\mathsf{r}_7$ (\texttt{99\,07\,EE\,0A\,7B\,$\ldots$})
      \\
    \cline{2-4}
  \end{tabular}
  };
  \foreach \x in {0,4,...,22} {
    \pgfmathsetmacro{\xc}{\x * 0.182 + 1.0}
    \pgfmathtruncatemacro{\xr}{\x + 1}
    \draw (\xc,0) -- (\xc,-0.1);
    \node[tick] at (\xc,-0.05) {\xr};
  }
  \foreach \x in {0,...,7} {
    \pgfmathsetmacro{\xc}{\x * 0.41 + 8.10}
    \pgfmathtruncatemacro{\xr}{\x + 1}
    \draw (\xc,0) -- (\xc,-0.1);
    \node[tick] at (\xc,-0.05) {\xr};
  }
\end{tikzpicture}
\end{table*}

\section{Background}

\subsection{Data Representation}



We use \emph{composite keys} to represent the paths and values of data
items in semi-structured hierarchical data.


\begin{definition}[Composite Key]
  A composite key $k$ is a two-dimensional key that consists of a path
  $k.P$ and a value $k.V$, and each key stores a reference $k.R$ as
  payload that points to the corresponding data item in the database.
\end{definition}


Given a dimension $D \in \{P,V\}$ we write $k.D$ to access $k$'s path
(if $D = P$) or value (if $D = V$). Composite keys can be extracted
from popular semi-structured hierarchical data formats, such as JSON
and XML.  In the context of SWH we use composite keys $k$ to represent
that a file with path $k.P$ is modified (i.e., added, changed, or
deleted) at time $k.V$ in revision $k.R$.

\begin{example}
  Table \ref{tab:ex2} shows the set
  $\mathsf{K}^{1..9} = \{ \mathsf{k}_1, \ldots, \mathsf{k}_9\}$ of
  composite keys (we use a \textsf{sans-serif} font to refer to
  concrete instances in our examples).  We write
  $\mathsf{K}^{2,5,6,7}$ to refer to
  $\{\mathsf{k}_2, \mathsf{k}_5, \mathsf{k}_6, \mathsf{k}_7\}$.
  Composite key $\mathsf{k}_2$ denotes that the file
  \texttt{/crypto/ecc.h\$} was modified on 2019-07-20 in revision
  $\mathsf{r}_2$. In the same revision, also file
  \texttt{/crypto/ecc.c\$} is modified, see key $\mathsf{k}_3$.
  $\hfill\Box$
\end{example}


We represent paths and values as byte strings that we access
byte-wise. We visualize them with one byte ASCII characters for the
path dimension and italic hexadecimal numbers for the value dimension,
see Table~\ref{tab:ex2}.  To guarantee that no path is a prefix of
another we append the end-of-string character \texttt{\$} (ASCII code
\texttt{0x00}) to each path.  Fixed-length byte strings (e.g., 64 bit
numbers) are prefix-free because of the fixed length. We assume that
the path and value dimensions are binary-comparable, i.e., two paths
or values are $<$,~$=$, or $>$ iff their corresponding byte strings
are $<$,~$=$, or $>$, respectively \cite{VL13}. For example,
big-endian integers are binary-comparable while little-endian integers
are not.

Let $s$ be a byte-string, then $|s|$ denotes the length of $s$ and
$s[i]$ denotes the $i$-th byte in $s$.  The left-most byte of a
byte-string is byte one.  $s[i] = \epsilon$ is the empty string if $i
> |s|$. $s[i,j]$ denotes the substring of $s$ from position $i$ to $j$
and $s[i,j] = \epsilon$ if $i > j$.

\begin{definition}[Longest Common Prefix] \label{def:lcp}
  The longest common prefix $\textsf{lcp}(K,D)$ of a set of keys $K$
  in dimension $D$ is the longest prefix $s$ that all keys $k \in K$
  share in dimension $D$, i.e.,
  {\begin{align*}
    \textsf{lcp}(K&,D) = s \,\,\,\text{iff}\,\,\,\\
      & \forall k \in K (k.D[1,|s|] = s)\ \wedge \\
      & \nexists l(l > |s|
      \wedge \forall k,k' \in K( \\
        & \hspace*{.6cm} l \leq \min(|k.D|, |k'.D|) \wedge
        k.D[1,l] = k'.D[1,l]))
  \end{align*}}
\end{definition}

\begin{example}
  The longest common prefix in the path and value dimensions of the
  nine keys in Table~\ref{tab:ex2} is
  $\textsf{lcp}(\mathsf{K}^{1..9},P) = \texttt{/}$ and
  $\textsf{lcp}(\mathsf{K}^{1..9},V) = \vtistr{00\,00\,00\,00}$.  If
  we narrow down the set of keys to $\mathsf{K}^{5,6}$ the longest
  common prefixes become longer:
  $\textsf{lcp}(\mathsf{K}^{5,6},P) = \texttt{/fs/ext}$ and
  $\textsf{lcp}(\mathsf{K}^{5,6},V) =
  \texttt{00\,00\,00\,00\,5E}$. $\hfill\Box$
\end{example}

\subsection{Content-and-Structure (CAS) Queries}
\label{sec:casqueries}

Content-and-structure (CAS) queries contain a path predicate and value
predicate \cite{CM15}.  The path predicate is expressed as a query
path $q$ that supports two wildcard symbols.  The descendant axis
\texttt{**} matches zero to any number of node labels, while the
\texttt{*} wildcard matches zero to any number of characters in a
single label.

\begin{definition}[Query Path] \label{def:querypath}
  A query path $q$ is denoted by $q =
  \texttt{/$\lambda_1$/$\lambda_2$/$\ldots$/$\lambda_{m}$}$.  Each
  label $\lambda_i$ is a string $\lambda_i \in (A \cup
  \{\texttt{*}\})^+$, where $A$ is an alphabet and \texttt{*} is a
  reserved wildcard symbol. The wildcard \texttt{*} matches zero to
  any number of characters in a label.  We call $\lambda_i =
  \texttt{**}$ the descendant axis that matches zero to any number of
  labels.
\end{definition}

\begin{definition}[CAS Query]
  CAS query $Q(q,[v_l,v_h])$ consists of a query path $q$ and a value
  predicate $[v_l,v_h]$.  Given a set $K$ of composite keys, CAS query
  $Q$ returns the revisions $k.R$ of all composite keys $k \in K$ for
  which $k.P$ matches $q$ and $v_l \leq k.V \leq v_h$.
\end{definition}

\begin{example} \label{ex:casquery} CAS query
  $Q(\texttt{/**/ext*/*.c},$ [2021-06-01, 2021-06-30]$)$ matches all
  revisions (a) committed in June 2021 that (b) modified a $C$ file
  located in a folder that begins with name \texttt{ext}, anywhere in
  the directory structure of a software repository.  $\hfill\Box$
\end{example}

\subsection{Interleaving of Composite Keys}

We integrate path $k.P$ and value $k.V$ of a key $k$ by interleaving
them.  Table~\ref{tab:interleaving} shows three common ways to
integrate $k.P$ and $k.V$ of key $\mathsf{k}_9$ from
Table~\ref{tab:ex2}.  Value bytes are written in italic and shown in
red, path bytes are shown in blue.  The first two rows show the
path-value and value-path concatenation ($I_{PV}$ and $I_{VP}$), respectively.
The byte-wise interleaving $I_{BW}$ in the
third row interleaves one value byte with one path byte.  Note that
none of these interleavings is well-balanced.  The byte-wise
interleaving is not well-balanced, since all value-bytes are
interleaved with a single label of the path (\texttt{/Sources}).

\begin{table}[htb]
\caption{Key $\mathsf{k}_9$ is interleaved using
different approaches.}
\label{tab:interleaving}
\centering
\begin{tikzpicture}
\node[anchor=north west] (table) at (0,0) {
${\scriptsize \begin{alignedat}{3}
  & \text{\footnotesize Approach} && &&
  \hspace{1.7cm}\text{\footnotesize
  Interleaving of Key} \\
  \hline
  \rule{0pt}{2.6ex}
  & I_{PV}(\mathsf{k}_9) && = &&
    \ptstr{/Sources/Scheduler.go\$}\,
    \vtstr{00\,00\,00\,00\,5D\,A8\,97\,8B}
    \\
  & I_{VP}(\mathsf{k}_9) && =\,\, &&
    \vtstr{00\,00\,00\,00\,5D\,A8\,97\,8B}\,
    \ptstr{/Sources/Scheduler.go\$}
    \\
  & I_{BW}(\mathsf{k}_9) && = &&
    \vtstr{00}\,
    \ptstr{/}\,
    \vtstr{00}\,
    \ptstr{S}\,
    \vtstr{00}\,
    \ptstr{o}\,
    \vtstr{00}\,
    \ptstr{u}\,
    \vtstr{5D}\,
    \ptstr{r}\,
    \vtstr{A8}\,
    \ptstr{c}\,
    \vtstr{97}\,
    \ptstr{e}\,
    \vtstr{8B}\,
    \ptstr{s}\,
    \ptstr{/Scheduler.go\$}
    \\
  \hline
\end{alignedat}}$
};
\end{tikzpicture}
\end{table}

\section{Theoretical Foundation -- Dynamic Interleaving}
\label{sec:theory}

We propose the dynamic interleaving to interleave the paths and values
of a set of composite keys $K$, and show how to build the dynamic
interleaving through a recursive partitioning that groups keys based
on the shortest prefixes that distinguish keys from one another.  We
introduce the partitioning in Section \ref{sec:psi} and highlight in
Section \ref{sec:psiprop} the properties that we use to construct the
interleaving. In Section \ref{sec:dynint} we define the dynamic
interleaving with a recursive partitioning and develop a cost model in
Section \ref{sec:costmodel} to analyze the efficiency of
interleavings.

The dynamic interleaving adapts to the specific characteristics of
paths and values, such as common prefixes, varying key lengths,
differing domain sizes, and the skew of the data.  To achieve this we
consider the \emph{discriminative bytes}.

\begin{definition}[Discriminative Byte] \label{def:dsc}
  The discriminative byte $\textsf{dsc}(K,D)$ of keys
  $K$ in dimension $D$ is the first byte for which the keys
  differ in dimension $D$, i.e., $\textsf{dsc}(K,D) =
  |\textsf{lcp}(K,D)| + 1$.  
\end{definition}

\begin{example}
  Table \ref{tab:discbyte} illustrates the position of the
  discriminative bytes for the path and value dimensions for various
  sets of composite keys $K$. Set $\mathsf{K}^{9}=\{\mathsf{k}_9\}$
  contains only a single key. In this case, the discriminative bytes
  are the first position after the end of $\mathsf{k}_9$'s
  byte-strings in the respective dimensions.  For example,
  $\mathsf{k}_9$'s value is eight bytes long, hence the discriminative
  value byte of $\{\mathsf{k}_9\}$ is the ninth byte. $\hfill\Box$
\end{example}

\begin{table}[htbp] \centering
  \caption{Illustration of the discriminative bytes for
    $\mathsf{K}^{1..9}$ from Table \ref{tab:ex2} and various subsets
    of it.}
  \label{tab:discbyte}
  \begin{tabular}{l|cc}
    \multicolumn{1}{l|}{Composite Keys $K$}
      & \multicolumn{1}{c}{$\textsf{dsc}(K,P)$}
      & \multicolumn{1}{c}{$\textsf{dsc}(K,V)$}
      \\ \hline
    $\mathsf{K}^{1..9}$
      & 2
      & 5
      \Tstrut
      \\
    $\mathsf{K}^{1,4,8,9}$
      & 10
      & 7 \\
    $\mathsf{K}^{4,8,9}$
      & 14
      & 7 \\
    $\mathsf{K}^{8,9}$
      & 18
      & 9 \\
    $\mathsf{K}^{9}$
      & 23
      & 9 \\
  \end{tabular}
\end{table}

Discriminative bytes are crucial during query evaluation since at
their positions the search space can be narrowed down.  We alternate
in a round-robin fashion between discriminative path and value bytes
in our interleaving. Each discriminative byte partitions a set of keys
into subsets, which we recursively partition further.

\subsection{$\psi$-Partitioning}
\label{sec:psi}

The $\psi$-partitioning of a set of keys $K$ groups composite keys
together that have the same value at the discriminative byte in
dimension $D$.  Thus, $K$ is split into at most $2^8$ non-empty
partitions, one partition for each value (\texttt{0x00} to
\texttt{0xFF}) of the discriminative byte in dimension $D$.

\begin{definition}[$\psi$-Partitioning] \label{def:partitioning}
  The $\psi$-partitioning of a set of keys $K$ in dimension $D$
  is $\psi(K,D) = \{ K_1, \ldots, K_m \}$ iff

\begin{enumerate} \setlength{\itemsep}{0pt plus 1pt}

    \item (\emph{Correctness}) All keys in a set $K_i$ have the same
      value at $K$'s discriminative byte in dimension $D$:

      \begin{itemize}[leftmargin=10pt]
      \item[--]
        $\forall k,k' \in K_i \left( k.D[\textsf{dsc}(K,D)] =
          k'.D[\textsf{dsc}(K,D)] \right) $
      \end{itemize}

    \item (\emph{Disjointness}) Keys from different sets $K_i \neq
      K_j$ do not have the same value at $K$'s discriminative byte in
      $D$:

      \begin{itemize}[leftmargin=10pt]
      \item[--]
        $\forall k \in K_i, k' \in K_j ($\\
          \hspace*{1cm}$k.D[\textsf{dsc}(K,D)] \neq k'.D[\textsf{dsc}(K,D)] ) $
      \end{itemize}

    \item (\emph{Completeness}) Every key in $K$ is assigned to a set
      $K_i$.  All $K_i$ are non-empty.

      \begin{itemize}[leftmargin=10pt]
      \item[--]
        $K = \bigcup_{1 \leq i \leq m} K_i \wedge \emptyset \notin
          \psi(K,D)$
      \end{itemize}

  \end{enumerate}
\end{definition}

Let $k \in K$ be a composite key.  We write $\psi_k(K, D)$ to denote
the $\psi$-\emph{partitioning of $k$} with respect to $K$ and dimension $D$,
i.e., the partition in $\psi(K, D)$ that contains key $k$.

\begin{example}
  Let $\mathsf{K}^{1..9}$ be the set of composite keys from
  Table~\ref{tab:ex2}.  The $\psi$-partitioning of selected sets of
  keys in dimension $P$ or $V$ is as follows:

  \begin{itemize}\itemsep=0pt
    \item $\psi(\mathsf{K}^{1..9},V)=\{\mathsf{K}^{1,4,8,9},
      \mathsf{K}^{5,6}, \mathsf{K}^{2,3,7}\}$
    \item $\psi(\mathsf{K}^{1,4,8,9},P)=\{\mathsf{K}^{1},
      \mathsf{K}^{4,8,9}\}$
    \item $\psi(\mathsf{K}^{4,8,9},V)=\{\mathsf{K}^{4},
      \mathsf{K}^{8,9}\}$
    \item $\psi(\mathsf{K}^{8,9},P) =\{\mathsf{K}^{8}, \mathsf{K}^{9}\}$
    \item $\psi(\mathsf{K}^{9},V) = \psi(\mathsf{K}^{9},P) =
      \{\mathsf{K}^{9}\}$
  \end{itemize}

  \noindent The $\psi$-partitioning of key $\mathsf{k}_9$
  with respect to sets of keys and dimensions is as follows:

  \begin{itemize}\itemsep=0pt
    \item $\psi_{\mathsf{k}_9}(\mathsf{K}^{1..9}, V) =
      \mathsf{K}^{1,4,8,9}$
    \item \MREV{$\psi_{\mathsf{k}_9}(\mathsf{K}^{1,4,8,9}, P) =
      \mathsf{K}^{4,8,9}$}
    \item \MREV{$\psi_{\mathsf{k}_9}(\mathsf{K}^{4,8,9}, V) =
      \mathsf{K}^{8,9}$}
    \item $\psi_{\mathsf{k}_9}(\mathsf{K}^{9}, V) =
      \psi_{\mathsf{k}_9}(\mathsf{K}^{9}, P) = \mathsf{K}^{9}$.
    $\hfill\Box$
\end{itemize}
\end{example}

\subsection{Properties of the $\psi$-Partitioning}
\label{sec:psiprop}

We work out four key properties of the $\psi$-partitioning.  The first
two properties, \emph{order-preserving} and \emph{prefix-preserving},
allow us to evaluate CAS queries efficiently while the other two
properties, \emph{guaranteed progress} and \emph{monotonicity}, help
us to construct the dynamic interleaving.

\begin{lemma}[Order-Preserving] \label{lemma:op}
  $\psi$-partitioning $\psi(K,D) = \{K_1, \ldots, K_m\}$ is
  order-preserving in dimension $D$, i.e., all keys in set $K_i$ are
  either strictly greater or smaller in dimension $D$ than all keys
  from another set $K_j$:
  \begin{align*}
    \forall 1 \leq i,j \leq m, i \neq j :\,
    &(\forall k \in K_i, \forall k' \in K_j : k.D < k'.D) \,\vee \\
    &(\forall k \in K_i, \forall k' \in K_j : k.D > k'.D)
  \end{align*}
\end{lemma}

\inRegularPaper{ The proofs of all lemmas and theorems can be found in
  the accompanying technical report \cite{KW22}.  }{ All proofs can be
  found in Appendix~\ref{app:proofs}.  }

\begin{example}
  The $\psi$-partitioning $\psi(\mathsf{K}^{1..9}, V)$ is equal to the
  partitions $\{\mathsf{K}^{1,4,8,9},$ $\mathsf{K}^{5,6},$
  $\mathsf{K}^{2,3,7} \}$.  It is order-preserving in dimension $V$.
  The partitions cover the following value ranges (denoted in seconds
  since the Unix epoch):
  \begin{itemize}
    \item {\small {[\vtistr{0x5D\,00\,00\,00},
      \vtistr{0x5D\,FF\,FF\,FF}]}; approx.~06/2019 -- 12/2019}
    \item {\small {[\vtistr{0x5E\,00\,00\,00},
      \vtistr{0x5E\,FF\,FF\,FF}]}; approx.~01/2020 -- 07/2020}
    \item {\small {[\vtistr{0x5F\,00\,00\,00},
      \vtistr{0x5F\,FF\,FF\,FF}]}; approx.~08/2020 -- 12/2021}
  \end{itemize}
  The value predicate $[\text{07/2019}, \text{09/2019})$ only needs to
  consider partition $\mathsf{K}^{1,4,8,9}$, which spans keys from
  June to December, 2019, since partitions do not
  overlap. $\hfill\Box$
\end{example}

\begin{lemma}[Prefix-Preserving] \label{lemma:pp}
  $\psi$-partitioning $\psi(K,D) = \{K_1, \ldots, K_m\}$ is
  prefix-preserving in dimension $D$, i.e., keys in the same set $K_i$
  have a longer common prefix in dimension $D$ than keys from
  different sets $K_i \neq K_j$:
  \begin{align*} \forall 1 \leq i,j \leq m, i \neq j :\,
    & |\textsf{lcp}(K_i,D)| > |\textsf{lcp}(K_i \cup
    K_j,D)|\ \wedge  \\
    & |\textsf{lcp}(K,D)| =   |\textsf{lcp}(K_i \cup
    K_j,D)|
  \end{align*}
\end{lemma}

\begin{example}
  The $\psi$-partitioning
  $\psi(\mathsf{K}^{1..9}, P) = \{\mathsf{K}^{1,4,8,9},$
  $\mathsf{K}^{2,3},$ $\mathsf{K}^{5,6,7}\}$ is prefix-preserving in
  dimension $P$.  For example, $\mathsf{K}^{1,4,8,9}$ has a longer
  common path prefix
  $\textsf{lcp}(\mathsf{K}^{1,4,8,9},P) = \texttt{/Source/}$ than keys
  across partitions, e.g.,
  $\textsf{lcp}(\mathsf{K}^{1,4,8,9} \cup \mathsf{K}^{2,3},P) =
  \texttt{/}$. Query path \texttt{/Source/S*.go} only needs to
  consider partition $\mathsf{K}^{1,4,8,9}$.  $\hfill\Box$
\end{example}

Lemmas \ref{lemma:op} and \ref{lemma:pp} guarantee a total ordering
among the sets $\psi(K,D) = \{K_1, \ldots, K_m\}$. In our RSCAS index
we order the nodes by the value at the discriminative byte such that
range and prefix queries can quickly choose the correct subtree.

The next two properties allow us to efficiently compute the dynamic
interleaving of composite keys.

\begin{lemma}[Guaranteed Progress] \label{lemma:gp}
  Let $K$ be a set of composite keys for which not all keys are equal
  in dimension $D$. $\psi(K,D)$ guarantees progress, i.e., $\psi$
  splits $K$ into at least two sets: $|\psi(K,D)| \ge 2$.
\end{lemma}


Guaranteed progress ensures that each step partitions the data and
when we repeatedly apply $\psi(K,D)$, we eventually narrow a set of
keys down to a single key.  For each set of keys that $\psi(K,D)$
creates, the position of the discriminative byte for dimension $D$
increases.  This property of the $\psi$-partitioning holds since each
set of keys is built based on the discriminative byte and to
$\psi$-partition an existing set of keys we need a discriminative byte
that is positioned further down in the byte-string.  For the alternate
dimension $\overline{D}$, i.e., $\overline{D} = P$ if $D = V$ and
$\overline{D} = V$ if $D = P$, the position of the discriminative byte
remains unchanged or increases.

\begin{lemma}[Monotonicity of Discriminative Bytes]
  \label{lemma:monotonicity}
  Let $K_i$ be one of the partitions of $K$ after partitioning in
  dimension $D$.  In dimension $D$, the position of the discriminative
  byte in $K_i$ is strictly greater than in $K$.  In dimension
  $\overline{D}$, the discriminative byte is equal or greater than in
  $K$, i.e.,
  \begin{align*}
  \begin{split}
    K_i &\in \psi(K,D) \land K_i \subset K
    \Rightarrow\\
    & \textsf{dsc}(K_i, D) > \textsf{dsc}(K, D)
    \land \textsf{dsc}(K_i, \overline{D}) \ge
    \textsf{dsc}(K, \overline{D})
  \end{split}
  \end{align*}
\end{lemma}


\begin{example}
  The discriminative path byte of $\mathsf{K}^{1..9}$ is $2$ while the
  discriminative value byte of $\mathsf{K}^{1..9}$ is $5$ as shown in
  Table~\ref{tab:discbyte}.  For partition $\mathsf{K}^{1,4,8,9}$,
  which is obtained by partitioning $\mathsf{K}^{1..9}$ in the value
  dimension, the discriminative path byte is $10$ while the
  discriminative value byte is $7$.  For partition
  $\mathsf{K}^{4,8,9}$, which is obtained by partitioning
  $\mathsf{K}^{1,4,8,9}$ in the path dimension, the discriminative
  path byte is $14$ while the discriminative value byte is still $7$.
  $\hfill\Box$
\end{example}

Monotonicity guarantees that each time we $\psi$-partition a set $K$
we advance the discriminative byte in at least one dimension.  Thus,
we make progress in at least one dimension when we dynamically
interleave a set of keys.

\begin{figure*}
  \centering
  $\rho(k,K,D) =
    \begin{cases}
      (K, D) \circ \rho(k,\psi_k(K,D),\overline{D})
        & \text{if }
        |K| > \tau \wedge \psi_k(K, D) \subset K \\[1pt]
      \rho(k,K,\overline{D})
        & \text{if }
        |K| > \tau \wedge
        \psi_k(K, D) = K \wedge
        \psi_k(K, \overline{D}) \subset K \\[1pt]
      (K, \bot)
        & \text{otherwise}
  \end{cases}$
  \caption{Definition of partitioning sequence $\rho(k,K,D)$ for a
  threshold $\tau \ge 1$. Operator $\circ$ denotes concatenation,
  e.g., $a \circ b = (a,b)$ and $a \,\circ\, (b,c) = (a,b,c)$.}
  \label{fig:partseq}
\end{figure*}

These four properties of the $\psi$-partitioning are true because we
partition $K$ at its discriminative byte. If we partitioned the data
\emph{before} this byte, we would not make progress and the
monotonicity would be violated, because every byte before the
discriminative byte is part of the longest common prefix.  If we
partitioned the data \emph{after} the discriminative byte, the
partitioning would no longer be order- and prefix-preserving.
Skipping some keys by sampling the set is not an option, as this could
lead to an (incorrect) partitioning using a byte located after the
actual discriminative byte.

\begin{example}
  $\mathsf{K}^{1..9}$'s discriminative value byte is byte five.  If we
  partitioned $\mathsf{K}^{1..9}$ at value byte four we would get
  $\{ \mathsf{K}^{1..9} \}$ and there is no progress since all keys
  have \vtistr{0x00} at value byte four.  The discriminative path and
  value bytes would remain unchanged.  If we partitioned
  $\mathsf{K}^{1..9}$ at value byte six we would get
  $\{ \mathsf{K}^{1,4,8,9}, \mathsf{K}^{2,3,6,7}, \mathsf{K}^5 \}$,
  which is neither order- nor prefix-preserving in $V$.  Consider keys
  $\mathsf{k}_3,\mathsf{k}_6 \in \mathsf{K}^{2,3,6,7}$ and
  $\mathsf{k}_5 \in \mathsf{K}^{5}$.  The partitioning is not
  order-preserving in $V$ since
  $\mathsf{k}_6.V < \mathsf{k}_5.V < \mathsf{k}_3.V$.  The
  partitioning is not prefix-preserving in $V$ since the longest
  common value prefix in $\mathsf{K}^{2,3,6,7}$ is
  \vtistr{00\,00\,00\,00}, which is not longer than the longest common
  value prefix of keys from different partitions since
  $\textsf{lcp}(\mathsf{K}^{2,3,6,7} \cup \mathsf{K}^{5}, V)=$
  $ \vtistr{00\,00\,00\,00}$.  $\hfill\Box$
\end{example}

\subsection{Dynamic Interleaving}
\label{sec:dynint}

To compute the dynamic interleaving of a composite key $k \in K$ we
recursively $\psi$-partition $K$ while alternating between dimension
$V$ and $P$.  In each step, we interleave a part of $k.P$ with a
part of $k.V$.  The recursive $\psi$-partitioning yields a
partitioning sequence $(K_1, D_1),$ $\ldots,$ $(K_n, D_n)$ for key $k$
with $K_1 \supset K_2 \supset \dots \supset K_n$.  We start with
$K_1 = K$ and $D_1 = V$.  Next, $K_2 = \psi_k(K_1, V)$ and
$D_2 = \overline{D}_1 = P$.  We continue with the general scheme
$K_{i+1} = \psi_k(K_i, D_i)$ and $D_{i+1} = \overline{D}_i$.  This
continues until we reach a set $K_n$ that contains at most $\tau$
keys, where $\tau$ is a threshold (explained later).  The recursive
$\psi$-partitioning alternates between dimensions $V$ and $P$ until we
run out of discriminative bytes in one dimension, which means
$\psi_k(K_i, D) = K_i$.  From then on, we can only $\psi$-partition in
dimension $\overline{D}$ until we run out of discriminative bytes in
this dimension as well, that is
$\psi_k(K_i, \overline{D}) = \psi_k(K_i, D) = K_i$, or we reach a
$K_n$ that contains at most $\tau$ keys.  The partitioning sequence is
finite due to the monotonicity of the $\psi$-partitioning (see
Lemma~\ref{lemma:monotonicity}), which guarantees that we make
progress in each step in at least one dimension.

\begin{definition}[Partitioning Sequence] \label{def:partseq}
  The partitioning sequence
  $\rho(k,K,D) = ((K_1, D_1), \ldots, (K_n, D_n))$ of a composite key
  $k \in K$ is the recursive $\psi$-partitioning of the sets to which
  $k$ belongs.  The pair $(K_i, D_i)$ denotes the partitioning of
  $K_i$ in dimension $D_i$.  The partitioning stops when $K_n$
  contains at most $\tau$ keys or $K_n$ cannot be further
  $\psi$-partitioned in any dimension ($K_n.D = \bot$ in this case).
  $\rho(k,K,D)$ is defined in Figure \ref{fig:partseq}.
\end{definition}

\begin{example} \label{ex:partseq}
  Below we illustrate the step-by-step expansion of
  $\rho(\mathsf{k}_9,\mathsf{K}^{1..9},V)$ to get $\mathsf{k}_9$'s
  partitioning sequence. We set $\tau = 2$.
  \begin{align*}
    \rho&(\mathsf{k}_9,\mathsf{K}^{1..9},V) \\
      & = (\mathsf{K}^{1..9},V)
          \circ \rho(\mathsf{k}_9, \mathsf{K}^{1,4,8,9},P) \\
      & = (\mathsf{K}^{1..9},V)
          \circ (\mathsf{K}^{1,4,8,9},P)
          \circ \rho(\mathsf{k}_9, \mathsf{K}^{4,8,9},V) \\
      & = (\mathsf{K}^{1..9},V)
          \circ (\mathsf{K}^{1,4,8,9},P)
          \circ (\mathsf{K}^{4,8,9},V)
          \circ \rho(\mathsf{k}_9, \mathsf{K}^{8,9},P) \\
      & = (\mathsf{K}^{1..9},V)
          \circ (\mathsf{K}^{1,4,8,9},P)
          \circ (\mathsf{K}^{4,8,9},V)
          \circ (\mathsf{K}^{8,9},\bot)
  \end{align*}
  \noindent Note the alternating partitioning in, respectively, $V$
  and $P$.  We only deviate from this if partitioning in one of the
  dimensions is not possible. Had we set $\tau = 1$,
  $\mathsf{K}^{8,9}$ would be partitioned once more in the path
  dimension. $\hfill\Box$
\end{example}

To compute the full dynamic interleaving of a key $k$ we set $\tau=1$
and continue until the final set $K_n$ contains a single key (i.e, key
$k$).  To interleave only a prefix of $k$ and keep a suffix
non-interleaved we increase $\tau$.  Increasing $\tau$ stops the
partitioning earlier and speeds up the computation.  An index
structure that uses dynamic interleaving can tune $\tau$ to trade the
time it takes to build the index and to query it.  In Section
\ref{sec:rscas} we introduce a memory-optimized and a disk-optimized
version of our RSCAS index.  They use different values of $\tau$ to
adapt to the underlying storage.

We determine the dynamic interleaving $I_{\text{DY}}(k,K)$ of a key
$k \in K$ via $k$'s partitioning sequence $\rho$.  For each element in
$\rho$, we generate a tuple with strings $s_P$ and $s_V$ and the
partitioning dimension of the element. The strings $s_P$ and $s_V$ are
composed of substrings of $k.P$ and $k.V$, ranging from the previous
discriminative byte up to, but excluding, the current discriminative
byte in the respective dimension. The order of $s_P$ and $s_V$ in a
tuple depends on the dimension used in the previous step: the
dimension that has been chosen for the partitioning comes first.
Formally, this is defined as follows:

\begin{definition}[Dynamic Interleaving] \label{def:dynint} %
  Let $k \in K$ be a composite key and let
  $\rho(k, K, V) = ((K_1, D_1),$ $\ldots,$ $(K_n, D_n))$ be the
  partitioning sequence of $k$.  The dynamic interleaving
  $I_{\text{DY}}(k,K) = (t_1, \ldots, t_n, t_{n+1})$ of $k$ is a
  sequence of tuples $t_i$, where $t_i = (s_P, s_V, D)$ if
  $D_{i-1} = P$ and $t_i = (s_V, s_P, D)$ if $D_{i-1} = V$.  The
  tuples $t_i$, $1 \leq i \leq n$, are determined as follows:
  \begin{align*}
    t_i.s_P & = k.P[\textsf{dsc}(K_{i-1},P),\textsf{dsc}(K_i,P)-1] \\
    t_i.s_V & = k.V[\textsf{dsc}(K_{i-1},V),\textsf{dsc}(K_i,V)-1] \\
    t_i.D & = D_i
  \end{align*}
  To correctly handle the first tuple we define $\textsf{dsc}(K_0,V) =
  1$, $\textsf{dsc}(K_0,P) = 1$ and $D_0 = V$.  The last tuple
  $t_{n+1} = (s_P,s_V,R)$ stores the non-interleaved suffixes along with revision
  $k.R$: \\[5pt]
  $\begin{aligned}[b]
    t_{n+1}.s_P & = k.P[\textsf{dsc}(K_{n},P),|k.P|] \\
    t_{n+1}.s_V & = k.V[\textsf{dsc}(K_{n},V),|k.P|] \\
    t_{n+1}.R & = k.R
  \end{aligned}\hfill\Box$
\end{definition}

\begin{example}
  We compute the tuples for the dynamic interleaving
  $I_{\text{DY}}(\mathsf{k}_9, \mathsf{K}^{1..9}) = (\mathsf{t}_1,
  \ldots, \mathsf{t}_5)$ of key $\mathsf{k}_9$ using the partitioning
  sequence $\rho(\mathsf{k}_9, \mathsf{K}^{1..9}, V)$ from Example
  \ref{ex:partseq}. The necessary discriminative path and value bytes
  can be found in Table \ref{tab:discbyte}.  Table \ref{tab:exdyn}
  shows the details of each tuple of $\mathsf{k}_9$'s dynamic
  interleaving with respect to $\mathsf{K}^{1..9}$.  The final dynamic
  interleavings of all keys from Table \ref{tab:ex2} are displayed in
  Table \ref{tab:dynints}.  We highlight in bold the values of the
  discriminative bytes at which the paths and values are interleaved,
  e.g., for key $\mathsf{k}_9$ these are bytes $\vtbstr{5D}$,
  $\ptbstr{S}$, and $\vtbstr{97}$. $\hfill\Box$
\end{example}

\begin{table}[htbp]\centering
  \caption{Computing $I_{\text{DY}}(\mathsf{k}_9, \mathsf{K}^{1..9})$.
    }
  \label{tab:exdyn}
  {\footnotesize
  \setlength{\tabcolsep}{4pt}
  \begin{tabular}{c|llc}
    {\normalsize $t$}
      & \multicolumn{1}{c}{\normalsize $s_V$}
      & \multicolumn{1}{c}{\normalsize $s_P$}
      & \multicolumn{1}{c}{\normalsize $D$} \\
    \hline
    $\mathsf{t}_1$
      & $\mathsf{k}_9.V[1,4] = \vtstr{00\,00\,00\,00}$
      & $\mathsf{k}_9.P[1,1] = \ptstr{/}$
      & $V$ \\
    $\mathsf{t}_2$
      & $\mathsf{k}_9.V[5,6] = \vtstr{5D\,A8}$
      & $\mathsf{k}_9.P[2,9] = \ptstr{Sources/}$
      & $P$ \\
    $\mathsf{t}_3$
      & $\mathsf{k}_9.V[7,6] = \vtstr{$\epsilon$}$
      & $\mathsf{k}_9.P[10,13] = \ptstr{Sche}$
      & $V$ \\
    $\mathsf{t}_4$
      & $\mathsf{k}_9.V[7,8] = \vtstr{97\,8B}$
      & $\mathsf{k}_9.P[14,17] = \ptstr{dule}$
      & $\bot$ \\
    $\mathsf{t}_5$
      & $\mathsf{k}_9.V[9,8] = \vtstr{$\epsilon$}$
      & $\mathsf{k}_9.P[18,22] = \ptstr{r.go\$}$
  \end{tabular}}
\end{table}

\begin{table*}[htb]
\caption{The dynamic interleaving of the composite keys
in $\mathsf{K}^{1..9}$. The values
at the discriminative bytes are written in bold.}
\label{tab:dynints}
\centering
{\normalsize
\begin{tabular}{l|l}
  \multicolumn{1}{c|}{\normalsize $k$}
    & \multicolumn{1}{c}{\normalsize Dynamic Interleaving
      $I_{\text{DY}}(k,\mathsf{K}^{1..9})$}
  \\ \hline
  $\mathsf{k}_1$
    &$((\vtstr{00\,00\,00\,00}, \ptstr{/}, V)$,\,%
      $(\vtbstr{5D}\,\vtstr{A8}, \ptstr{Sources/}, P)$,\,%
      $(\ptbstr{M}\ptstr{ap.go\$}, \vtstr{94\,2A}, \bot)$,\,
      $(\ptstr{$\epsilon$}, \vtstr{$\epsilon$}, \mathsf{r}_1))$ \\
  $\mathsf{k}_4$
    &$((\vtstr{00\,00\,00\,00}, \ptstr{/}, V)$,\,%
      $(\vtbstr{5D}\,\vtstr{A8}, \ptstr{Sources/}, P)$,\,%
      $(\ptbstr{S}\ptstr{che}, \vtstr{$\epsilon$}, V)$,\,%
      $(\vtbstr{94}\,\vtstr{8C}, \ptstr{ma.go\$}, \bot))$,\,%
      $(\ptstr{$\epsilon$}, \vtstr{$\epsilon$}, \mathsf{r}_3))$ \\
  $\mathsf{k}_8$
    &$((\vtstr{00\,00\,00\,00}, \ptstr{/}, V)$,\,%
      $(\vtbstr{5D}\,\vtstr{A8}, \ptstr{Sources/}, P)$,\,%
      $(\ptbstr{S}\ptstr{che}, \vtstr{$\epsilon$}, V)$,\,%
      $(\vtbstr{97}\,\vtstr{8B}, \ptstr{dule}, \bot)$,\,%
      $(\ptstr{.go\$}, \vtstr{$\epsilon$}, \mathsf{r}_7))$ \\
  $\mathsf{k}_9$
    &$((\vtstr{00\,00\,00\,00}, \ptstr{/}, V)$,\,%
      $(\vtbstr{5D}\,\vtstr{A8}, \ptstr{Sources/}, P)$,\,%
      $(\ptbstr{S}\ptstr{che}, \vtstr{$\epsilon$}, V)$,\,%
      $(\vtbstr{97}\,\vtstr{8B}, \ptstr{dule}, \bot)$,\,%
      $(\ptstr{r.go\$}, \vtstr{$\epsilon$}, \mathsf{r}_7))$ \\
  $\mathsf{k}_5$
    &$((\vtstr{00\,00\,00\,00}, \ptstr{/}, V)$,\,%
      $(\vtbstr{5E}, \ptstr{fs/ext}, \bot)$,\,%
      $(\ptbstr{3}\ptstr{/inode.c\$}, \vtstr{F2\,9C\,59}, \mathsf{r}_4))$ \\
  $\mathsf{k}_6$
    &$((\vtstr{00\,00\,00\,00}, \ptstr{/}, V)$,\,%
      $(\vtbstr{5E}, \ptstr{fs/ext}, \bot)$,\,%
      $(\ptbstr{4}\ptstr{/inode.h\$}, \vtstr{BD\,23\,C2}, \mathsf{r}_5))$ \\
  $\mathsf{k}_2$
    &$((\vtstr{00\,00\,00\,00}, \ptstr{/}, V)$,\,%
      $(\vtbstr{5F}\,\vtstr{BD}, \ptstr{$\epsilon$}, P)$,\,%
      $(\ptbstr{c}\ptstr{rypto/ecc.}, \vtstr{8D\,C4}, \bot)$,\,%
      $(\ptstr{h\$}, \vtstr{$\epsilon$}, \mathsf{r}_2))$ \\
  $\mathsf{k}_3$
    &$((\vtstr{00\,00\,00\,00}, \ptstr{/}, V)$,\,%
      $(\vtbstr{5F}\,\vtstr{BD}, \ptstr{$\epsilon$}, P)$,\,%
      $(\ptbstr{c}\ptstr{rypto/ecc.}, \vtstr{8D\,C4}, \bot)$,\,%
      $(\ptstr{c\$}, \vtstr{$\epsilon$}, \mathsf{r}_2))$ \\
  $\mathsf{k}_7$
    &$((\vtstr{00\,00\,00\,00}, \ptstr{/}, V)$,\,%
      $(\vtbstr{5F}\,\vtstr{BD}, \ptstr{$\epsilon$}, P)$,\,%
      $(\ptbstr{f}\ptstr{s/ext4/inode.c\$}, \vtstr{3D\,5A}, \bot)$,\,%
      $(\ptstr{$\epsilon$}, \vtstr{$\epsilon$}, \mathsf{r}_6))$ \\
\end{tabular}}
\end{table*}

Unlike static interleavings $I(k)$ that interleave a key $k$ in
isolation, the dynamic interleaving $I_{\text{DY}}(k,K)$ of $k$
depends on the set of all keys $K$ to adapt to the data. The result is
a well-balanced interleaving (compare Tables \ref{tab:interleaving}
and \ref{tab:dynints}).

\inRegularPaper{}{\MREV{In Section \ref{sec:algorithms} we propose
    efficient algorithms to dynamically interleave composite keys and
    analyze them for different key distributions.}}

\subsection{Efficiency of Interleavings}
\label{sec:costmodel}

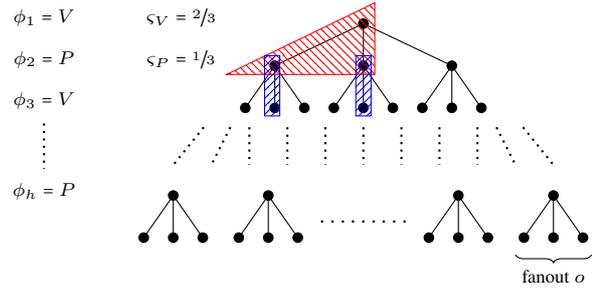
\begin{figure}[htb]
\centering
\begin{tikzpicture}
  \node at (0,0) {
  \begin{forest}
    casindex,
    for tree={
      parent anchor=north,
      l=0mm,
      s sep+=5pt,
    },
    [{},name=root,
      [{},bbnode,
        [{},bbnode,
        ]
        [{},bbnode,
        ]
        [{},bbnode,
        ]
      ]
      [{},bbnode,
        [{},bbnode,
        ]
        [{},bbnode,
        ]
        [{},bbnode,
        ]
      ]
      [{},bbnode,
        [{},bbnode,
        ]
        [{},bbnode,
        ]
        [{},bbnode,
        ]
      ]
    ]
    \fill[black] (root.parent anchor) circle[radius=2pt];
  \end{forest}
  };
  \foreach \x in {-2.5,-1.25,1.25,2.5} {
    \node (leaf-\x) at (\x,-2) {
    \begin{forest}
      searchindex,
      [{},name=root,
        []
        []
        []
      ]
      \fill[black] (root.parent anchor) circle[radius=2pt];
    \end{forest}
    };
  }
  %
  %
  \node at (0,-2) {$\ldots\ldots\ldots$};
  \draw[thick,dotted] (-2.1,-0.75)  -- (-2.5,-1.25);
  \draw[thick,dotted] (-1.75,-0.75) -- (-2,-1.25);
  \draw[thick,dotted] (+1.75,-0.75) -- (+2,-1.25);
  \draw[thick,dotted] (+2.1,-0.75)  -- (+2.5,-1.25);
  \foreach \x in {-1.5,-1.0,-0.5,0,0.5,1,1.5} {
    \draw[thick,dotted] (\x,-0.75) -- (\x,-1.25);
  }
  \draw[decorate,decoration={brace,amplitude=3pt,raise=4pt,mirror},yshift=0pt]
    (2.0,-2.3) -- (3.0,-2.3)
    node[midway,yshift=-12pt,font={\scriptsize}]
    {fanout $o$};
  \node[font=\scriptsize] at (-4.2,0.7)  {$\phi_1 = V$};
  \node[font=\scriptsize] at (-4.2,0.15) {$\phi_2 = P$};
  \node[font=\scriptsize] at (-4.2,-0.4) {$\phi_3 = V$};
  \node[font=\scriptsize] at (-4.2,-1.6) {$\phi_h = P$};
  \draw[thick,dotted] (-4.2,-0.7)  -- (-4.2,-1.3);
  %
  %
  \draw[draw=red,pattern=north west lines, pattern color=red]
    (0.15,0.9) -- (-1.8,-0.05) -- (0.15,-0.05) -- cycle;
  \node[font=\scriptsize] at (-2.4,0.7) {$\varsigma_V = \sfrac{2}{3}$};
  %
  \fill[draw=blue,pattern=north east lines, pattern color=blue]
    (-1.30,+0.2) -- (-1.30,-0.6) -- (-1.10,-0.6) -- (-1.10,0.2) -- cycle;
  \fill[draw=blue,pattern=north east lines, pattern color=blue]
    (0.1,+0.2) -- (0.1,-0.6) -- (-0.1,-0.6) -- (-0.1,0.2) -- cycle;
  \node[font=\scriptsize] at (-2.4,0.15) {$\varsigma_P = \sfrac{1}{3}$};
\end{tikzpicture}
\caption{The search structure in our cost model is a complete tree of
height $h$ and fanout $o$.}
\label{fig:searchscheme}
\end{figure}

We propose a cost model to measure the efficiency of interleavings
that organize the interleaved keys in a tree-like search structure.
Each node represents the $\psi$-partitioning of the composite keys by
either path or value, and the node branches for each different value
of a discriminative path or value byte.  We simplify the cost model by
assuming that the search structure is a complete tree with fanout $o$
where every root-to-leaf path contains $h$ edges ($h$ is the height).
Further, we assume that all nodes on one level represent a
partitioning in the same dimension $\phi_i \in \{P,V\}$ and we use a
vector $\phi(\phi_1, \ldots, \phi_h)$ to specify the partitioning
dimension on each level.  We assume that the number of $P$s and $V$s
in each $\phi$ are equal.  Figure~\ref{fig:searchscheme} visualizes
this scheme.

To answer a query we start at the root and traverse the search
structure to determine the answer set. In the case of range queries,
more than one branch must be followed.  A search follows a fraction of
the outgoing branches $o$ originating at this node.  We call this the
selectivity of a node (or just selectivity).  We assume that every
path node has a selectivity of $\varsigma_P$ and every value node has
a selectivity of $\varsigma_V$.  The cost $\widehat{C}$ of a search,
measured in the number of visited nodes during the search, is as
follows:
\begin{equation*}
  \widehat{C}(o,h,\phi,\varsigma_P,\varsigma_V) = 1 + \sum_{l=1}^{h}
  \prod_{i=1}^{l} (o \cdot \varsigma_{\phi_i})
\end{equation*}

\MREV{If a workload is well-known and consists of a small set of
  specific queries, it is highly likely that an index adapted to this
  workload will outperform RSCAS. For instance, if
  $\varsigma_V \ll \varsigma_P$ for all queries, then a VP-index shows
  better performance than an RSCAS-index. However, it performs badly
  for queries deviating from that workload
  ($\varsigma_V > \varsigma_P$).}  Our goal is an access method that
can deal with a wide range of queries in a dynamic environment in a
robust way, i.e., avoiding a bad performance for any particular query
type.  This is motivated by the fact that modern data analytics
utilizes a large number of ad-hoc queries to do exploratory
analysis. For example, in the context of building a robust
partitioning for ad-hoc query workloads, Shanbhag et
al.~\cite{Shanbhag17} found that after analyzing the first 80\% of a
real-world workload the remaining 20\% still contained 57\% completely
new queries.  We aim for a good average performance across all
queries.

\begin{definition}[Robustness] \label{def:robustness}
  A CAS-index is \emph{robust} if it optimizes the average performance
  and minimizes the variability over all queries.
\end{definition}

State-of-the-art CAS-indexes are not robust because they favor either
path or value predicates.  As a result they show a very good
performance for one type of query but run into problems for other
types of queries.  To illustrate this problem we define the notion of
\emph{complementary queries}.

\begin{definition}[Complementary Query] %
  Given a query $Q = (\varsigma_P, \varsigma_V)$ with path selectivity
  $\varsigma_P$ and value selectivity $\varsigma_V$, there is a
  \emph{complementary query} $Q' = (\varsigma'_P, \varsigma'_V)$ with
  path selectivity $\varsigma'_P = \varsigma_V$ and value selectivity
  $\varsigma'_V = \varsigma_P$
\end{definition}

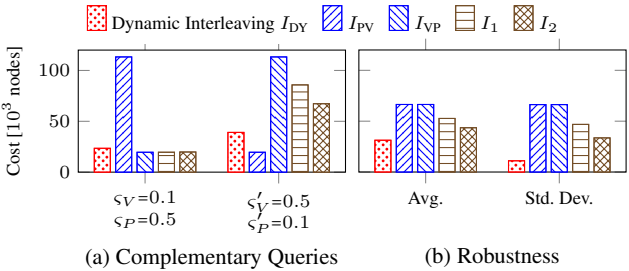
\begin{figure}[htb]
\begin{tikzpicture}
  \begin{groupplot}[
    height=32mm,
    width=145pt,
    ticklabel style={font=\scriptsize},
    xlabel near ticks,
    ylabel near ticks,
    ylabel/.append style={font=\scriptsize},
    scaled y ticks=base 10:-3,
    ytick scale label code/.code={},
    symbolic x coords={k1,k2},
    xtick=data,
    xticklabel style={
      align=center,
    },
    group style={
      group size=2 by 1,
      horizontal sep=5pt,
      yticklabels at=edge left,
      ylabels at=edge left,
    },
    ybar,
    enlarge x limits=0.5,
    bar cycle list,
  ]
  \nextgroupplot[
    bar width=6pt,
    ylabel={Cost [$10^3$ nodes]},
    xticklabels={
      {$\varsigma_V{=}0.1$\\$\varsigma_P{=}0.5$},
      {$\varsigma'_V{=}0.5$\\$\varsigma'_P{=}0.1$}
    },
    legend style={
      at={(0.0,1.05)},
      anchor=south west,
      legend columns=-1,
      draw=none,
      /tikz/every even column/.append style={column sep=3pt},
    },
    xlabel={\small (a) Complementary Queries},
    ymax=120000,
    ymin=0,
  ]
  \addplot[idy]  coordinates {(k1, 23436) (k2, 39060)};
  \addplot[ipv]  coordinates {(k1,113280) (k2, 19536)};
  \addplot[ivp]  coordinates {(k1, 19536) (k2,113280)};
  \addplot[ione] coordinates {(k1, 19564) (k2, 85780)};
  \addplot[itwo] coordinates {(k1, 19808) (k2, 67280)};
  \legend{Dynamic Interleaving $I_\text{DY}$, $I_{\text{PV}}$,
    $I_{\text{VP}}$, $I_1$, $I_2$}
  \nextgroupplot[
    bar width=6pt,
    xticklabels={Avg., Std.~Dev.},
    xlabel={\small (b) Robustness},
    x label style={yshift=-9pt},
    ymax=120000,
    ymin=0,
  ]
  \addplot[idy]  coordinates {(k1, 31248) (k2, 11047)};
  \addplot[ipv]  coordinates {(k1, 66408) (k2, 66287)};
  \addplot[ivp]  coordinates {(k1, 66408) (k2, 66287)};
  \addplot[ione] coordinates {(k1, 52672) (k2, 46821)};
  \addplot[itwo] coordinates {(k1, 43544) (k2, 33567)};
  \end{groupplot}
\end{tikzpicture}
\caption{Robustness of dynamic interleaving.}
\label{fig:interleavecost}
\end{figure}

\begin{example} \label{ex:cost}
  Figure~\ref{fig:interleavecost}a shows the costs for a query $Q$ and
  its complementary query $Q'$ for different interleavings in terms of
  the number of visited nodes during the search.  We assume parameters
  $o=10$ and $h=12$ for the search structure and a dynamic
  interleaving $I_{\text{DY}}$ with $\tau = 1$.  $I_{\text{PV}}$
  stands for path-value concatenation with $\phi_i = P$ for
  $1 \leq i \leq 6$ and $\phi_i = V$ for $7 \leq i \leq 12$.
  $I_{\text{VP}}$ is a value-path concatenation (with an inverse
  $\phi$ compared to $I_{\text{PV}}$).  We also consider two
  additional permutations: $I_1$ uses a vector
  $\phi = (V,V,V,V,P,V,P,V,P,P,P,P)$ and $I_2$ a vector equal to
  $(V,V,V,P,P,V,P,V,V,P,P,P)$.  They resemble byte-wise interleavings,
  which usually exhibit irregular alternation patterns with a
  clustering of, respectively, discriminative path and value bytes.
  Figure \ref{fig:interleavecost}b shows the average costs and the
  standard deviation.  The numbers demonstrate the robustness of our
  dynamic interleaving: it performs best in terms of average costs and
  standard deviation.
\end{example}

In the previous example we used our cost model to show that a
perfectly alternating interleaving exhibits the best overall
performance and standard deviation when evaluating complementary
queries.  We prove that this is always the case.

\begin{theorem}\label{theo:avg}
  Consider a query $Q$ with selectivities $\varsigma_P$ and
  $\varsigma_V$ and its complementary query $Q'$ with selectivities
  $\varsigma'_P = \varsigma_V$ and $\varsigma'_V = \varsigma_P$.
  There is no interleaving that on average performs better than the
  dynamic interleaving with a perfectly alternating vector
  $\phi_{\text{DY}}$, i.e.,
  $\forall \phi:
  \widehat{C}(o,h,\phi_{\text{DY}},\varsigma_P,\varsigma_V) +
  \widehat{C}(o,h,\phi_{\text{DY}},\varsigma'_P,\varsigma'_V) \leq
  \widehat{C}(o,h,\phi,\varsigma_P,\varsigma_V) +
  \widehat{C}(o,h,\phi,\varsigma'_P,\varsigma'_V)$.%
\end{theorem}

Theorem \ref{theo:avg} shows that the dynamic interleaving has the
best query performance for complementary queries.  It follows that for
any set of complementary queries $\mathbf{Q}$, the dynamic
interleaving has the best performance.

\begin{theorem}\label{theo:robustness}
  Let $\mathbf{Q}$ be a set of complementary queries, i.e.,
  $(\varsigma_P,\varsigma_V) \in \mathbf{Q} \Leftrightarrow
  (\varsigma_V,\varsigma_P) \in \mathbf{Q}$.  There is no interleaving
  $\phi$ that in total performs better than the dynamic interleaving
  over all queries $\mathbf{Q}$, i.e.,
  \begin{align*}
    \forall \phi:
     \sum_{(\varsigma_P, \varsigma_V) \in \mathbf{Q}}
      &\widehat{C}(o,h,\phi_{\text{DY}},\varsigma_P,\varsigma_V)
      \\ & \leq
     \sum_{(\varsigma_P, \varsigma_V) \in \mathbf{Q}}
      \widehat{C}(o,h,\phi,\varsigma_P,\varsigma_V)
  \end{align*}
\end{theorem}

This also holds for the set of all queries, since for every query
there exists a complementary query.  Thus, the dynamic interleaving
optimizes the average performance over all queries and, as a result, a
CAS index that uses dynamic interleaving is robust.

\begin{corollary}
  Let $\mathbf{Q} = \{ (\varsigma_P, \varsigma_V) \mid 0 \leq
  \varsigma_P, \varsigma_V \leq 1 \}$ be the set of all possible
  queries. There is no interleaving $\phi$ that in total performs
  better than the dynamic interleaving $\phi_{\text{DY}}$ over all
  queries $\mathbf{Q}$.
\end{corollary}

We now turn to the variability of the search costs and show that they
are minimal for dynamic interleavings.

\begin{theorem}\label{theo:dev}
  Given a query $Q$ (with $\varsigma_P$ and $\varsigma_V$) and its
  complementary query $Q'$ (with $\varsigma'_P = \varsigma_V$ and
  $\varsigma'_V = \varsigma_P$), there is no interleaving that has a smaller
  variability than the dynamic interleaving with a perfectly alternating vector
  $\phi_{\text{DY}}$, i.e., $\forall \phi:
  |\widehat{C}(o,h,\phi_{\text{DY}},\varsigma_P,\varsigma_V) -
  \widehat{C}(o,h,\phi_{\text{DY}},\varsigma'_P,\varsigma'_V)| \leq$ \\
  $|\widehat{C}(o,h,\phi,\varsigma_P,\varsigma_V) -
  \widehat{C}(o,h,\phi,\varsigma'_P,\varsigma'_V)|$.%
\end{theorem}

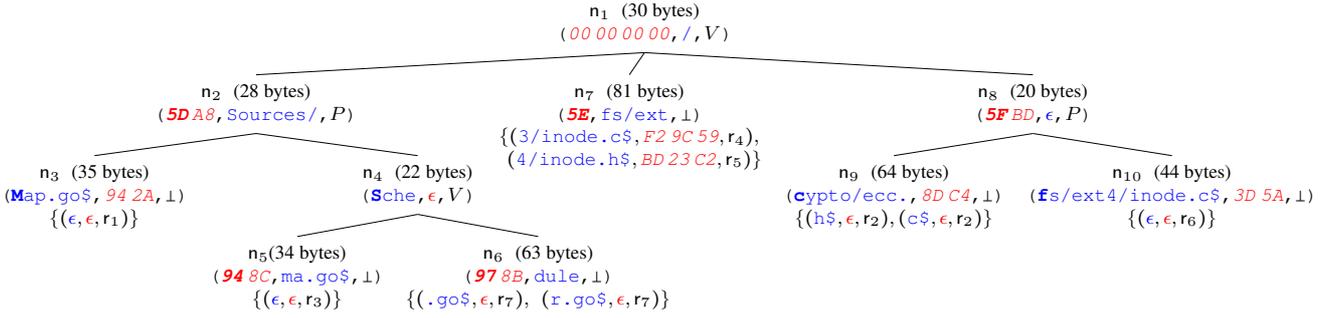
\begin{figure*}[htbp]\centering
\scalebox{0.85}{
\begin{forest}
  casindex,
  for tree={
    font={\ttfamily \small},
  },
  for tree={
    s sep+=1mm,
  },
  [{$\mathsf{n}_{1}$ \textnormal{(30 bytes)}}\\
   {(\vtstr{00\,00\,00\,00},\pstr{/},$V$)},name=n1
    [{$\mathsf{n}_{2}$ \textnormal{(28 bytes)}}\\
     {(\vtbstr{5D}\,\vtstr{A8},\pstr{Sources/},$P$)},name=n2
      [{{$\mathsf{n}_{3}$ \textnormal{(35 bytes)}}\\
       (\ptbstr{M}\pstr{ap.go\$},\vtstr{94\,2A},$\bot$)\\
       $\{(\ptstr{$\epsilon$},\vtstr{$\epsilon$},\mathsf{r}_{1})\}$},name=n3
      ]
      [{{$\mathsf{n}_{4}$ \textnormal{(22 bytes)}}\\
       (\ptbstr{S}\pstr{che},\vtstr{$\epsilon$},$V$)},name=n4
        [{{$\mathsf{n}_{5} \textnormal{(34 bytes)}$}\\
          (\vtbstr{94}\,\vtstr{8C},\pstr{ma.go\$},$\bot$)\\
          $\{(\ptstr{$\epsilon$},\vtstr{$\epsilon$},\mathsf{r}_{3})\}$},name=n5
        ]
        [{$\mathsf{n}_{6}$ \textnormal{(63 bytes)}}\\
         {(\vtbstr{97}\,\vtstr{8B},\pstr{dule},$\bot$)}\\
         {$\{(\ptstr{.go\$}, \vtstr{$\epsilon$}, \mathsf{r}_{7}),$}
         {$(\ptstr{r.go\$}, \vtstr{$\epsilon$}, \mathsf{r}_{7})\}$},name=n6
        ]
      ]
    ]
    [{$\mathsf{n}_{7}$ \textnormal{(81 bytes)}}\\
     {(\vtbstr{5E},\pstr{fs/ext},$\bot$)}\\
     {$\{(\ptstr{3/inode.c\$},\vtstr{F2\,9C\,59},\mathsf{r}_{4}),$}\\
     {\phantom{\{}$(\ptstr{4/inode.h\$},\vtstr{BD\,23\,C2},\mathsf{r}_{5})\}$},name=n7
    ]
    [{$\mathsf{n}_{8}$ \textnormal{(20 bytes)}}\\
     {(\vtbstr{5F}\,\vtstr{BD},\pstr{$\epsilon$},$P$)},name=n8
      [{$\mathsf{n}_{9}$ \textnormal{(64 bytes)}}\\
       {(\ptbstr{c}\pstr{ypto/ecc.},\vtstr{8D\,C4},$\bot$)}\\
       {$\{(\ptstr{h\$},\vtstr{$\epsilon$},\mathsf{r}_{2}),(\ptstr{c\$},\vtstr{$\epsilon$},\mathsf{r}_{2})\}$},name=n9
      ]
      [{$\mathsf{n}_{10}$ \textnormal{(44 bytes)}}\\
       {(\ptbstr{f}\pstr{s/ext4/inode.c\$},\vtstr{3D\,5A},$\bot$)}\\
       {$\{(\ptstr{$\epsilon$}, \vtstr{$\epsilon$}, \mathsf{r}_{6})\}$},name=n10
      ]
    ]
  ]
\end{forest}
}
\caption{The RSCAS trie for the composite keys $\mathsf{K}^{1..9}$.}
\label{fig:solution}
\end{figure*}

Similar to the results for the average performance,
Theorem~\ref{theo:dev} can be generalized to the set of all queries.

Note that in practice the search structure is not a complete tree and
the fraction $\varsigma_P$ and $\varsigma_V$ of children that are
traversed at each node is not constant.  We previously evaluated the
cost model experimentally on real-world datasets \cite{KW20} and
showed that the estimated and true cost of a query are off by a factor
of two on average, which is a good estimate for the cost of a query.

\section{Robust and Scalable CAS (RSCAS) Index}
\label{sec:rscas}

Data-intensive applications require indexing techniques that make it
possible to efficiently index, \MREV{insert}, and query large amounts of
data.  The SWH archive, for example, stores billions of revisions and
every day millions of revisions are crawled from popular software
forges.  We propose the Robust and Scalable Content-And-Structure
(RSCAS) index to provide support for querying and updating the content
and structure of big hierarchical data.  For robustness, the RSCAS
index uses our dynamic interleaving to integrate the paths and values
of composite keys in a trie structure. For scalability, RSCAS
implements log-structured merge trees (LSM trees) that combine a
memory-optimized trie with a series of disk-optimized tries (see
Figure~\ref{fig:logmethod}).

\subsection{Structure of an RSCAS Trie}

RSCAS tries support CAS queries with range and prefix searches.  Each
node $n$ in an RSCAS trie includes a dimension $n.D$, a path substring
$n.s_P$, and a value substring $n.s_V$.  They correspond to fields
$t.D$, $t.s_P$ and $t.s_V$ in the dynamic interleaving of a key (see
Definition \ref{def:dynint}).  Substrings $n.s_P$ and $n.s_V$ are
variable-length strings.  Dimension $n.D$ is $P$ or $V$ for inner
nodes and $\bot$ for leaf nodes. Leaf nodes additionally store a set
of suffixes, denoted by $n.\textsf{suffixes}$. This set contains
non-interleaved path and value suffixes along with references to data
items in the database.  Each dynamically interleaved key corresponds
to a root-to-leaf path in the RSCAS trie.

\begin{definition}[RSCAS Trie] \label{def:rscas} %
  Let $K$ be a set of composite keys and let $R$ be a trie.
  Trie $R$ is the RSCAS trie for $K$ iff the following
  conditions are satisfied.
  \begin{enumerate}
  \item $I_\text{DY}(k,K) = (t_1,\ldots,t_m,t_{m+1})$ is the dynamic
    interleaving of a key $k \in K$ iff there is a
    root-to-leaf path $(n_1, \ldots, n_m)$ in $R$ such that
    $t_i.s_P = n_i.s_P$, $t_i.s_V = n_i.s_V$, and $t_i.D = n_i.D$ for
      $1 \leq i \leq m$. Suffix $t_{m+1}$ is stored in leaf node
      $n_m$, i.e.,  $t_{m+1} \in n_m.\textsf{suffixes}$.
  \item $R$ does not include duplicate siblings, i.e., no two sibling
    nodes $n$ and $n'$, $n \neq n'$, in $R$ have the same values for
    $s_P$, $s_V$, and $D$, respectively.
  \end{enumerate}
\end{definition}

\begin{example}
  Figure \ref{fig:solution} shows the RSCAS trie for keys
  $\mathsf{K}^{1..9}$.  The values at the discriminative bytes are
  highlighted in bold.  The dynamic interleaving
  $I_{\text{DY}}(\mathsf{k}_9,\mathsf{K}^{1..9}) =$
  $(\mathsf{t}_1, \mathsf{t}_2, \mathsf{t}_3, \mathsf{t}_4,
  \mathsf{t}_5)$ from Table \ref{tab:dynints} is mapped to the
  root-to-leaf path $(\mathsf{n}_1, \mathsf{n}_2, \mathsf{n}_4,$
  $\mathsf{n}_6)$ in the RSCAS trie.  Tuple $\mathsf{t}_5$ is stored
  in $\mathsf{n}_6.\textsf{suffixes}$. Key $\mathsf{k}_8$ is stored in
  the same root-to-leaf path.  For key $\mathsf{k}_1$, the first two
  tuples of $I_{\text{DY}}(\mathsf{k}_1,\mathsf{K}^{1..9})$ are mapped
  to $\mathsf{n}_1$ and $\mathsf{n}_2$, respectively, while the third
  tuple is mapped to $\mathsf{n}_3$. $\hfill\Box$
\end{example}

\subsection{RSCAS Index}
\label{sec:scaling}

The RSCAS index combines a memory-optimized RSCAS trie for in-place
\MREV{insertions} with a sequence of disk-based RSCAS tries for
out-of-place \MREV{insertions} to get good \MREV{insertion}
performance for large data-intensive applications.  LSM trees
\cite{PO96,OP03} have pioneered combining memory- and disk-resident
components, and are now the de-facto standard to build scalable index
structures (see, e.g., \cite{SA14,FC08,GD07}).

\MREV{We implement RSCAS as an LSM trie that fixes the size ratio
  between two consecutive tries at $T=2$ and uses the leveling merge
  policy with full merges (this combination is also known as the
  logarithmic method in \cite{OP03}). Leveling optimizes query
  performance and space utilization in comparison to the tiering merge
  policy at the expense of a higher merge cost~\cite{CL19,CL20}. Luo
  and Carey show that a size ratio of $T=2$ achieves the maximum write
  throughput for leveling, but may have a negative impact on the
  latency~\cite{CL19}. Since query performance and space utilization
  are important to us, while latency does not play a large role (due
  to batched updates in the background), we choose the setup described
  above.  If needed the LSM trie can be improved with the techniques
  presented by Luo and Carey \cite{CL19,CL20}.  For example, one such
  improvement is partitioned merging where multiple tries with
  non-overlapping key ranges can exist at the same level and when a
  trie overflows at level $i$, this trie needs only to be merged with
  overlapping tries at level $i+1$. Partitioned merges reduce the I/O
  during merging since not all data at level $i$ needs to be merged
  into level $i+1$.}

\MREVB{Our focus is to show how to integrate a CAS index with LSM
  trees.  We do not address aspects related to recovery and multi-user
  synchronization.  These challenges, however, exist and must be
  handled by the system.  Typical KV-stores use write-ahead logging
  (WAL) to make their system recoverable and multi-version concurrency
  control (MVCC) to provide concurrency.  These techniques are also
  applicable to the RSCAS index.}

\begin{figure}[htbp]
\begin{tikzpicture}
  \begin{scope}[shift={(0pt,0mm)}]
    \draw (0,0) -- (0.75,0) -- (0.375,0.75) -- cycle;
    \node[anchor=north,align=center] at (0.375,0) {
      Trie $R^M_0$\\{\scriptsize $(0,M]$ keys}\\{\scriptsize or empty}};
  \end{scope}
  \begin{scope}[shift={(15mm,0mm)}]
    \draw (0,0) -- (0.75,0) -- (0.375,0.75) -- cycle;
    \node[anchor=north,align=center] at (0.375,0) {
      Trie $R_0$\\{\scriptsize $(0,M]$ keys}\\{\scriptsize or empty}};
  \end{scope}
  \begin{scope}[shift={(30mm,0mm)}]
    \draw (0,0) -- (1,0) -- (0.5,1) -- cycle;
    \node[anchor=north,align=center] at (0.5,0) {
      Trie $R_1$\\{\scriptsize $(M,2M]$ keys}\\{\scriptsize or empty}};
  \end{scope}
  \node at (50mm,0) {$\ldots\ldots$};
  \begin{scope}[shift={(60mm,0mm)}]
    \draw (0,0) -- (2,0) -- (1,2) -- cycle;
    \node[anchor=north,align=center] at (1,0) {
      Trie $R_k$\\{\scriptsize $(2^{k-1}M,2^kM]$ keys}\\{\scriptsize or empty}};
  \end{scope}
  \draw[cmbrace] (-2mm,-12mm) -- (10mm,-12mm) node
    [black,midway,anchor=north,yshift=-3pt] {Memory};
  \draw[cmbrace] (15mm,-12mm) -- (80mm,-12mm) node
    [black,midway,anchor=north,yshift=-3pt] {Disk};
  \draw[cmbrace] (-2mm,-18mm) -- (80mm,-18mm) node
    [black,midway,anchor=north,yshift=-3pt] {RSCAS Index};
\end{tikzpicture}
\caption{The RSCAS index combines memory- and disk-based RSCAS tries
  for scalability.}
\label{fig:logmethod}
\end{figure}
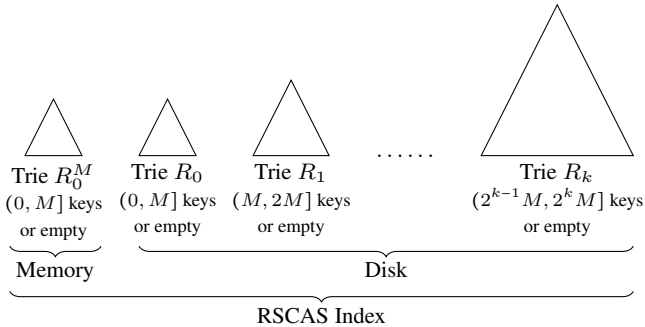

The in-memory RSCAS trie $R^M_0$ is combined with a sequence of
disk-based RSCAS tries $R_0, \ldots, R_k$ that grow in size as
illustrated in Figure \ref{fig:logmethod}. \MREV{The most recently
  inserted keys are accumulated in the in-memory RSCAS trie $R^M_0$
  where insertions can be performed efficiently}. When $R^M_0$ grows
too big, the keys are migrated to a disk-based RSCAS trie $R_i$. A
query is executed on each trie individually and the result sets are
combined.  We only consider insertions since deletions do not occur in
the SWH archive.

The size of each trie is bounded. $R^M_0$ and $R_0$ contain up to $M$
keys, where $M$ is chosen according to the memory capacity of the
system.  With an average key length of 80 bytes in the SWH archive,
reasonable values of $M$ range from tens of millions to a few billion
keys (e.g., with $M = 10^8$, $R^M_0$ requires about 8\,GB of memory).
Each disk-based trie $R_i$, $i\ge1$, is either empty or contains
between $2^{i-1}M$ keys (exclusive) and $2^iM$ keys (inclusive).

When $R^M_0$ is full, we look for the first disk-based trie $R_i$ that
is empty.  We (a) collect all keys in tries $R^M_0$ and $R_j$,
$0 \leq j < i$, (b) bulk-load trie $R_i$ from these keys, and (c)
delete all previous tries.

\begin{example}
  Assume we set the number of keys that fit in memory to $M=10$
  million, which is the number of new keys that arrive every day in
  the SWH archive, on average. When $R^M_0$ overflows after one day we
  redirect incoming insertions to a new in-memory trie and look for
  the first non-empty trie $R_i$.  Assuming this is $R_0$, the
  disk-resident trie $R_0$ is bulk-loaded with the keys in $R^M_0$.
  After another day, $R^M_0$ overflows again and this time the first
  non-empty trie is $R_1$.  Trie $R_1$ is created from the keys in
  $R^M_0$ and $R_0$.  At the end $R_1$ contains $20M$ keys, and
  $R^M_0$ and $R_0$ are deleted.  $\hfill\Box$
\end{example}

An overflow in $R^M_0$ does not stall continuous indexing since we
immediately redirect all incoming insertions to a new in-memory trie
$R^{M'}_0$ while we bulk-load $R_i$ in the background.  \MREV{In order
  for this to work, $R^M_0$ cannot allocate all of the available
  memory.  We need to reserve a sufficient amount of memory for
  $R^{M'}_0$ (in the SWH archive scenario we allowed $R^M_0$ to take
  up at most half of the memory).}  During bulk-loading we keep the
old tries $R^M_0$ and $R_0, \ldots, R_{i-1}$ around such that queries
have access to all indexed data. As soon as $R_i$ is complete, we
replace $R^M_0$ with $R^{M'}_0$ and $R_0, \ldots, R_{i-1}$ with $R_i$.
In practice neither insertions nor queries stall \MREV{as long as the
  insertion rate is bounded}.  If the insertion rate is too high and
$R^{M'}_0$ overflows before we finish bulk-loading $R_i$, we block and
do not accept more insertions.  This does not happen in the SWH
archive since with our default of $M = 10^8$ keys (about 8\,GB memory)
trie $R^{M'}_0$ overflows every ten days and bulk-loading the trie on
our biggest dataset takes about four hours.

\subsection{Storage Layout}

The RSCAS index consists of a \emph{mutable} in-memory trie $R^M_0$
and a series of \emph{immutable} disk-based tries $R_i$.  For $R^M_0$
we use a node structure that is easy to update in-place, while we
design $R_i$ for compact storage on disk.

\subsubsection{Memory-Optimized RSCAS Trie}

\begin{figure}[htb]
\centering
\begin{tikzpicture}[
  box/.style={
    draw,
    minimum height=15pt,
    anchor=west,
  },
  ptr/.style={
    fill=colxf,
  },
]
  \node[box] (n0) {$(D,l_P,l_V,m)$};
  \node[box] (n1) at (n0.east) {$s_P$};
  \node[box] (n2) at (n1.east) {$s_V$};
  \node[box,ptr] (n00) at (n2.east) {\pcodes{00}};
  \node[box,ptr] (n01) at (n00.east) {\pcodes{01}};
  \node[box,ptr] (n02) at (n01.east) {\pcodes{02}};
  \node[box,ptr] (nXX) at (n02.east) {$\ldots\ldots$};
  \node[box,ptr] (nFD) at (nXX.east) {\pcodes{FD}};
  \node[box,ptr] (nFE) at (nFD.east) {\pcodes{FE}};
  \node[box,ptr] (nFF) at (nFE.east) {\pcodes{FF}};
  \draw[fill=black] (n00.south) circle (1pt) edge[->] ([yshift=-8pt] n00.south) ;
  \draw[fill=black] (n01.south) circle (1pt) edge[->] ([yshift=-8pt] n01.south) ;
  \draw[fill=black] (n02.south) circle (1pt) edge[->] ([yshift=-8pt] n02.south) ;
  \draw[fill=black] (nFD.south) circle (1pt) edge[->] ([yshift=-8pt] nFD.south) ;
  \draw[fill=black] (nFE.south) circle (1pt) edge[->] ([yshift=-8pt] nFE.south) ;
  \draw[fill=black] (nFF.south) circle (1pt) edge[->] ([yshift=-8pt] nFF.south) ;
  \draw[cbrace] (n0.north west) -- (n0.north east) node
    [black,midway,anchor=south,yshift=2pt] {\scriptsize header};
\end{tikzpicture}
\caption{Structure of an inner node with 256 pointers.}
\label{fig:node}
\end{figure}
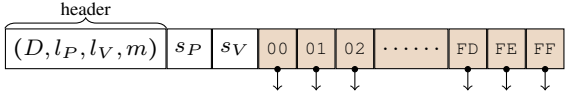

The memory-optimized RSCAS trie $R^M_0$ provides fast in-place
\MREV{insertions} for a small number of composite keys that fit into
memory.  Since all insertions are buffered in $R^M_0$ before they are
migrated in bulk to disk, $R^M_0$ is in the critical path of our
indexing pipeline and must support efficient \MREV{insertions}.  We
reuse the memory-optimized trie \cite{KW20} that is based on the
memory-optimized Adaptive Radix Tree (ART) \cite{VL13}. ART implements
four node types that are optimized for the hardware's memory hierarchy
and that have a physical fanout of 4, 16, 48, and 256 child pointers,
respectively. A node uses the smallest node type that can accommodate
the node's child pointers. \MREV{Insertions} add node pointers and
when a node becomes too big, the node is resized to the next
appropriate node type.  This ensures that not every \MREV{insertion}
requires resizing, e.g., a node with ten children can sustain six
deletions or seven insertions before it is resized.  Figure
\ref{fig:node} illustrates the node type with 256 child pointers; for
the remaining node types we refer to Leis et al.~\cite{VL13}.  The
node header stores the dimension $D$, the lengths $l_P$ and $l_V$ of
substrings $s_P$ and $s_V$, and the number of children $m$. Substrings
$s_P$ and $s_V$ are implemented as variable-length byte vectors.  The
remaining space of an inner node (beige-colored in Figure
\ref{fig:node}) is reserved for child pointers.  For each possible
value $b$ of the discriminative byte there is a pointer (possibly
\pcode{NULL}) to the subtree where all keys have value $b$ at the
discriminative byte in dimension $D$.

The structure of leaf nodes is similar, except that leaf nodes contain
a variable-length vector with references $k.R$ instead of child
pointers.

For the memory-optimized RSCAS trie we set the partitioning threshold
$\tau=1$ meaning that $R^M_0$ dynamically interleaves keys completely.
This provides fast and fine-grained access to the indexed keys.

\subsubsection{Disk-Optimized RSCAS Trie}
\label{sec:diskbased}

We propose a disk-resident RSCAS trie to compactly store
dynamically-interleaved keys on disk.  Since a disk-resident RSCAS
trie is immutable, we optimize it for compact storage. To that end we
store nodes gapless on disk and we increase the granularity of leaf
nodes by setting $\tau > 1$.  We look at these techniques in turn.  We
store nodes gapless on disk since we do not have to reserve space for
future in-place \MREV{insertions}.  This means a node can cross page
boundaries but we found that in practice this is not a problem.  We
tested various node clustering techniques to align nodes to disk
pages. The most compact node clustering algorithm \cite{CCK06b}
produced a trie that was 30\% larger than with gapless storage as it
kept space empty on a page if it could not add another node without
exceeding the page size.
\inRegularPaper{}{Besides being simpler to implement and more
compact, the gapless storage yields better query performance because
less data needs to be read from disk.}
In addition to the gapless
storage, we increase the granularity of leaf nodes by setting
$\tau > 1$.  As a result the RSCAS index contains fewer nodes but the
size of leaf nodes increases.  We found that by storing fewer but
bigger nodes we save space because we store less meta-data like node
headers, child pointers, etc.  In Section~\ref{sec:calibration} we
determine the optimal value for $\tau$.

\begin{figure}[htbp]
  \centering
  \begin{tikzpicture}[
    box/.style={
      draw,
      minimum height=16pt,
      inner sep=2.25pt,
      anchor=west,
      font={\scriptsize},
    },
    file/.style={
      box,
      fill=colxy,
    },
    page/.style={
      box,
      fill=colxe,
    },
    none/.style={
      box,
      fill=white,
    },
    ptrs/.style={
      box,
      fill=colxf,
    },
    suffix/.style={
      box,
      fill=colxe,
    },
    tree/.style={
      box,
      fill=colxi,
    },
    legend/.style={
      anchor=north,
      font={\tiny},
    },
  ]
  \begin{scope}[shift={(0mm,0mm)}]
    \node[none] (n01) {$(D, l_P, l_V, m)$};
    \node[none] (n02) at (n01.east) {$s_P$};
    \node[none] (n03) at (n02.east) {$s_V$};
    \node[ptrs] (n04) at (n03.east) {$b_1$};
    \node[ptrs] (n05) at (n04.east) {$\text{ptr}_1$};
    \node[ptrs] (n06) at (n05.east) {$~\ldots~$};
    \node[ptrs] (n07) at (n06.east) {$b_m$};
    \node[ptrs] (n08) at (n07.east) {$\text{ptr}_m$};
    \node[tree] (n09) at (n08.east) {$\text{node}_1$};
    \node[tree] (n10) at (n09.east) {$~\ldots~$};
    \node[tree] (n11) at (n10.east) {$\text{node}_m$};
    \node[legend] at (n01.south) {{\scriptsize header}$\quad$4B};
    \node[legend] at (n02.south) {$l_P$};
    \node[legend] at (n03.south) {$l_V$};
    \node[legend] at (n04.south) {1B};
    \node[legend] at (n05.south) {6B};
    \node[legend] at (n07.south) {1B};
    \node[legend] at (n08.south) {6B};
    \draw[->] (n05.north) -- ([yshift= 4pt] n05.north) -| (n09.north);
    \draw[->] (n08.north) -- ([yshift= 8pt] n08.north) -| (n11.north);
    \node[anchor=south west] at (n01.north west) {\scriptsize
    \textbf{\colorbox{colxi}{Inner Node:}}};
  \end{scope}
  \begin{scope}[shift={(0pt,-16mm)}]
    \node[none] (n01) {$(D, l_P, l_V, m)$};
    \node[none] (n02) at (n01.east) {$s_P$};
    \node[none] (n03) at (n02.east) {$s_V$};
    \node[suffix] (n04) at (n03.east) {$l^1_P$};
    \node[suffix] (n05) at (n04.east) {$l^1_V$};
    \node[suffix] (n06) at (n05.east) {$s^1_P$};
    \node[suffix] (n07) at (n06.east) {$s^1_V$};
    \node[suffix] (n08) at (n07.east) {$r^1$};
    \node[suffix] (n09) at (n08.east) {$\ldots$};
    \node[suffix] (n10) at (n09.east) {$l^m_P$};
    \node[suffix] (n11) at (n10.east) {$l^m_V$};
    \node[suffix] (n12) at (n11.east) {$s^m_P$};
    \node[suffix] (n13) at (n12.east) {$s^m_V$};
    \node[suffix] (n14) at (n13.east) {$r^m$};
    \node[legend] at (n01.south) {{\scriptsize header}$\quad$4B};
    \node[legend] at (n02.south) {$l_P$};
    \node[legend] at (n03.south) {$l_V$};
    \node[legend] at (n04.south) {1B};
    \node[legend] at (n05.south) {1B};
    \node[legend] at (n06.south) {$l^1_P$};
    \node[legend] at (n07.south) {$l^1_V$};
    \node[legend] at (n08.south) {20B};
    \node[legend] at (n10.south) {1B};
    \node[legend] at (n11.south) {1B};
    \node[legend] at (n12.south) {$l^m_P$};
    \node[legend] at (n13.south) {$l^m_V$};
    \node[legend] at (n14.south) {20B};
    \draw[cbrace] (n04.north west) -- (n08.north east) node
      [black,midway,anchor=south,yshift=2pt] {\scriptsize 1st suffix};
    \draw[cbrace] (n10.north west) -- (n14.north east) node
      [black,midway,anchor=south,yshift=2pt] {\scriptsize $m$th suffix};
    \node[anchor=south west] at (n01.north west) {\scriptsize
    \textbf{\colorbox{colxi}{Leaf Node:}}};
  \end{scope}
  \end{tikzpicture}
  \caption{Serializing nodes on disk.}
  \label{fig:pagelayout}
\end{figure}
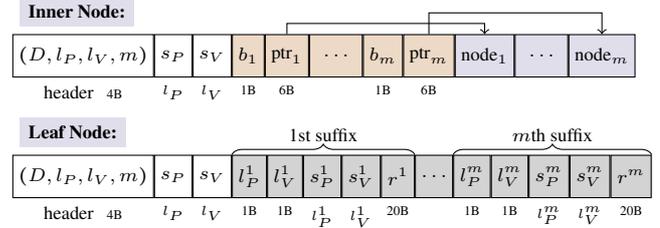

Figure~\ref{fig:pagelayout} shows how to compactly serialize nodes on
disk.  Inner nodes point to other nodes, while leaf nodes store a set
of suffixes.  Both node types store the same four-byte header that
encodes dimension $D \in \{P,V,\bot\}$, the lengths $l_P$ and $l_V$ of
the substrings $s_P$ and $s_V$, and a number $m$.  For inner nodes $m$
denotes the number of children, while for leaf nodes it denotes the
number of suffixes.  Next we store substrings $s_P$ and $s_V$ (exactly
$l_P$ and $l_V$ bytes long, respectively).  After the header, inner
nodes store $m$ pairs $(b_i,\text{ptr}_i)$, where $b_i$ (1 byte long)
is the value at the discriminative byte that is used to descend to
this child node and $\text{ptr}_i$ (6 bytes long) is the position of
this child in the trie file.  Leaf nodes, instead, store $m$ suffixes
and for each suffix we record substrings $s_P$ and $s_V$ along with
their lengths and the revision $r$ (20 byte SHA1 hash).

\begin{example}
  The size of $\mathsf{n}_1$ in Figure~\ref{fig:solution} is 30 bytes:
  4 bytes for the header, 4 bytes for $s_V$, 1 byte for $s_P$, and
  $3\times(1+6) = 21$ bytes for the three child pointers and their
  discriminative bytes.  $\hfill\Box$
\end{example}

\section{Algorithms}
\label{sec:algorithms}

We propose algorithms for querying, \MREV{inserting}, bulk-loading,
and merging RSCAS tries. Queries are executed independently on all
in-memory and disk-based RSCAS tries and the results are combined.
\MREV{Insertions} are directed at the in-memory RSCAS trie
alone. Merging is used whenever the in-memory RSCAS trie overflows and
applies bulk-loading to create a large disk-optimized RSCAS trie.

\subsection{Querying RSCAS}
\label{sec:querying}

We traverse an RSCAS trie in pre-order to evaluate a CAS query,
skipping subtrees that cannot match the query. Starting at the root
node, we traverse the trie and evaluate at each node part of the
query's path and value predicate. Evaluating a predicate on a node
returns \texttt{MATCH} if the full predicate has been matched,
\texttt{MISMATCH} if it has become clear that no node in the current
node's subtree can match the predicate, and \texttt{INCOMPLETE} if we
need more information. In case of a \texttt{MISMATCH}, we can safely
skip the entire subtree. If both predicates return \texttt{MATCH}, we
collect all revisions $r$ in the leaf nodes of this subtree.
Otherwise, we traverse the trie further to reach a decision.

\subsubsection{Query Algorithm}

Algorithm \ref{alg:query} shows the pseudocode for evaluating a CAS
query on a RSCAS trie. It takes the following parameters: the current
node $n$ (initially the root node of the trie), a query path $q$, and
a range $[v_l, v_h]$ for the value predicate. Furthermore, we need two
buffers $\texttt{buff}_{P}$ and $\texttt{buff}_{V}$ (initially empty)
that hold, respectively, all path and value bytes from the root to the
current node $n$.  Finally, we require state information $s$ to
evaluate the path and value predicates (we provide details as we go
along) and an answer set $W$ to collect the results.

\begin{algorithm2e}[htb]
\scriptsize
\SetInd{0.65em}{0.65em} 
\DontPrintSemicolon
\SetKwProg{Fn}{Function}{}{end}
\SetArgSty{textnormal}
$\texttt{UpdateBuffers}(n.s_V,n.s_P,\texttt{buff}_V,\texttt{buff}_P)$\;
\eIf{$n$ is an inner node}{%
  $\texttt{match}_V \gets \texttt{MatchValue}(\texttt{buff}_V,v_l,v_h,s,n)$\;
  $\texttt{match}_P \gets \texttt{MatchPath}(\texttt{buff}_P,q,s,n)$\;
  \If{$\texttt{match}_V \neq \texttt{MISMATCH} \wedge \texttt{match}_P
        \neq\texttt{MISMATCH}$} {%
    \For{each matching child $c$ of $n$} {
      $\textsf{CasQuery}(c,q,[v_l,v_h],\texttt{buff}_V,\texttt{buff}_P,s,W)$\;
    }
  }
}{
  \ForEach{$t \in n.\textsf{suffixes}$}{
    $\texttt{UpdateBuffers}(t.s_V,t.s_P,\texttt{buff}_V,\texttt{buff}_P)$\;
    $\texttt{match}_V \gets \texttt{MatchValue}(\texttt{buff}_V,v_l,v_h,s,n)$\;
    $\texttt{match}_P \gets \texttt{MatchPath}(\texttt{buff}_P,q,s,n)$\;
    \If{$\texttt{match}_V = \texttt{MATCH} \wedge
      \texttt{match}_P = \texttt{MATCH}$} {
      $W \gets W \cup \{ t.R \}$\;
    }
  }
}
%
\caption{\small$\textsf{CasQuery}(n, q, [v_l, v_h], \texttt{buff}_V,
\texttt{buff}_P, s, W)$}
\label{alg:query}
\end{algorithm2e}

First, we update $\texttt{buff}_{V}$ and $\texttt{buff}_{P}$ by adding
the information in $s_V$ and $s_P$ of the current node $n$ (line 1).

For inner nodes, we match the query predicates against the current
node.  \texttt{MatchValue} computes the longest common prefix between
$\texttt{buff}_{V}$ and $v_l$ and between $\texttt{buff}_{V}$ and
$v_h$.  The position of the first byte for which $\texttt{buff}_{V}$
and $v_l$ differ is \texttt{lo} and the position of the first byte for
which $\texttt{buff}_{V}$ and $v_h$ differ is \texttt{hi}.  If
$\texttt{buff}_{V}[\texttt{lo}] < v_l[\texttt{lo}]$, we know that the
node's value lies outside of the range, hence we return
\texttt{MISMATCH}. If
$\texttt{buff}_{V}[\texttt{hi}] > v_h[\texttt{hi}]$, the node's value
lies outside of the upper bound and we return \texttt{MISMATCH} as
well.  If $\texttt{buff}_{V}$ contains a complete value (e.g., all
eight bytes of a 64 bit integer) and
$v_l \leq \texttt{buff}_{V} \leq v_h$, we return \texttt{MATCH}.  If
$\texttt{buff}_{V}$ is incomplete, but
$v_l[\texttt{lo}] < \texttt{buff}_{V}[\texttt{lo}]$ and
$\texttt{buff}_{V}[\texttt{hi}] < v_h[\texttt{hi}]$, we know that all
values in the subtree rooted at $n$ match and we also return
\texttt{MATCH}. In all other cases we cannot make a decision yet and
return \texttt{INCOMPLETE}. The values of \texttt{lo} and \texttt{hi}
are kept in the state to avoid recomputing the longest common prefix
from scratch for each node.  Instead we resume the search from the
previous values of \texttt{lo} and \texttt{hi}.

Function \texttt{MatchPath} matches the query path $q$ against the
current path prefix $\texttt{buff}_P$. It supports symbols \texttt{*}
and \texttt{**} to match any number of characters in a node label,
respectively any number of node labels in a path.  As long as we do
not encounter any wildcards in the query path $q$, we directly compare
(a prefix of) $q$ with the current content of $\texttt{buff}_{P}$ byte
by byte. As soon as a byte does not match, we return
\texttt{MISMATCH}.  If we successfully match the complete query path
$q$ against a complete path in $\texttt{buff}_{P}$ (both terminated by
\texttt{\$}), we return \texttt{MATCH}.  Otherwise, we return
\texttt{INCOMPLETE}. When we encounter wildcard \texttt{*} in $q$, we
match it successfully to the corresponding label in
$\texttt{buff}_{P}$ and continue with the next label. A wildcard
\texttt{*} itself will not cause a mismatch (unless we try to match it
against the terminator \texttt{\$}), so we either return
\texttt{MATCH} if it is the final label in $q$ and $\texttt{buff}_{P}$
or \texttt{INCOMPLETE}.  Matching the descendant-axis \texttt{**} is
more complicated. We store in state $s$ the current position where we
are in $\texttt{buff}_{P}$ and continue matching the label after
\texttt{**} in $q$. If at any point we find a mismatch, we backtrack
to the next path separator after the noted position, thus skipping a
label in $\texttt{buff}_{P}$ and restarting the search from there.
Once $\texttt{buff}_{P}$ contains a complete path, we can make a
decision between \texttt{MATCH} or \texttt{MISMATCH}.

The algorithm continues by checking the outcomes of the value and path
matching (line 5).  If one of the outcomes is \texttt{MISMATCH}, we
stop the search since no descendant can match the query. Otherwise, we
continue with the matching children of $n$ (lines 6--8).  Finding the
matching children follows the same logic as described above for
\texttt{MatchValue} and \texttt{MatchPath}. If node $n.D = P$ and we
have seen a descendant axis in the query path, all children of the
current node match.

As soon as we reach a leaf node, we iterate over each suffix $t$ in
the leaf to check if it matches the query using the same functions as
explained above (lines 10--14). If the current buffers indeed match
the query, we add the reference $t.R$ to the result set.

\begin{example}\label{example:querying}
  Consider a CAS query that looks for revisions in 2020 that modified
  a C file in the \texttt{ext3} or \texttt{ext4} filesystem.  Thus,
  the query path is $q=\ptstr{/fs/ext*/*.c\$}$ and the value range is
  $v_l = \textsf{2020-01-01}$ (\vtstr{00\,00\,00\,00\,5E\,0B\,E1\,00})
  and $v_h = \textsf{2020-12-31}$
  (\vtstr{00\,00\,00\,00\,5F\,EE\,65\,FF}).  We execute the query on
  the trie in Figure~\ref{fig:solution}.

  \begin{itemize}[leftmargin=15pt]

  \item Starting at $\mathsf{n}_1$, we update $\texttt{buff}_V$ to
    \vtstr{00\,00\,00\,00} and $\texttt{buff}_P$ to \ptstr{/}.
    \texttt{MatchValue} matches four value bytes and returns
    \texttt{INCOMPLETE}.  \texttt{MatchPath} matches one path byte and
    also returns \texttt{INCOMPLETE}.  Both functions return
    \texttt{INCOMPLETE}, so we have to traverse all matching children.
    Since $\mathsf{n}_1$ is a value node, we look for all matching
    children whose value for the discriminative value byte is between
    \vtstr{5E} and \vtstr{5F}.  Nodes $\mathsf{n}_7$ and
    $\mathsf{n}_8$ satisfy this condition.

  \item Node $\mathsf{n}_7$ is a leaf. We iterate over each suffix
    (there are two) and update the buffers accordingly. For the first
      suffix with path substring \ptstr{3/inode.c\$} we find that
      \texttt{MatchPath} and \texttt{MatchValue} both return
      \texttt{MATCH}. Hence, revision $\mathsf{r}_4$ is added to $W$.
      The next suffix matches the value predicate but not the path
      predicate and is therefore discarded.

    \item Next we look at node $\mathsf{n}_8$. We find that
      $v_l[5] = \vtstr{5E} < \vtstr{5F} = \texttt{buff}_V[5] = v_h[5]$
      and $\texttt{buff}_V[6] = \vtstr{BD} < \vtstr{EE} = v_h[6]$,
      thus all values of $\mathsf{n}_9$'s descendants are within the
      bounds $v_l$ and $v_h$, and \texttt{MatchValue} returns
      \texttt{MATCH}.  Since $\mathsf{n}_8.s_P$ is the empty string,
      \texttt{MatchPath} still returns \texttt{INCOMPLETE} and we
      descend further.  According to the second byte in the query
      path, $q[2] = \ptstr{f}$, we must match letter \ptstr{f}, hence
      we descend to node $\mathsf{n}_{10}$, where both predicates
      match.  Therefore, revision $\mathsf{r}_6$ is added to $W$.

  \end{itemize}
\end{example}

\subsection{Updating Memory-Based RSCAS Trie}

All insertions are performed in the in-memory RSCAS trie $R^M_0$ where
they can be executed efficiently.  Inserting a new key into $R^M_0$
usually changes the position of the discriminative bytes, which means
that the dynamic interleaving of all keys that are located in the
node's subtree is invalidated.

\begin{example}
  We insert the key $\mathsf{k}_{10} = (\ptstr{/crypto/rsa.c\$},$
  $\vtstr{00\,00\,00\,00\,5F\,83\,B9\,AC}, \mathsf{r}_{8})$ into the
  RSCAS trie in Figure~\ref{fig:solution}.  First we traverse the trie
  starting from root $\mathsf{n}_1$.  Since $\mathsf{n}_1$'s
  substrings completely match $\mathsf{k}_{10}$'s path and value we
  traverse to child $\mathsf{n}_8$. In $\mathsf{n}_8$ there is a
  mismatch in the value dimension: $\mathsf{k}_{10}$'s sixth byte is
  \vtstr{83} while for node $\mathsf{n}_8$ the corresponding byte is
  \vtstr{BD}. This invalidates the dynamic interleaving of keys
  $\mathsf{K}^{2,3,7}$ in $\mathsf{n}_8$'s subtree.  $\hfill\Box$
\end{example}

\subsubsection{Lazy Restructuring}
\label{sec:lazy-restructuring}

If we want to preserve the dynamic interleaving, we need to re-compute
the dynamic interleaving of all affected keys, which is expensive.
Instead, we relax the dynamic interleaving using \emph{lazy
  restructuring} \cite{KW21}.  Lazy restructuring resolves the
mismatch by adding exactly two new nodes, $n_{\textsf{par}}$ and
$n_{\textsf{sib}}$, to RSCAS instead of restructuring large parts of
the trie.  The basic idea is to add a new intermediate node
$n_{\textsf{par}}$ between node $n$ where the mismatch happened and
$n$'s new sibling node $n_{\textsf{sib}}$ that represents the newly
inserted key. We put all bytes leading up to the position of the
mismatch into $n_{\textsf{par}}$, and all bytes starting from this
position move to nodes $n$ and $n_{\textsf{sib}}$.  After that, we
insert node $n_{\textsf{par}}$ between node $n$ and its previous
parent node $n_p$.

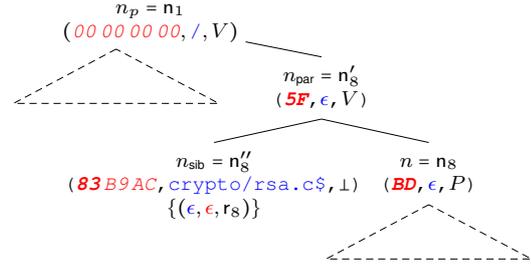
\begin{figure}[tbh]\centering
\scalebox{0.90}{
\begin{forest}
  casindex,
  for tree={
    font={\ttfamily \small},
  },
  for tree={
  },
  [{$n_{\textsf{par}} = \mathsf{n}'_8$}\\
   {(\vtbstr{5F},\pstr{$\epsilon$},$V$)},name=npar,
    [{$n_{\textsf{sib}} = \mathsf{n}''_8$}\\
     {(\vtbstr{83}\,\vtstr{B9\,AC},\pstr{crypto/rsa.c\$},$\bot$)} \\
     {$\{(\ptstr{$\epsilon$}, \vtstr{$\epsilon$}, \mathsf{r}_{8})\}$}
    ]
    [{$n = \mathsf{n}_{8}$}\\
     {(\vtbstr{BD},\pstr{$\epsilon$},$P$)},name=n8,l sep+=3mm
    ]
  ]
  \node[align=center] (root) at ([xshift=-25mm,yshift=5mm] npar.north) {
    $n_p = \mathsf{n}_1$ \\
    $(\vtstr{00\,00\,00\,00}, \ptstr{/}, V)$
  };
  \draw[] (npar.north) -- (root);
  \draw[densely dashed] (n8.south) --
    ([xshift=-15mm,yshift=-8mm] n8.south) --
    ([xshift=+15mm,yshift=-8mm] n8.south) -- cycle;
  \draw[densely dashed] ([xshift=-5mm]root.south) --
    ([xshift=-20mm,yshift=-8mm] root.south) --
    ([xshift=+10mm,yshift=-8mm] root.south) -- cycle;
\end{forest}
}
\caption{The rightmost subtree of Figure \ref{fig:solution} after
inserting key $\mathsf{k}_{10}$ with lazy restructuring.}
\label{fig:insertion}
\end{figure}

\begin{example}
  Figure \ref{fig:insertion} shows the rightmost subtree of Figure
  \ref{fig:solution} after it is lazily restructured when
  $\mathsf{k}_{10}$ is inserted. Two new nodes are created, parent
  $n_{\textsf{par}} = \mathsf{n}'_8$ and sibling $n_{\textsf{sib}} =
  \mathsf{n}''_8$.  Additionally, $\mathsf{n}_8.s_V$ is updated.
  $\hfill\Box$
\end{example}

Lazy restructuring is efficient: it adds exactly two new nodes to
$R^M_0$, thus the main cost is traversing the trie.  However, while
efficient, lazy restructuring introduces small irregularities that are
limited to the dynamic interleaving of the keys in the subtree where
the mismatch occurred. These irregularities do not affect the
correctness of CAS queries, but they slowly \emph{separate} (rather
than \emph{interleave}) paths and values if insertions repeatedly
force the algorithm to split the same subtree in the same dimension.
Since $R^M_0$ is memory-based and small in comparison to the
disk-based tries, the overall effect on query performance is
negligible.

\begin{example}
  After inserting $\mathsf{k}_{10}$, root node $\mathsf{n}_1$ and its
  new child $\mathsf{n}'_8$ both $\psi$-partition the data in the
  value dimension, violating the strictly alternating property of the
  dynamic interleaving, see Figure \ref{fig:insertion}.  $\hfill\Box$
\end{example}

\subsubsection{Inserting Keys with Lazy Restructuring}

Algorithm \ref{alg:insertion} inserts a key $k$ in $R^M_0$ rooted at
node $n$. If $R^M_0$ is empty (i.e., $n$ is $\nil$) we create a new
root node in lines 1-3.  Otherwise, we traverse the trie to $k$'s
insertion position.  We compare the key's path and value with the
current node's path and value by keeping track of positions
$g_P, g_V, i_P, i_V$ in strings $k.P, k.V, n.s_P, n.s_V$, respectively
(lines 8--11). As long as the substrings at their corresponding
positions coincide we descend.  If we completely matched key $k$, it
means that we reached a leaf node and we add $k.R$ to the current
node's suffixes (lines 12--14).  If during the traversal we cannot
find the next node to descend to, the key has a new value at a
discriminative byte that did not exist before in the data.  We create
a new leaf node and set its substrings $s_P$ and $s_V$ to the still
unmatched bytes in $k.P$ and $k.V$, respectively (lines 20--22). If we
find a mismatch between the key and the current node in at least one
dimension, we lazily restructure the trie (lines 15--17).

\begin{algorithm2e}[htb]
  \scriptsize
  \DontPrintSemicolon
  \SetInd{0.65em}{0.65em} 
  \If(\tcp*[f]{RSCAS is empty; create new root}){\textnormal{$n = \nil$}} {
    Install new root node: $\textsf{leaf}(k.P, k.V, k.R)$\;
    \Return
  }
  $n_p \gets \nil$\;
  $g_P, g_V \gets 1$\;
  \While{\textnormal{\textbf{true}}}{
    $i_P, i_V \gets 1$\;
    \While{\textnormal{$i_P \leq |n.s_P|
        \wedge g_P \leq |k.P|
        \wedge n.s_P[i_P] = k.P[g_P]
        $}}{
      $g_P\text{++};\quad i_P\text{++}$\;
    }
    \While{\textnormal{$i_V \leq |n.s_V|
        \wedge g_V \leq |k.V|
        \wedge n.s_V[i_V] = k.V[g_V]
        $}}{
      $g_V\text{++};\quad i_V\text{++}$\;
    }
    \uIf{\textnormal{$g_P > |k.P| \wedge g_V > |k.V|$}} {
      $n.\textsf{suffixes} \gets n.\textsf{suffixes} \cup \{
        (\epsilon, \epsilon, k.R) \}$\;
      \Return
    }
    \ElseIf{\textnormal{$i_P \leq |n.s_P| \vee i_V \leq |n.s_V|$}} {%
      $\textsf{LazyRestructuring}(k, n, n_p, g_P, g_V, i_P, i_V)$\;
      \Return
    }
    \leIf{$n.D=P$}{$b \gets k.P[g_P]$}{$b \gets k.V[g_V]$}
    $(n_p, n) \gets (n, n.\textsf{children}[b])$ \;
    \If{\textnormal{$n = \nil$}} {%
      $n_p.\textsf{children}[b] \gets \textsf{leaf}(
          k.P[g_P, |k.P|], k.V[g_V, |k.V|], k.R)$\;
      \Return
    }
  }
  \caption{$\textsf{Insert}(k, n)$}
  \label{alg:insertion}
\end{algorithm2e}

Algorithm~\ref{alg:restructure} implements lazy restructuring.  Lines
1--4 determine the dimension in which $n_{\textsf{par}}$ partitions
the data.  If only a path mismatch occurred between $n$ and $k$, we
have to use dimension $P$.  In case of only a value mismatch, we have
to use $V$.  If we have mismatches in both dimensions, then we take
the opposite dimension of parent node $n_p$ to keep up an alternating
interleaving as long as possible.  In lines 5--6 we create nodes
$n_{\textsf{par}}$ and $n_{\textsf{sib}}$.  Node $n_{\textsf{par}}$ is
an inner node of type \textsf{node4}, which is the node type with the
smallest fanout in ART \cite{VL13}. In lines 9--12 we install $n$ and
$n_{\textsf{sib}}$ as children of $n_{\textsf{par}}$. Finally, in
lines 13--15, we place the new parent node $n_{\textsf{par}}$ between
$n$ and its former parent node $n_p$.

\begin{algorithm2e}[htb]
  \scriptsize
  \DontPrintSemicolon
  \SetInd{0.65em}{0.65em} 
  \lIf{\textnormal{$i_P \leq |n.s_P| \wedge i_V > |n.s_V|$}}{%
    $D \gets P$%
    \tcp*[f]{mismatch in $P$}
  }
  \lElseIf{\textnormal{$i_P > |n.s_P| \wedge i_V \leq |n.s_V|$}}{%
    $D \gets V$%
    \tcp*[f]{mismatch in $V$}
  }
  \lElseIf{\textnormal{$n_p \neq \nil$}}{%
    $D \gets n_p.\overline{D}$%
    \tcp*[f]{mismatch in $P$ and $V$}
  }
  \lElse{%
    $D \gets V$
  }
  \BlankLine
  $n_{\textsf{par}} \gets \textsf{node4}(D, n.s_P[1, i_P-1], n.s_V[1, i_V-1])$\;
  $n_{\textsf{sib}} \gets \textsf{leaf}(k.P[g_P, |k.P|], k.V[g_V, |k.V|], k.R)$\;
  \BlankLine
  $n.s_P \gets n.s_P[i_P, |n.s_P|]$\;
  $n.s_V \gets n.s_V[i_V, |n.s_V|]$\;
  \BlankLine
  \leIf{$D=P$}{$b_1 \gets n_{\textsf{sib}}.s_P[1]$}{$b_1 \gets n_{\textsf{sib}}.s_V[1]$}
  \leIf{$D=P$}{$b_2 \gets n.s_P[1]$}{$b_2 \gets n.s_V[1]$}
  $n_{\textsf{par}}.\textsf{children}[b_1] \gets n_{\textsf{sib}}$\;
  $n_{\textsf{par}}.\textsf{children}[b_2] \gets n$\;
  \BlankLine
  \lIf{\textnormal{$n_p = \nil$}}{%
    install $n_{\textsf{par}}$ as new root node%
  }
  \lElseIf{\textnormal{$n_p.D = P$}}{%
    $n_p.\textsf{children}[n_{\textsf{par}}.s_P[1]] \gets n_{\textsf{par}}$%
  }
  \lElseIf{\textnormal{$n_p.D = V$}}{%
    $n_p.\textsf{children}[n_{\textsf{par}}.s_V[1]] \gets n_{\textsf{par}}$%
  }
  \caption{$\textsf{LazyRestructuring}(k, n, n_p, g_P, g_V, i_P, i_V)$}
  \label{alg:restructure}
\end{algorithm2e}

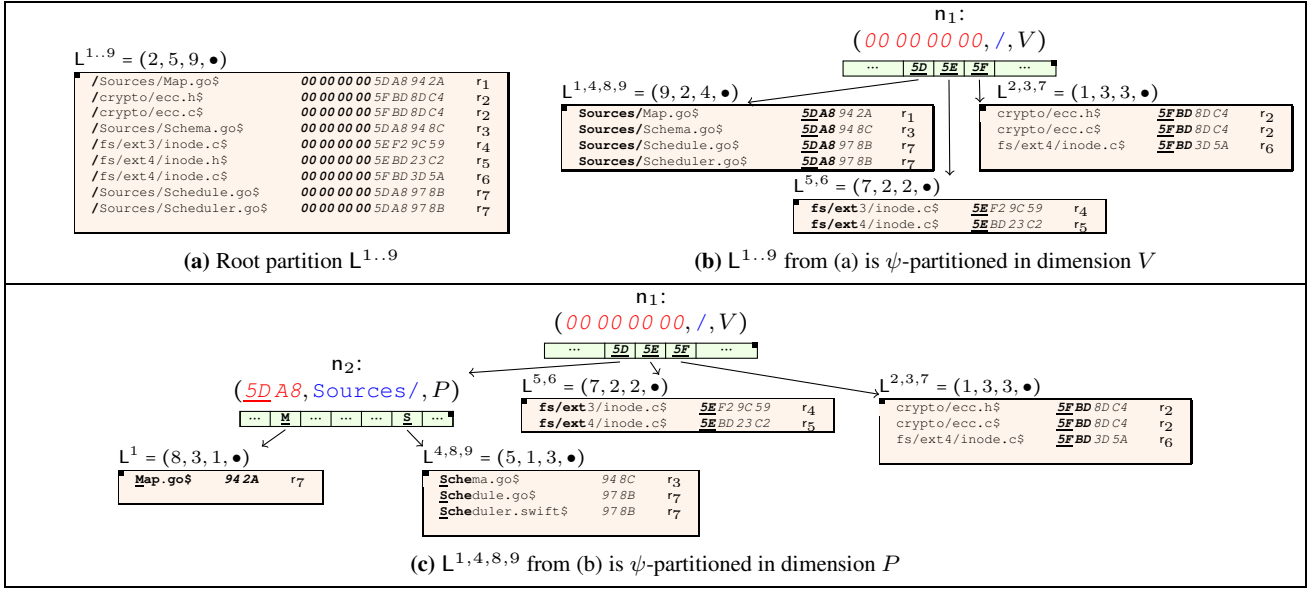
\begin{figure*}[h]
\centering
\begin{tikzpicture}
  \node[draw,anchor=west,fill=coltbd,minimum width=15pt,minimum height=7pt] (legend1) at (0,0)
    {};
    \node[anchor=west] (legend1l) at (legend1.east) {\footnotesize
    Partition $L=(g_P, g_V, \textsf{mptr}, \textsf{fptr})$};
  \node[anchor=north west,littlemarker] at (legend1.north west) {};
  \node[draw,anchor=west,fill=colptb,minimum width=15pt,minimum height=7pt] (legend2) at ([xshift=0.5cm] legend1l.east)
    {};
    \node[anchor=north east,littlemarker] at (legend2.north east) {};
    \node[anchor=west] (legend2l) at (legend2.east) {\footnotesize Partition Table $T$};
  \node[anchor=west] (legend3) at ([xshift=0.5cm] legend2l.east)
    {\footnotesize$n:$};
    \node[anchor=west] (legend3l) at (legend3.east) {\footnotesize Node
    $n$ in RSCAS};
\end{tikzpicture}\\
{
\centering
\begin{tabular}{@{}|c|@{}}\hline
\setlength\tabcolsep{6pt}
\begin{subfigure}[t]{0.40\textwidth}
\centering
  \begin{tikzpicture}
  \node[partition] (p1) at (-12,0.3) {
  {\scriptsize $\mathsf{L}^{1..9} = (2, 5, 9, \bullet)$}
  \\
  \begin{tabular}{|lcc|}
    \hline
    \cellcolor{coltbd}\texttt{\textbf{/}Sources/Map.go\$}
      & \cellcolor{coltbd}\vtistr{\textbf{00\,00\,00\,00}\,5D\,A8\,94\,2A}
      & \cellcolor{coltbd}$\mathsf{r}_1$
      \\
    \cellcolor{coltbd}\texttt{\textbf{/}crypto/ecc.h\$}
      & \cellcolor{coltbd}\vtistr{\textbf{00\,00\,00\,00}\,5F\,BD\,8D\,C4}
      & \cellcolor{coltbd}$\mathsf{r}_2$
      \\
    \cellcolor{coltbd}\texttt{\textbf{/}crypto/ecc.c\$}
      & \cellcolor{coltbd}\vtistr{\textbf{00\,00\,00\,00}\,5F\,BD\,8D\,C4}
      & \cellcolor{coltbd}$\mathsf{r}_2$
      \\
    \cellcolor{coltbd}\texttt{\textbf{/}Sources/Schema.go\$}
      & \cellcolor{coltbd}\vtistr{\textbf{00\,00\,00\,00}\,5D\,A8\,94\,8C}
      & \cellcolor{coltbd}$\mathsf{r}_3$
      \\
    \cellcolor{coltbd}\texttt{\textbf{/}fs/ext3/inode.c\$}
      & \cellcolor{coltbd}\vtistr{\textbf{00\,00\,00\,00}\,5E\,F2\,9C\,59}
      & \cellcolor{coltbd}$\mathsf{r}_4$
      \\
    \cellcolor{coltbd}\texttt{\textbf{/}fs/ext4/inode.h\$}
      & \cellcolor{coltbd}\vtistr{\textbf{00\,00\,00\,00}\,5E\,BD\,23\,C2}
      & \cellcolor{coltbd}$\mathsf{r}_5$
      \\
    \cellcolor{coltbd}\texttt{\textbf{/}fs/ext4/inode.c\$}
      & \cellcolor{coltbd}\vtistr{\textbf{00\,00\,00\,00}\,5F\,BD\,3D\,5A}
      & \cellcolor{coltbd}$\mathsf{r}_6$
      \\
    \cellcolor{coltbd}\texttt{\textbf{/}Sources/Schedule.go\$}
      & \cellcolor{coltbd}\vtistr{\textbf{00\,00\,00\,00}\,5D\,A8\,97\,8B}
      & \cellcolor{coltbd}$\mathsf{r}_7$
      \\
    \cellcolor{coltbd}\texttt{\textbf{/}Sources/Scheduler.go\$}
      & \cellcolor{coltbd}\vtistr{\textbf{00\,00\,00\,00}\,5D\,A8\,97\,8B}
      & \cellcolor{coltbd}$\mathsf{r}_7$
      \\
    \cellcolor{coltbd}
      & \cellcolor{coltbd}
      & \cellcolor{coltbd}
      \\
    \hline
  \end{tabular}
  };
  \node[anchor=north west,littlemarker] at ([yshift=-9pt] p1.north west) {};
  %
  %
\end{tikzpicture}
\caption{Root partition $\mathsf{L}^{1..9}$}
\end{subfigure}
\hfill
\begin{subfigure}[t]{0.55\textwidth}
\begin{tikzpicture}
  \node[rscasnode] (p1) at (0.1,0.35) {
    $\mathsf{n}_1$:\\
    $(\vtstr{00\,00\,00\,00}, \ptstr{/}, V)$%
  };
  \node[ptable] (m1) at (p1.south) {};
  \foreach \i in {8,12,16,20} {
    \draw ([xshift=\i mm] m1.north west) -- ([xshift=\i mm] m1.south west);
  }
  \node[ptablenode] (m1l1)
    at ([xshift=4mm,yshift=1mm] m1.south west) {$\cdots$};
  \node[ptablenode] (m1l2)
    at ([xshift=10mm,yshift=1mm] m1.south west) {\vtistr{\underline{5D}}};
  \node[ptablenode] (m1l3)
    at ([xshift=14mm,yshift=1mm] m1.south west) {\vtistr{\underline{5E}}};
  \node[ptablenode] (m1l4)
    at ([xshift=18mm,yshift=1mm] m1.south west) {\vtistr{\underline{5F}}};
  \node[ptablenode] (m1l5)
    at ([xshift=24mm,yshift=1mm] m1.south west) {$\cdots$};
  \node[anchor=north east,littlemarker] at (m1.north east) {};
  \node[partition] (p2) at (-5.0,-0.7) {
  {\scriptsize $\mathsf{L}^{1,4,8,9} = (9, 2, 4, \bullet)$}
  \\
  \begin{tabular}{|p{2.5cm}p{0.9cm}p{0.2cm}|}
    \hline
    \cellcolor{coltbd}\texttt{\textbf{Sources/}Map.go\$}
    & \cellcolor{coltbd}\vtistr{\textbf{\underline{5D}\,A8}\,94\,2A}
    & \cellcolor{coltbd}$\mathsf{r}_1$
    \\
    \cellcolor{coltbd}\texttt{\textbf{Sources/}Schema.go\$}
    & \cellcolor{coltbd}\vtistr{\textbf{\underline{5D}\,A8}\,94\,8C}
    & \cellcolor{coltbd}$\mathsf{r}_3$
    \\
    \cellcolor{coltbd}\texttt{\textbf{Sources/}Schedule.go\$}
    & \cellcolor{coltbd}\vtistr{\textbf{\underline{5D}\,A8}\,97\,8B}
    & \cellcolor{coltbd}$\mathsf{r}_7$
    \\
    \cellcolor{coltbd}\texttt{\textbf{Sources/}Scheduler.go\$}
    & \cellcolor{coltbd}\vtistr{\textbf{\underline{5D}\,A8}\,97\,8B}
    & \cellcolor{coltbd}$\mathsf{r}_7$
    \\
    \hline
  \end{tabular}
  };
  \node[anchor=north west,littlemarker] at ([yshift=-9pt] p2.north west) {};
  \node[partition] (p3) at (-1.95,-1.95) {
  {\scriptsize $\mathsf{L}^{5,6} = (7, 2, 2, \bullet)$}
  \\
  \begin{tabular}{|lcc|}
    \hline
    \cellcolor{coltbd}\texttt{\textbf{fs/ext}3/inode.c\$}
    & \cellcolor{coltbd}\vtistr{\underline{\textbf{5E}}\,F2\,9C\,59}
    & \cellcolor{coltbd}$\mathsf{r}_4$
    \\
    \cellcolor{coltbd}\texttt{\textbf{fs/ext}4/inode.c\$}
    & \cellcolor{coltbd}\vtistr{\underline{\textbf{5E}}\,BD\,23\,C2}
    & \cellcolor{coltbd}$\mathsf{r}_5$
    \\
    \hline
  \end{tabular}
  };
  \node[anchor=north west,littlemarker] at ([yshift=-9pt] p3.north west) {};
  \node[partition] (p4) at (0.5,-0.7) {
  $\quad$
  {\scriptsize $\mathsf{L}^{2,3,7} = (1, 3, 3, \bullet)$}
  \\
  \begin{tabular}{|lcc|}
    \hline
    \cellcolor{coltbd}\texttt{crypto/ecc.h\$}
    & \cellcolor{coltbd}\vtistr{\textbf{\underline{5F}\,BD}\,8D\,C4}
    & \cellcolor{coltbd}$\mathsf{r}_2$
    \\
    \cellcolor{coltbd}\texttt{crypto/ecc.c\$}
    & \cellcolor{coltbd}\vtistr{\textbf{\underline{5F}\,BD}\,8D\,C4}
    & \cellcolor{coltbd}$\mathsf{r}_2$
    \\
    \cellcolor{coltbd}\texttt{fs/ext4/inode.c\$}
    & \cellcolor{coltbd}\vtistr{\textbf{\underline{5F}\,BD}\,3D\,5A}
    & \cellcolor{coltbd}$\mathsf{r}_6$
    \\
    \cellcolor{coltbd}
    & \cellcolor{coltbd}
    & \cellcolor{coltbd}
    \\
    \hline
  \end{tabular}
  };
  \node[anchor=north west,littlemarker] at ([yshift=-9pt] p4.north west) {};
  \draw[->] (m1l2.south) -- ([yshift=-8pt]p2.north);
  \draw[->] (m1l3.south) -- (p3);
  \draw[->] (m1l4.south) -- ([yshift=-8pt]p4.north west);
\end{tikzpicture}
\caption{$\mathsf{L}^{1..9}$ from (a) is $\psi$-partitioned in dimension $V$}
\end{subfigure}
\\[0.5cm]\hline
\begin{subfigure}{0.9\textwidth}
\centering
\begin{tikzpicture}
  \node[rscasnode] (p1) at (0,0.35) {
    $\mathsf{n}_1$:\\
    $(\vtstr{00\,00\,00\,00}, \ptstr{/}, V)$%
  };
  \node[ptable] (m1) at (p1.south) {};
  \foreach \i in {8,12,16,20} {
    \draw ([xshift=\i mm] m1.north west) -- ([xshift=\i mm] m1.south west);
  }
  \node[ptablenode] (m1l1)
    at ([xshift=4mm,yshift=1mm] m1.south west) {$\cdots$};
  \node[ptablenode] (m1l2)
    at ([xshift=10mm,yshift=1mm] m1.south west) {\vtistr{\underline{5D}}};
  \node[ptablenode] (m1l3)
    at ([xshift=14mm,yshift=1mm] m1.south west) {\vtistr{\underline{5E}}};
  \node[ptablenode] (m1l4)
    at ([xshift=18mm,yshift=1mm] m1.south west) {\vtistr{\underline{5F}}};
  \node[ptablenode] (m1l5)
    at ([xshift=24mm,yshift=1mm] m1.south west) {$\cdots$};
  \node[anchor=north east,littlemarker] at (m1.north east) {};
  \node[rscasnode] (p2) at (-4,-0.55) {
    $\mathsf{n}_2$:\\
    $(\vtstr{\underline{5D}\,A8}, \ptstr{Sources/}, P)$
  };
  \node[ptable] (m2) at (p2.south) {};
  \foreach \i in {0,4,...,28} {
    \draw ([xshift=\i mm] m2.north west) -- ([xshift=\i mm] m2.south west);
  }
  \node[ptablenode] (m2l1)
    at ([xshift=2mm,yshift=1mm] m2.south west) {$\cdots$};
  \node[ptablenode] (m2l2)
    at ([xshift=6mm,yshift=1mm] m2.south west) {\underline{M}};
  \node[ptablenode] (m2l3)
    at ([xshift=10mm,yshift=1mm] m2.south west) {$\cdots$};
  \node[ptablenode] (m2l4)
    at ([xshift=14mm,yshift=1mm] m2.south west) {$\cdots$};
  \node[ptablenode] (m2l5)
    at ([xshift=18mm,yshift=1mm] m2.south west) {$\cdots$};
  \node[ptablenode] (m2l6)
    at ([xshift=22mm,yshift=1mm] m2.south west) {\underline{S}};
  \node[ptablenode] (m2l7)
    at ([xshift=26mm,yshift=1mm] m2.south west) {$\cdots$};
  \node[anchor=north east,littlemarker] at (m2.north east) {};
  \node[partition] (p3) at (-1.7,-0.85) {
  {\scriptsize $\mathsf{L}^{5,6} = (7, 2, 2, \bullet)$}
  \\
  \begin{tabular}{|lcc|}
    \hline
    \cellcolor{coltbd}\texttt{\textbf{fs/ext}3/inode.c\$}
    & \cellcolor{coltbd}\vtistr{\underline{\textbf{5E}}\,F2\,9C\,59}
    & \cellcolor{coltbd}$\mathsf{r}_4$
    \\
    \cellcolor{coltbd}\texttt{\textbf{fs/ext}4/inode.c\$}
    & \cellcolor{coltbd}\vtistr{\underline{\textbf{5E}}\,BD\,23\,C2}
    & \cellcolor{coltbd}$\mathsf{r}_5$
    \\
    \cline{1-3}
  \end{tabular}
  };
  \node[anchor=north west,littlemarker] at ([yshift=-9pt] p3.north west) {};
  \node[partition] (p4) at (3,-0.85) {
    {\scriptsize $\mathsf{L}^{2,3,7} = (1, 3, 3, \bullet)$}
  \\
  \begin{tabular}{|lcc|}
    \hline
    \cellcolor{coltbd}\texttt{crypto/ecc.h\$}
    & \cellcolor{coltbd}\vtistr{\textbf{\underline{5F}\,BD}\,8D\,C4}
    & \cellcolor{coltbd}$\mathsf{r}_2$
    \\
    \cellcolor{coltbd}\texttt{crypto/ecc.c\$}
    & \cellcolor{coltbd}\vtistr{\textbf{\underline{5F}\,BD}\,8D\,C4}
    & \cellcolor{coltbd}$\mathsf{r}_2$
    \\
    \cellcolor{coltbd}\texttt{fs/ext4/inode.c\$}
    & \cellcolor{coltbd}\vtistr{\textbf{\underline{5F}\,BD}\,3D\,5A}
    & \cellcolor{coltbd}$\mathsf{r}_6$
    \\
    \cellcolor{coltbd}
    & \cellcolor{coltbd}
    & \cellcolor{coltbd}
    \\
    \hline
  \end{tabular}
  };
  \node[anchor=north west,littlemarker] at ([yshift=-9pt] p4.north west) {};
  \node[partition] (p5) at (-7,-1.8) {
    {\scriptsize $\mathsf{L}^{1} = (8, 3, 1, \bullet)$}
  \\
  \begin{tabular}{|lcc|}
    \hline
    \cellcolor{coltbd}\texttt{\textbf{\underline{M}ap.go\$}}
    & \cellcolor{coltbd}\vtistr{\textbf{94\,2A}}
    & \cellcolor{coltbd}$\mathsf{r}_7$
    \\
    \cellcolor{coltbd}
    & \cellcolor{coltbd}
    & \cellcolor{coltbd}
    \\
    \hline
  \end{tabular}
  };
  \node[anchor=north west,littlemarker] at ([yshift=-9pt] p5.north west) {};
  \node[partition] (p7) at (-3.00,-1.8) {
    {\scriptsize $\mathsf{L}^{4,8,9} = (5, 1, 3, \bullet)$}
  \\
  \begin{tabular}{|lcc|}
    \hline
    \cellcolor{coltbd}\texttt{\textbf{\underline{S}che}ma.go\$}
    & \cellcolor{coltbd}\vtistr{94\,8C}
    & \cellcolor{coltbd}$\mathsf{r}_3$
    \\
    \cellcolor{coltbd}\texttt{\textbf{\underline{S}che}dule.go\$}
    & \cellcolor{coltbd}\vtistr{97\,8B}
    & \cellcolor{coltbd}$\mathsf{r}_7$
    \\
    \cellcolor{coltbd}\texttt{\textbf{\underline{S}che}duler.swift\$}
    & \cellcolor{coltbd}\vtistr{97\,8B}
    & \cellcolor{coltbd}$\mathsf{r}_7$
    \\
    \cellcolor{coltbd}
    & \cellcolor{coltbd}
    & \cellcolor{coltbd}
    \\
    \hline
  \end{tabular}
  };
  \node[anchor=north west,littlemarker] at ([yshift=-9pt] p7.north west) {};
  \draw[->] (m1l2.south) -- (p2);
  \draw[->] (m1l3.south) -- (p3);
  \draw[->] (m1l4.south) -- (p4);
  \draw[->] (m2l2.south) -- (p5);
  \draw[->] (m2l6.south) -- (p7.north west);
\end{tikzpicture}
\caption{$\mathsf{L}^{1,4,8,9}$ from (b) is $\psi$-partitioned in dimension $P$}
\end{subfigure}
\\[0.5cm]\hline
\end{tabular}}
\caption{The keys are recursively $\psi$-partitioned depth-first,
creating new RSCAS nodes in pre-order.  A node represents the longest
common path and value prefixes of its corresponding partition.}
\label{fig:recpart}
\end{figure*}

\subsection{Bulk-Loading a Disk-Based RSCAS Trie}
\label{sec:bulk}

We create and bulk-load a new disk-based RSCAS trie whenever the
in-memory trie $R^M_0$ overflows.  The bulk-loading algorithm
constructs RSCAS while, at the same time, dynamically interleaving a
set of keys.  Bulk-loading RSCAS is difficult because all keys must be
considered together to dynamically interleave them.  The bulk-loading
algorithm starts with all non-interleaved keys in the \emph{root
  partition}.  We use partitions during bulk-loading to temporarily
store keys along with their discriminative bytes. Once a partition has
been processed, it is deleted.

\begin{definition}[Partition]
  A partition $L = (g_P,g_V,\textsf{size},\textsf{ptr})$ stores a set
  $K$ of composite keys.  $g_P = \textsf{dsc}(K,P)$ and
  $g_V = \textsf{dsc}(K,V)$ denote the discriminative path and value
  byte, respectively. $\textsf{size} = |K|$ denotes the number of keys
  in the partition.  $L$ is either memory-resident or disk-resident,
  and $\textsf{ptr}$ points to the keys in memory or on disk.
  $\hfill\Box$
\end{definition}

\begin{example}
  Root partition $\mathsf{L}^{1..9} = (2,5,9,\bullet)$ in Figure
  \ref{fig:recpart}a stores keys $\mathsf{K}^{1..9}$ from Table
  \ref{tab:ex2}.  The longest common prefixes of $\mathsf{L}^{1..9}$
  are type-set in bold-face. The first bytes after these prefixes are
  $\mathsf{L}^{1..9}$'s discriminative bytes $g_P = 2$ and $g_V = 5$.
  We use placeholder $\bullet$ for pointer $\textsf{ptr}$; we describe
  later how to decide if partitions are stored on disk or in memory.
  $\hfill\Box$
\end{example}

Bulk-loading starts with root partition $L$ and breaks it into smaller
partitions using the $\psi$-partitioning until a partition contains at
most $\tau$ keys.  The $\psi$-partitioning $\psi(L,D)$ groups keys
together that have the same prefix in dimension $D$, and returns a
\emph{partition table} where each entry in this table points to a new
partition $L_i$.  We apply $\psi$ alternatingly in dimensions $V$ and
$P$ to interleave the keys at their discriminative bytes.  In each
call, the algorithm adds a new node to RSCAS with $L$'s longest common
path and value prefixes.

\begin{example}
  Figure \ref{fig:recpart} shows how the RSCAS from
  Figure~\ref{fig:solution} is built.  In Figure \ref{fig:recpart}b we
  extract $\mathsf{L}^{1..9}$'s longest common path and value prefixes
  and store them in the new root node $\mathsf{n}_1$.  Then, we
  $\psi$-partition $\mathsf{L}^{1..9}$ in dimension $V$ and obtain a
  partition table (light green) that points to three new partitions:
  $\mathsf{L}^{1,4,8,9}$, $\mathsf{L}^{5,6}$, and
  $\mathsf{L}^{2,3,7}$.  We drop $\mathsf{L}^{1..9}$'s longest common
  prefixes from these new partitions.  We proceed recursively with
  $\mathsf{L}^{1,4,8,9}$. In Figure \ref{fig:recpart}c we create node
  $\mathsf{n}_2$ as before and this time we $\psi$-partition in
  dimension $P$ and obtain two new partitions.  Given $\tau = 2$,
  $\mathsf{L}^1$ is not partitioned further, but in the next recursive
  step, $\mathsf{L}^{4,8,9}$ would be partitioned one last time in
  dimension $V$.  $\hfill\Box$
  %
\end{example}

To avoid scanning $L$ twice (first to compute the discriminative byte;
second to compute $\psi(L,D)$) we make the $\psi$-partitioning
\emph{proactive} by exploiting that $\psi(L,D)$ is applied
hierarchically. This means we pre-compute the discriminative bytes of
every new partition $L_i \in \psi(L,D)$ as we $\psi$-partition $L$.
As a result, by the time $L_i$ itself is $\psi$-partitioned, we
already know its discriminative bytes and can directly compute the
partitioning.  Algorithm~\ref{alg:merge} in Section~\ref{sec:merging}
shows how to compute the root partition's discriminative bytes; the
discriminative bytes of all subsequent partitions are computed
proactively during the partitioning itself.  This halves the scans
over the data during bulk-loading.

\subsubsection{Bulk-Loading Algorithm}

The bulk-loading algorithm (Algorithm \ref{alg:bulk}) takes three
parameters: a partition $L$ (initially the root partition), the
partitioning dimension $D$ (initially dimension $V$), and the position
in the trie file where the next node is written to (initially 0).
Each invocation adds a node $n$ to the RSCAS trie and returns the
position in the trie file of the first byte after the subtree rooted
in $n$.  Lines 1--3 create node $n$ and set its longest common
prefixes $n.s_P$ and $n.s_V$, which are extracted from a key $k \in L$
from the first byte up to, but excluding, the positions of $L$'s
discriminative bytes $L.g_P$ and $L.g_V$.  If the number of keys in
the current partition exceeds the partitioning threshold $\tau$ and
$L$ can be $\psi$-partitioned, we break $L$ further up.  In lines 5--6
we check if we can indeed $\psi$-partition $L$ in $D$ and switch to
the alternate dimension $\overline{D}$ otherwise.  In line 8 we apply
$\psi(L,D)$ and obtain a partition table $T$, which is a $2^8$-long
array that maps the $2^8$ possible values $b$ of a discriminative byte
($\texttt{0x00} \leq b \leq \texttt{0xFF}$) to partitions. We write
$T[b]$ to access the partition for value $b$ ($T[b] = \nil$ if no
partition exists for value $b$).  $\psi(L,D)$ drops $L$'s longest
common prefixes from each key $k \in L$ since we store these prefixes
already in node $n$.  We apply Algorithm \ref{alg:bulk} recursively on
each partition in $T$ with the alternate dimension $\overline{D}$,
which returns the position where the next child is written to on disk.
We terminate if partition $L$ contains no more than $\tau$ keys or
cannot be partitioned further.  We iterate over all remaining keys in
$L$ and store their non-interleaved suffixes in the set
$n.\textsf{suffixes}$ of leaf node $n$ (lines 16--19).  Finally, in
line 22 we write node $n$ to disk at the given offset in the trie
file.

\begin{algorithm2e}[htb] \scriptsize
  \SetInd{0.65em}{0.65em} 
  Let $n$ be a new node, $k$ a key in $L$\;
  $n.s_P \gets k.P[1, L.g_P - 1]$\;
  $n.s_V \gets k.V[1, L.g_V - 1]$\;
  \eIf{\textnormal{$L.\textsf{size} > \tau \wedge (L.g_P > |k.P| \vee L.g_V > |k.V|)$}} {
    \lIf{\textnormal{$D=P \wedge L.g_P > |k.P|$}}{%
      $D \gets V$%
    }
    \lElseIf{\textnormal{$D=V \wedge L.g_V > |k.V|$}}{%
      $D \gets P$%
    }
    $n.D \gets D$\;
    $T \gets \psi(L,D)$\;
    $\text{pos} \gets \textsf{preorderPos} + \textsf{size}(n)$\;
    \For{\textnormal{$b \gets \texttt{0x00}$ \KwTo $\texttt{0xFF}$}} {
      \If{\textnormal{$T[b] \neq \nil$}} {
        $n.\textsf{children}[b] \gets \text{pos}$\;
        $\text{pos} \gets \textsf{BulkLoad}(T[b],
        \overline{D}, \text{pos})$\;
      }
    }
  }{
    $n.D \gets \bot$\;
    \ForEach{\textnormal{key $k \in L$}}{%
      $s_P \gets k.P[L.g_P, |k.P|]$\;
      $s_V \gets k.V[L.g_V, |k.V|]$\;
      $n.\textsf{suffixes} \gets n.\textsf{suffixes} \cup \{ (s_P, s_V, k.R) \}$\;
    }
    Delete $L$\;
    $\text{pos} \gets \textsf{preorderPos} + \textsf{size}(n)$\;
  }
  Write node $n$ to disk from position \textsf{preorderPos} to
  $\textsf{preorderPos}+\textsf{size}(n)$\;
  \Return{\textnormal{\text{pos}}}\;
  \caption{$\textsf{BulkLoad}(L, D, \textsf{preorderPos})$}
  \label{alg:bulk}
\end{algorithm2e}

Algorithm \ref{alg:psi} implements $\psi(L,D)$. We organize the keys
in a partition $L$ at the granularity of pages so that we can
seamlessly transition between memory- and disk-resident partitions.  A
page is a fixed-length buffer that contains a variable number of
keys. If $L$ is disk-resident, $L.\textsf{ptr}$ points to a
page-structured file on disk and if $L$ is memory-resident,
$L.\textsf{mptr}$ points to the head of a singly-linked list of pages.
Algorithm \ref{alg:psi} iterates over all pages in $L$ and for each
key in a page, line 6 determines the partition $T[b]$ to which $k$
belongs by looking at its value $b$ at the discriminative byte.  Next
we drop the longest common path and value prefixes from $k$ (lines
7--8).  We proactively compute $T[b]$'s discriminative bytes whenever
we add a key $k$ to $T[b]$ (lines 10--17). Two cases can arise. If $k$
is $T[b]$'s first key, we initialize partition $T[b]$. If $L$ fits
into memory, we make $T[b]$ memory-resident, else disk-resident.  We
initialize $g_P$ and $g_V$ with one past the length of $k$ in the
respective dimension (lines 9--12).  These values are valid
upper-bounds for the discriminative bytes since keys are prefix-free.
We store $k$ as a reference key for partition $T[b]$ in
$\textsf{refkeys}[b]$. If $k$ is not the first key in $T[b]$, we
update the upper bounds (lines 13--17) as follows.  Starting from the
first byte, we compare $k$ with reference key $\textsf{refkeys}[b]$
byte-by-byte in both dimension until we reach the upper-bounds
$T[b].g_P$ and $T[b].g_V$, or we find new discriminative bytes and
update $T[b].g_P$ and $T[b].g_V$.

\begin{algorithm2e}[htbp] \scriptsize
  \SetInd{0.65em}{0.65em} 
  \SetKw{KwDownTo}{down to}
  \tikzmk{A}
  \tikz[remember picture,overlay]{%
    \draw[fill=blue,opacity=.25] ($(A)+(+0.6,-2.18)$) rectangle ($(A)+(8.0,-4.9)$);%
    \node[colxb,anchor=north east,align=left,font=\small] at (8.0,-2.9) {%
      proactively compute\\
      discriminative bytes
    };
  }%
  Let $T$ be a new partition table\;
  Let $\textsf{outpages}$ be an array of $2^8$ pages for output buffering\;
  Let $\textsf{refkeys}$ be an array to store $2^8$ composite keys\;
  \ForEach{\textnormal{$\textsf{page} \in L.\textsf{ptr}$}}{
    \ForEach{\textnormal{key $k \in \textsf{page}$}}{
      \leIf{$D=P$}{$b \gets k.P[L.g_P]$}{$b \gets k.V[L.g_V]$}
      $k.P \gets k.P[L.g_P, |k.P|]$\;
      $k.V \gets k.V[L.g_V, |k.V|]$\;
      \eIf{\textnormal{$T[b] = \nil$}}{
        \leIf{\textnormal{$L$ fits into memory}}%
          {$\textsf{ptr} \gets$ new linked list}%
          {$\textsf{ptr} \gets$ new file}
        $T[b] \gets (|k.P|{+}1, |k.V|{+}1, 0, \textsf{ptr})$\;
        $\textsf{refkeys}[b] \gets k$\;
      }{
        $k', g_P, g_V \gets (\textsf{refkeys}[b], 1, 1)$\;
        \lWhile{\textnormal{$g_P < T[b].g_P \wedge k.P[g_P] = k'.P[g_P]$}}{%
          $g_P\text{++}$%
        }
        \lWhile{\textnormal{$g_V < T[b].g_V \wedge k.V[g_V] = k'.V[g_V]$}}{%
          $g_V\text{++}$%
        }
        $T[b].g_P, T[b].g_V \gets (g_P,\, g_V)$\;
      }
      \If{\textnormal{$\textsf{outpages}[b]$ is full}} {
        $\textsf{Push}(T[b].\textsf{ptr},\textsf{outpages}[b])$\;
        Clear contents of page $\textsf{outpages}[b] $\;
      }
      Add $k$ to $\textsf{outpages}[b]$\;
      $T[b].\textsf{size}\text{++}$\;
    }
    Delete $\textsf{page}$\;
  }
  \For{\textnormal{$b \gets \texttt{0x00}$ \KwTo $\texttt{0xFF}$}} {
    \lIf{\textnormal{$T[b] \neq \nil$}}{%
      $\textsf{Push}(T[b].\textsf{ptr},\textsf{outpages}[b])$%
    }
  }
  Delete $L$\;
  \Return{$T$}\;
  \caption{$\psi(L, D)$}
  \label{alg:psi}
\end{algorithm2e}

\subsection{Merging RSCAS Tries Upon Overflow}
\label{sec:merging}

When the memory-resident trie $R^M_0$ reaches its maximum size of $M$
keys, we move its keys to the first disk-based trie $R_i$ that is
empty using Algorithm \ref{alg:merge}.  We keep pointers to the root
nodes of all tries in an array.  Algorithm~\ref{alg:merge} first
collects all keys from tries $R^M_0$, $R_0,\ldots,R_{i-1}$ and stores
them in a new partition $L$ (lines 2--4).  Next, in lines 5--11, we
compute $L$'s discriminative bytes $L.g_P$ and $L.g_V$ from the
substrings $s_P$ and $s_V$ of the root nodes of the $i$ tries.
Finally, in lines 12--14, we bulk-load trie $R_i$ and delete all
previous tries.

\begin{algorithm2e}[htb] \scriptsize
  \SetInd{0.65em}{0.65em} 
  Let $i$ be the smallest number such that index $R_i$ is empty\;
  Let $L$ be a new disk-resident partition\;
  \ForEach{\textnormal{trie $R \in \{ R^M_0,R_0,\ldots,R_{i-1} \}$}}{
    Collect all composite keys in $R$ and store them in $L.\textsf{ptr}$\;
  }
  $\{n^M_0, n_0, \ldots, n_{i-1}\} \gets$ root nodes of all tries
     $R^M_0,R_0,\ldots,R_{i-1}$\;
  $L.g_P, L.g_V \gets (|n^M_0.s_P|+1, |n^M_0.s_V|+1)$\;
  \ForEach{\textnormal{root node $n \in \{n_0, \ldots, n_{i-1}\}$}}{
    $g_P, g_V \gets (1, 1)$\;
    \lWhile{\textnormal{$g_P < L.g_P \wedge n^M_0.s_P[g_P] = n.s_P[g_P]$}}{%
      $g_P\text{++}$%
    }
    \lWhile{\textnormal{$g_V < L.g_V \wedge n^M_0.s_P[g_V] = n.s_V[g_V]$}}{%
      $g_V\text{++}$%
    }
    $L.g_P, L.g_V \gets (g_P, g_V)$\;
  }
  Create new trie file $R_i$\;
  $\textsf{BulkLoad}(L, V, \text{position 0 in $R_i$'s trie file})$\;
  Delete tries $R^M_0,R_0,\ldots,R_{i-1}$\;
  \caption{$\textsf{HandleOverflow}$}
  \label{alg:merge}
\end{algorithm2e}

\subsection{Analytical Evaluation}
\label{sec:analysis}

\subsubsection{Total I/O Overhead During Bulk-Loading}

The I/O overhead is the number of page I/Os without reading the input
and writing the output.  We use $N$, $M$, and $B$ for the number of
input keys, the number of keys that fit into memory, and the number of
keys that fit into a page, respectively \cite{AA88}.  We analyze the
I/O overhead of Algorithm \ref{alg:bulk} for a uniform data
distribution with a balanced RSCAS and for a maximally skewed
distribution with an unbalanced RSCAS.  The $\psi$-partitioning splits
a partition into equally sized partitions.  Thus, with a fixed fanout
$f$ the $\psi$-partitioning splits a partition into $f$,
$2 \leq f \leq 2^8$, partitions.

\begin{lemma} \label{lemma:bestcase}
  The I/O overhead to build RSCAS with Algorithm \ref{alg:bulk} from
  uniformly distributed data is
  $$ 2 \times \ceil[\big]{\log_f \ceil[\big]{\frac{N}{M}}} \times
  \ceil[\Big]{\frac{N}{B}} $$
\end{lemma}

\begin{example}
  We compute the I/O overhead for $N = 16$, $M = 4$, $B = 2$, and
  $f=2$.  There are
  $\lceil \log_2 \lceil \frac{16}{4} \rceil \rceil = 2$ intermediate
  levels with the data on disk.  On each level we read and write
  $\frac{16}{2} = 8$ pages.  In total, the disk I/O is
  $2 \times 2 \times 8 = 32$.  $\hfill\Box$
\end{example}

For maximally skewed data RSCAS deteriorates to a trie whose height is
linear in the number of keys in the dataset.

\begin{lemma} \label{lemma:worstcase}
  The I/O overhead to build RSCAS with Algorithm \ref{alg:bulk} from
  maximally skewed data is
  {$$ 2 \times {\sum_{i=1}^{N - \ceil{\frac{M}{B}}B}}
  \Big(\ceil[\Big]{\frac{N-i}{B}} + 1\Big)
  $$}
\end{lemma}

\begin{example}
  We use the same parameters as in the previous example but assume
  maximally skewed data.  There are $16-\ceil{\frac{4}{2}}2 = 12$
  levels before the partitions fit into memory. For example, at level
  $i=1$ we write and read $\ceil{\frac{16-1}{2}} = 8$ pages for
  $L_{1,2}$. In total, the I/O overhead is $144$ pages. $\hfill\Box$
\end{example}

\begin{theorem} \label{theorem:iooverhead}
  The I/O overhead to build RSCAS with Algorithm~\ref{alg:bulk}
  depends on the data distribution.  It is lower bounded by
  $O(\log(\frac{N}{M}) \frac{N}{B})$ and upper bounded by
  $O((N-M)\frac{N}{B})$.
\end{theorem}

Note that, since RSCAS is trie-based and keys are encoded by the path
from the root to the leaves, the height of the trie is bounded by the
length of the keys.  The worst-case is very unlikely in practice
because it requires that the lengths of the keys is linear in the
number of keys. Typically, the key length is at most tens or hundreds
of bytes. We show in Section~\ref{sec:experiments} that building RSCAS
performs close to the best case on real world data.

\subsubsection{Amortized I/O Overhead During Insertions}

Next, we consider the amortized I/O overhead of a single insertion
during a series of $N$ insertions into an empty trie.  Note that $M-1$
out of $M$ consecutive insertions incur no disk I/O since they are
handled by the in-memory trie $R^M_0$.  Only the $M$th insertion
bulk-loads a new disk-based trie.

\begin{lemma} \label{lemma:amortized}
  Let $\textsf{cost}(N,M,B)$ be the I/O overhead of bulk-loading RSCAS.
  The amortized I/O overhead of one insertion out of $N$ insertions
  into an initially empty trie is $O(\frac{1}{N} \times
  \log_2(\frac{N}{M}) \times \textsf{cost}(N,M,B)) $.
\end{lemma}

\section{Experimental Evaluation}
\label{sec:experiments}

\subsection{Setup}

\textbf{Environment.} We use a Debian 10 server with 80 cores and
400\,GB main memory. The machine has six hard disks, each 2\,TB big,
that are configured in a RAID 10 setup. The code is written in C++ and
compiled with g++ 8.3.0. \smallskip

\noindent\textbf{Datasets.} We use three real-world datasets and one synthetic
dataset. Table \ref{tab:datasets} provides an overview.

\begin{itemize}

\item \emph{GitLab.} The GitLab data from SWH contains archived copies
  of all publicly available GitLab repositories up to 2020-12-15. The
  dataset contains \num{914593} archived repositories, which
  correspond to a total of \num{120071946} unique revisions and
  \num{457839953} unique files.  For all revisions in the GitLab
  dataset we index the commit time and the modified files (equivalent
  to ``commit diffstats'' in version control system terminology). In
  total, we index \num{6.9} billion composite keys similar to Table
  \ref{tab:ex2}.

\item \emph{ServerFarm.} The ServerFarm dataset \cite{KW20} mirrors
  the file systems of 100 Linux servers.  For each server we installed
  a default set of packages and randomly picked a subset of optional
  packages.  In total there are 21 million files. For each file we
  record the file's full path and size.

\item \emph{Amazon.} The Amazon dataset \cite{RH16} contains
  hierarchically categorized products.  For each product its location
  in the hierarchical categorization (the path) and its price in cents
  (the value) are recorded.  For example, the shoe `evo' has path
  \texttt{/sports/outdoor/running/evo} and its price is \num{10000}
  cents.

\item \emph{XMark.} The XMark dataset \cite{AS02} is a synthetic
  dataset that models a database for an internet auction site. It
  contains information about people, regions (subdivided by
  continent), etc.  We generated the dataset with scale factor 500 and
  we index the numeric attribute `category'.

\end{itemize}

\begin{table}[htb]\centering\setlength{\tabcolsep}{4pt}
\caption{Dataset Statistics}
\label{tab:datasets}
{\scriptsize\begin{tabular}{l|rrrr}
  & \multicolumn{1}{c}{\textbf{GitLab}}
  & \multicolumn{1}{c}{\textbf{ServerFarm}}
  & \multicolumn{1}{c}{\textbf{Amazon}}
  & \multicolumn{1}{c}{\textbf{XMark}}
    \\
  \hline
  Origin
    & SWH
    & \cite{KW20}
    & \cite{RH16}
    & \cite{AS02}
    \\
  Attribute
    & Commit time
    & Size
    & Price
    & Category
    \\
  Type
    & real-world
    & real-world
    & real-world
    & synthetic
    \\
  Size
    &  1.6\,TB
    &  3.0\,GB
    & 10.5\,GB
    & 58.9\,GB
    \\
  Nr.~Keys
    & \num{6891972832}
    & \num{21291019}
    &  \num{6707397}
    &  \num{60272422}
    \\
  Nr.~Unique Keys
    & \num{5849487576}
    & \num{   9345668}
    & \num{   6461587}
    & \num{   1506408}
    \\
  Nr.~Unique Paths
    & \num{340614623}
    & \num{  9331389}
    & \num{  6311076}
    & \num{        7}
    \\
  Nr.~Unique Values
    & \num{81829152}
    & \num{  234961}
    & \num{   47852}
    & \num{  389847}
    \\
  Avg.~Key Size
    & 79.8\,B
    & 79.8\,B
    & 119.3\,B
    & 54.8\,B
    \\
  Total Size of Keys
    & 550.2\,GB
    & 1.7\,GB
    & 0.8\,GB
    & 3.3\,GB
\end{tabular}}
\end{table}

\textbf{Previous Results.} In our previous work \cite{KW21} we
compared RSCAS to state-of-the-art research solutions.  We compared
RSCAS to the CAS index by Mathis et al.~\cite{CM15}, which indexes
paths and values in separate index structures.  We also compared RSCAS
to a trie-based index where composite keys are integrated with four
different methods: (i) the $z$-order curve with surrogate functions to
map variable-length keys to fixed-length keys, (ii) a label-wise
interleaving where we interleave one path label with one value byte,
(iii) the path-value concatenation, and (iv) value-path concatenation.
Our experiments showed that the approaches do not provide robust CAS
query performance because they may create large intermediate results.

\textbf{Compared Approaches.} This paper compares RSCAS to scalable
state-of-the-art industrial-strengths systems.  First, we compare
RSCAS to Apache Lucene \cite{lucene}, which builds separate indexes
for the paths and values. Lucene creates an FST on the paths and a
Bkd-tree \cite{OP03} on the values. Lucene evaluates CAS queries by
decomposing queries into their two predicates, evaluating the
predicates on the respective indexes, and intersecting the sorted
posting lists to produce the final result. Second, we compare RSCAS to
composite B-trees in Postgres. This simulates the two possible
$c$-order curves that concatenate the paths and values (or vice
versa).  We create a table $\texttt{data}(P,V,R)$, similar to Table
\ref{tab:ex2}, and create two composite B+ trees on attributes $(P,V)$
and $(V,P)$, respectively.

\textbf{Parameters.} Unless otherwise noted, we set the partitioning
threshold $\tau = 100$ based on experiments in Section
\ref{sec:calibration}.  The number of keys $M$ that the main-memory
RSCAS trie $R^M_0$ can hold is $M = 10^8$.

\textbf{Artifacts.} The code and the datasets used for our experiments
are available
online.\footnote{\url{https://github.com/k13n/scalable_rcas}}

\subsection{Impact of Datasets on RSCAS's Structure}
\label{sec:expstructure}

\begin{figure}[htb]
\begin{tikzpicture}
  \begin{groupplot}[
    col2,
    width=46mm,
    xticklabel style={
      align=center,
    },
    group style={
      columns=2,
      rows=4,
      horizontal sep=25pt,
      vertical sep=20pt,
    },
    legend style={
      at={(0.5,1.00)},
      anchor=south west,
      legend columns=-1,
      draw=none,
      /tikz/every even column/.append style={column sep=3pt},
    },
    ymin=0,
    xtick pos=left,
  ]
  \node[anchor=north] at (1.40,2.20)
    {\small \textbf{Node Depth}};
  \node[anchor=north] at (5.50,2.20)
    {\small \textbf{Node Fanout}};
  \nextgroupplot[
    ylabel={\textbf{GitLab}\\[2pt] Frequency [\%]},
    ybar,
    bar width=1pt,
    ymax=42,
    xmax=30,
  ]
  \pgfplotstableread[col sep=semicolon]{experiments/03_structure/swh_depth.csv}\plot
  \addplot[barRCASsolid] table[x=value,y=percent] {\plot};
  \draw[densely dotted] (9.6,0) -- (9.6,42);
  \node[anchor=west] at (9.6,25) {Avg: 9.6\\\MREV{Exp: 8.7}};
  \node[boxlabel] at (rel axis cs:1,1) {(a)};
  \nextgroupplot[
    ybar,
    bar width=1pt,
    ymax=42,
    xmax=40,
  ]
  \pgfplotstableread[col sep=semicolon]{experiments/03_structure/swh_fanout.csv}\plot
  \addplot[barRCASsolid] table[x=value,y=percent] {\plot};
  \draw[densely dotted] (8.0,0) -- (8.0,105);
  \node[anchor=south west] at (8.0,25) {Avg: 8.0};
  \node[boxlabel] at (rel axis cs:1,1) {(b)};
  \nextgroupplot[
    ylabel={\textbf{ServerFarm}\\[2pt] Frequency [\%]},
    ybar,
    bar width=1pt,
    ymax=42,
    xmax=30,
  ]
  \pgfplotstableread[col sep=semicolon]{experiments/03_structure/serverfarm_depth.csv}\plot
  \addplot[barRCASsolid] table[x=value,y=percent] {\plot};
  \draw[densely dotted] (10.6,0) -- (10.6,105);
  \node[anchor=west,align=left] at (10.6,25) {Avg: 10.6\\\MREV{Exp: 6.1}};
  \node[boxlabel] at (rel axis cs:1,1) {(c)};
  \nextgroupplot[
    ybar,
    bar width=1pt,
    ymax=42,
    xmax=40,
  ]
  \pgfplotstableread[col sep=semicolon]{experiments/03_structure/serverfarm_fanout.csv}\plot
  \addplot[barRCASsolid] table[x=value,y=percent] {\plot};
  \draw[densely dotted] (7.6,0) -- (7.6,105);
  \node[anchor=south west] at (7.6,25) {Avg: 7.6};
  \node[boxlabel] at (rel axis cs:1,1) {(d)};
  \nextgroupplot[
    ylabel={\textbf{Amazon}\\[2pt] Frequency [\%]},
    ybar,
    bar width=1pt,
    ymax=42,
    xmax=30,
  ]
  \pgfplotstableread[col sep=semicolon]{experiments/03_structure/amazon_depth.csv}\plot
  \addplot[barRCASsolid] table[x=value,y=percent] {\plot};
  \draw[densely dotted] (6.9,0) -- (6.9,105);
  \node[anchor=west,align=left] at (6.9,25) {Avg: 6.9\\\MREV{Exp: 4}};
  \node[boxlabel] at (rel axis cs:1,1) {(e)};
  \nextgroupplot[
    ybar,
    bar width=1pt,
    ymax=42,
    xmax=40,
  ]
  \pgfplotstableread[col sep=semicolon]{experiments/03_structure/amazon_fanout.csv}\plot
  \addplot[barRCASsolid] table[x=value,y=percent] {\plot};
  \draw[densely dotted] (16.9,0) -- (16.9,105);
  \node[anchor=south west] at (16.9,23) {Avg: 16.9};
  \node[boxlabel] at (rel axis cs:1,1) {(f)};
  \nextgroupplot[
    ylabel={\textbf{XMark}\\[2pt] Frequency [\%]},
    ybar,
    xtick=data,
    bar width=4pt,
  ]
  \pgfplotstableread[col sep=semicolon]{experiments/03_structure/xmark_depth.csv}\plot
  \addplot[barRCASsolid] table[x=value,y=percent] {\plot};
  \draw[densely dotted] (4.7,0) -- (4.7,105);
  \node[anchor=east] at (4.8,40) {Avg: 4.8\\\MREV{Exp: 5.1}};
  \node[boxlabel] at (rel axis cs:1,1) {(g)};
  \nextgroupplot[
    ybar,
    bar width=1pt,
    xmax=40,
    ymin=-1,
  ]
  \pgfplotstableread[col sep=semicolon]{experiments/03_structure/xmark_fanout.csv}\plot
  \addplot[barRCASsolid] table[x=value,y=percent] {\plot};
  \draw[densely dotted] (32.6,0) -- (32.6,105);
  \node[anchor=south east] at (32.6,35) {Avg: 32.6};
  \node[boxlabel] at (rel axis cs:1,1) {(h)};
  \end{groupplot}
\end{tikzpicture}
\caption{Structure of the RSCAS trie}
\label{exp:shape}
\end{figure}
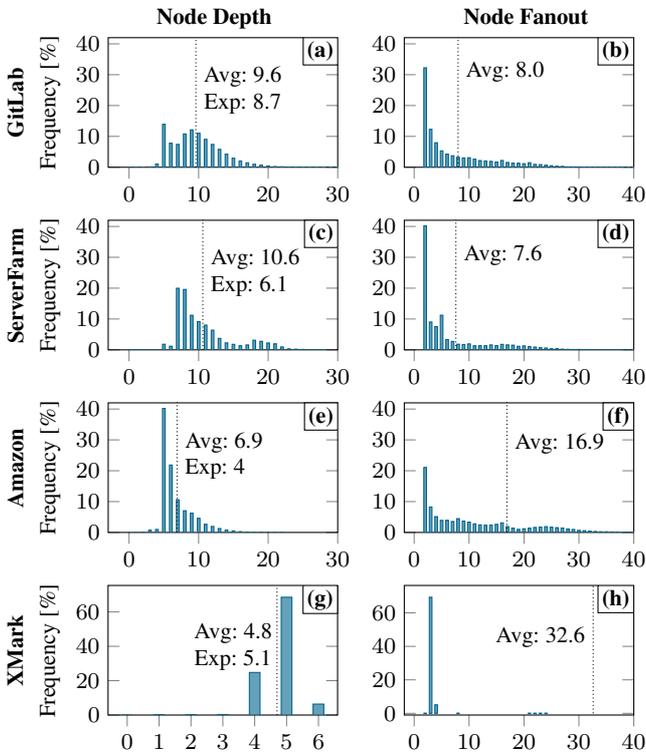

In Figure \ref{exp:shape} we show how the shape (depth and width) of the RSCAS
trie adapts to the datasets.  Figure \ref{exp:shape}a shows the distribution
of the node depths in the RSCAS trie for the GitLab dataset. Because of its
trie-based structure not every root-to-leaf path in RSCAS has the same length
(see also Figure \ref{fig:solution}).  The average node depth is about 10,
with 90\% of all nodes occurring no deeper than level 14. \MREV{The expected
depth is $\log_{\bar{f}}\lceil{\frac{N}{\tau}}\rceil =
\log_{8}\lceil{\frac{\text{6.9B}}{100}}\rceil=8.7$, where $N$ is the number of
keys, $\tau$ is the partitioning threshold that denotes the maximum size of a
leaf partition, and $\bar{f}$ is the average fanout. The actual average depth
is higher than the expected depth since the GitLab dataset is skewed and the
expected depth assumes a uniformly distributed dataset.} In the GitLab dataset
the average key length is 80 bytes, but the average node depth is 10, meaning
that RSCAS successfully extracts common prefixes.  Figure \ref{exp:shape}b
shows how the fanout of the nodes is distributed. Since RSCAS
$\psi$-partitions the data at the granularity of bytes, the fanout of a node
is upper-bounded by $2^8$, but in practice most nodes have a smaller fanout
(we cap the x-axis in Figure \ref{exp:shape}b at fanout 40, because there is a
long tail of high fanouts with low frequencies).  Nodes that $\psi$-partition
the data in the path dimension typically have a lower fanout because most
paths contain only printable ASCII characters (of which there are about 100),
while value bytes span the entire available byte spectrum.

\begin{table*}
\caption{CAS queries with their result size and the number of keys that
match the path, respectively value  predicate.}
\label{tab:queries}
\begin{center}
  {\scriptsize \begin{tabular}{|llrr|rrr|}%
  \hline
    & \multicolumn{1}{c}{Query path $q$}
    & \multicolumn{1}{c}{$v_l$}
    & \multicolumn{1}{c}{$v_h$}
    & \multicolumn{1}{c}{Result size $(\sigma)$}
    & \multicolumn{1}{c}{Path matches $(\sigma_P)$}
    & \multicolumn{1}{c|}{Value matches $(\sigma_V)$}
    \\
  \hline
  \multicolumn{7}{|l|}{Dataset: \textbf{GitLab} (the values are commit times
    that are stored as 64 bit Unix timestamps)} \\
  \hline
\Tstrut
  $Q_1$ & \texttt{/drivers/android/binder.c} & 09/10/17 & 09/10/17
    & \num{29}        $(4.2 \cdot 10^{-9})$
    & \num{125849}    $(1.8 \cdot 10^{-5})$
    & \num{328603}    $(4.8 \cdot 10^{-5})$
    \\
  $Q_2$ & \texttt{/drivers/android/binder.c} & 01/10/17 & 01/11/17
    & \num{3236}      $(4.7 \cdot 10^{-7})$
    & \num{125849}    $(1.8 \cdot 10^{-5})$
    & \num{72883667}  $(1.1 \cdot 10^{-2})$
    \\
  $Q_3$ & \texttt{/drivers/gpu/**} & 07/08/14 & 08/08/14
    & \num{60344}     $(8.8 \cdot 10^{-6})$
    & \num{151871503} $(2.2 \cdot 10^{-2})$
    & \num{3503076}   $(5.1 \cdot 10^{-4})$
    \\
  $Q_4$ & \texttt{/Documentation/**/arm/**/*.txt} & 06/05/13 & 22/05/13
    & \num{11720}     $(1.7 \cdot 10^{-6})$
    & \num{5927129}   $(8.6 \cdot 10^{-4})$
    & \num{22221892}  $(3.2 \cdot 10^{-3})$
    \\
  $Q_5$ & \texttt{/**/Makefile} & 22/05/12 & 04/06/12
    & \num{263754}    $(3.8 \cdot 10^{-5})$
    & \num{112037140} $(1.6 \cdot 10^{-2})$
    & \num{10932756}  $(1.6 \cdot 10^{-3})$
    \\
  $Q_6$ & \texttt{/**/ext*/inode.*} & 07/08/18 & 29/08/18
    & \num{5080}      $(7.4 \cdot 10^{-7})$
    & \num{529875}    $(7.7 \cdot 10^{-5})$
    & \num{70971382}  $(1.0 \cdot 10^{-2})$
    \\
  \hline
  \multicolumn{7}{|l|}{Dataset: \textbf{ServerFarm} (the values are
    the file sizes in bytes)} \\
  \hline
  $Q_7$ & \texttt{/usr/lib/**} & 0\,kB & 1\,kB
    & \num{512497}    $(2.4 \cdot 10^{-2})$
    & \num{2277518}   $(1.1 \cdot 10^{-1})$
    & \num{8403809}   $(3.9 \cdot 10^{-1})$
    \\
  $Q_8$ & \texttt{/usr/share/doc/**/README} & 4\,kB & 5\,kB
    & \num{521}       $(2.4 \cdot 10^{-5})$
    & \num{24698}     $(1.2 \cdot 10^{-3})$
    & \num{761513}    $(3.6 \cdot 10^{-2})$
    \\
  \hline
  \multicolumn{7}{|l|}{Dataset: \textbf{Amazon} (the values are product prices in cents)} \\
  \hline
  $Q_{9}$ & \texttt{/CellPhones\&Accessories/**} & 100\,\$ & 200\,\$
    & \num{2758}    $(4.1 \cdot 10^{-4})$
    & \num{291625}  $(4.3 \cdot 10^{-2})$
    & \num{324272}  $(4.8 \cdot 10^{-2})$
    \\
  $Q_{10}$ & \texttt{/Clothing/Women/*/Sweaters/**} & 70\,\$ & 100\,\$
    & \num{239}     $(3.6 \cdot 10^{-5})$
    & \num{4654}    $(6.9 \cdot 10^{-4})$
    & \num{269936}  $(4.0 \cdot 10^{-2})$
    \\
  \hline
  \multicolumn{7}{|l|}{Dataset: \textbf{XMark} (the values denote the
    numeric attribute category)} \\
  \hline
  $Q_{11}$ & \texttt{/site/people/**/interest} & 0 & 50000
    & \num{1910524}  $(3.2 \cdot 10^{-2})$
    & \num{19009723} $(3.2 \cdot 10^{-1})$
    & \num{6066546}  $(1.0 \cdot 10^{-1})$
    \\
  $Q_{12}$ & \texttt{/site/regions/africa/**} & 0 & 50000
    & \num{104500}   $(1.7 \cdot 10^{-3})$
    & \num{1043247}  $(1.7 \cdot 10^{-2})$
    & \num{6066546}  $(1.0 \cdot 10^{-1})$
    \\
  \hline
\end{tabular}}
\end{center}
\end{table*}

The shape of the RSCAS tries on the ServerFarm and Amazon datasets
closely resemble that of the trie on the GitLab dataset, see the
second and third row in Figure \ref{exp:shape}.  This is to be
expected since all three datasets contain a large number of unique
paths and values, see Table \ref{tab:datasets}. As a result, the data
contains a large number of discriminative bytes that are needed to
distinguish keys from one another. The paths in these datasets are
typically longer than the values and contain more discriminative
bytes. In addition, as seen above, the discriminative path bytes
typically $\psi$-partition the data into fewer partitions than the
discriminative value bytes. As a consequence, the RSCAS trie on these
three datasets is narrower and deeper than the RSCAS trie on the XMark
dataset, which has only seven unique paths and about 390k unique
values in a dataset of 60M keys. Since the majority of the
discriminative bytes in the XMark dataset are value bytes, the trie is
flatter and wider on average, see the last row in Figure
\ref{exp:shape}.

\subsection{Query Performance}

Table \ref{tab:queries} shows twelve typical CAS queries with their
query path $q$ and the value range $[v_l, v_h]$. We show for each
query the final result size and the number of keys that match the
individual predicates. In addition, we provide the selectivities of
the queries.  The selectivity $\sigma$ ($\sigma_P$) [$\sigma_V$] is
computed as the fraction of all keys that match the CAS query (path
predicate) [value predicate].  A salient characteristic of the queries
is that the final result is orders of magnitude smaller than the
results of the individual predicates.  Queries $Q_1$ through $Q_6$ on
the GitLab dataset increase in complexity.  $Q_1$ looks up all
revisions that modify one specific file in a short two-hour time
frame.  Thus, $Q_1$ is similar to a point query with very low
selectivity in both dimensions.  The remaining queries have a higher
selectivity in at least one dimension.  $Q_2$ looks up all revisions
that modify one specific file in a one-month period. Thus, its path
selectivity is low but its value selectivity is high. Query $Q_3$ does
the opposite: its path predicate matches all changes to GPU drivers
using the \texttt{**} wildcard, but we only look for revisions in a
very narrow one-day time frame.  $Q_4$ mixes the \texttt{*} and
\texttt{**} wildcards multiple times and puts them in different
locations of the query path (in the middle and towards the end). $Q_5$
looks for changes to all Makefiles, using the \texttt{**} wildcard at
the front of the query path.  Similarly, $Q_6$ looks for all changes
to files named \texttt{inode} (all file extensions are accepted with
the \texttt{*} wildcard). The remaining six queries on the other three
datasets are similar.

\begin{figure}[htb]
\centering
\begin{tikzpicture}
  \begin{groupplot}[
    col2,
    width=39mm,
    height=32mm,
    ytick scale label code/.code={},
    xticklabel shift={-7pt},
    group style={
      columns=3,
      rows=2,
      horizontal sep=0pt,
      vertical sep=20pt,
      yticklabels at=edge left,
      ylabels at=edge left,
    },
    ybar,
    ymode=log,
    ymin=1,
    ymax=10000000,
    ytick={1,1000,1000000},
    ylabel={Runtime [ms]},
    legend style={
      at={(-0.2,1.10)},
      anchor=south west,
      legend columns=-1,
      draw=none,
      /tikz/every even column/.append style={column sep=3pt},
    },
  ]
  \nextgroupplot[
    bar width=8pt,
    symbolic x coords={q},
    xticklabels={},
    xlabel={(a) Query $Q_1$},
    enlarge x limits=0.05,
  ]
  \addplot[barRCAS]         coordinates {(q, 8300)};
  \addplot[barLucene]       coordinates {(q, 22000)};
  \addplot[barVP]           coordinates {(q, 1100)};
  \addplot[barPV]           coordinates {(q, 1100)};
  \legend{
    RSCAS,
    Lucene,
    Postgres (VP),
    Postgres (PV),
  }
  \nextgroupplot[
    bar width=8pt,
    symbolic x coords={q},
    xticklabels={},
    xlabel={(b) Query $Q_2$},
    enlarge x limits=0.05,
  ]
  \addplot[barRCAS]         coordinates {(q, 13155)};
  \addplot[barLucene]       coordinates {(q, 83029)};
  \addplot[barVP]           coordinates {(q, 58738)};
  \addplot[barPV]           coordinates {(q, 42667)};
  %
  %
  \nextgroupplot[
    bar width=8pt,
    symbolic x coords={q},
    xticklabels={},
    xlabel={(c) Query $Q_3$},
    enlarge x limits=0.05,
  ]
  \addplot[barRCAS]         coordinates {(q, 11255)};
  \addplot[barLucene]       coordinates {(q, 740335)};
  \addplot[barVP]           coordinates {(q, 18042)};
  \addplot[barPV]           coordinates {(q, 97206)};
  \nextgroupplot[
    bar width=8pt,
    symbolic x coords={q},
    xticklabels={},
    xlabel={(d) Query $Q_4$},
    enlarge x limits=0.05,
  ]
  \addplot[barRCAS]         coordinates {(q, 7089)};
  \addplot[barLucene]       coordinates {(q, 219884)};
  \addplot[barVP]           coordinates {(q, 77589)};
  \addplot[barPV]           coordinates {(q, 390855)};
  \nextgroupplot[
    bar width=8pt,
    symbolic x coords={q},
    xticklabels={},
    xlabel={(e) Query $Q_5$},
    enlarge x limits=0.05,
  ]
  \addplot[barRCAS]         coordinates {(q, 22908)};
  \addplot[barLucene]       coordinates {(q, 1897284)};
  \addplot[barVP]           coordinates {(q, 633033)};
  \addplot[barPV]           coordinates {(q, 3853407)};
  \nextgroupplot[
    bar width=8pt,
    symbolic x coords={q},
    xticklabels={},
    xlabel={(f) Query $Q_6$},
    enlarge x limits=0.05,
  ]
  \addplot[barRCAS]         coordinates {(q, 112942)};
  \addplot[barLucene]       coordinates {(q, 909362)};
  \addplot[barVP]           coordinates {(q, 2067303)};
  \addplot[barPV]           coordinates {(q, 5261819)};
  \end{groupplot}
\end{tikzpicture}%
\caption{Runtime of queries $Q_1, \ldots, Q_6$ on cold caches}
\label{exp:q1-6-cold}
\end{figure}
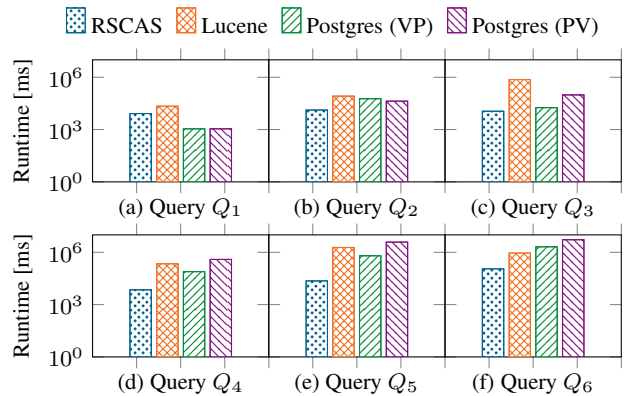

Figure \ref{exp:q1-6-cold} shows the runtime of the six queries on the
GitLab dataset (note the logarithmic y-axis).  We clear the
operating system's page cache before each query (later we repeat the
same experiment on warm caches). We start with the runtime of query
$Q_1$ in Figure \ref{exp:q1-6-cold}a.  This point query is well suited
for existing solutions because both predicates have low selectivities
and produce small intermediate results. Therefore, the composite VP
and PV indexes perform best. No matter what attribute is ordered first
in the composite index (the paths or the values), the index can
quickly narrow down the set of possible candidates. Lucene on the
other hand evaluates both predicates and intersects the results, which
is more expensive. RSCAS is in between Lucene and the two composite
indexes.  $Q_2$ has a low path but high value selectivity. Because of
this, the composite PV index outperforms the composite VP index, see
Figure \ref{exp:q1-6-cold}b.  Evaluating this query in Lucene is
costly since Lucene must fully iterate over the large intermediate
result produced by the value predicate. RSCAS, on the other hand, uses
the selective path predicate to prune subtrees early during query
evaluation.  For query $Q_3$ in Figure \ref{exp:q1-6-cold}c, RSCAS
performs best but it is closely followed by the composite VP index,
for which $Q_3$ is the best case since $Q_3$ has a very low value
selectivity.  While $Q_3$ is the best case for VP, it is the worst
case for PV and indeed its query performance is an order of magnitude
higher.  For Lucene the situation is similar to query $Q_2$, except
that the path predicate produces a large intermediate result (rather
than the value predicate).  Query $Q_4$ uses the \texttt{*} and
\texttt{**} wildcards at the end of its query path.  The placement of
the wildcards is important for all approaches. Query paths that have
wildcards in the middle or at the end can be evaluated efficiently
with prefix searches.  As a result, RSCAS's query performance remains
stable and is similar to that for queries $Q_1, \ldots, Q_3$. Queries
$Q_5$ and $Q_6$ are more difficult for all approaches because they
contain the descendant axis at the beginning of the query path.
Normally, when the query path does not match a path in the trie, the
node that is not matched and its subtrees do not need to be considered
anymore because no path suffix can match the query path.  The
\texttt{**} wildcard, however, may skip over mismatches and the query
path's suffix may match.  For this reason, Lucene must traverse its
entire FST that is used to evaluate path predicates.  Likewise, the
composite PV index must traverse large parts of the index because the
keys are ordered first by the paths and in the index.  The VP index
can use the value predicate to prune subtrees that do not match the
value predicate before looking at the path predicate. RSCAS uses the
value predicate to prune subtrees when the path predicate does not
prune anymore because of wildcards and therefore delivers the best
query runtime.

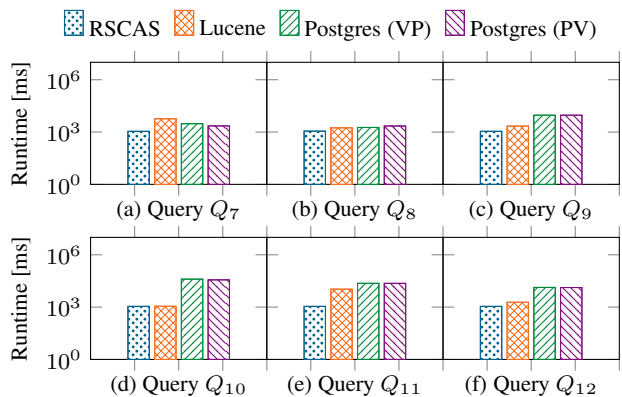
\begin{figure}[htb]
\centering
\begin{tikzpicture}
  \begin{groupplot}[
    col2,
    width=39mm,
    height=32mm,
    ytick scale label code/.code={},
    xticklabel shift={-7pt},
    group style={
      columns=3,
      rows=2,
      horizontal sep=0pt,
      vertical sep=20pt,
      yticklabels at=edge left,
      ylabels at=edge left,
    },
    ybar,
    ymode=log,
    ymin=1,
    ymax=10000000,
    ytick={1,1000,1000000},    
    ylabel={Runtime [ms]},
    legend style={
      at={(-0.2,1.10)},
      anchor=south west,
      legend columns=-1,
      draw=none,
      /tikz/every even column/.append style={column sep=3pt},
    },
  ]
  \nextgroupplot[
    bar width=8pt,
    symbolic x coords={q},
    xticklabels={},
    xlabel={(a) Query $Q_7$},
    enlarge x limits=0.05,
  ]
  \addplot[barRCAS]         coordinates {(q, 1100)};
  \addplot[barLucene]       coordinates {(q, 5788)};
  \addplot[barVP]           coordinates {(q, 3015)};
  \addplot[barPV]           coordinates {(q, 2250)};
  \legend{
    RSCAS,
    Lucene,
    Postgres (VP),
    Postgres (PV),
  }
  \nextgroupplot[
    bar width=8pt,
    symbolic x coords={q},
    xticklabels={},
    xlabel={(b) Query $Q_8$},
    enlarge x limits=0.05,
  ]
  \addplot[barRCAS]         coordinates {(q, 1127)};
  \addplot[barLucene]       coordinates {(q, 1742)};
  \addplot[barVP]           coordinates {(q, 1831)};
  \addplot[barPV]           coordinates {(q, 2227)};
  \nextgroupplot[
    bar width=8pt,
    symbolic x coords={q},
    xticklabels={},
    xlabel={(c) Query $Q_9$},
    enlarge x limits=0.05,
  ]
  \addplot[barRCAS]         coordinates {(q, 1100)};
  \addplot[barLucene]       coordinates {(q, 2204)};
  \addplot[barVP]           coordinates {(q, 9272)};
  \addplot[barPV]           coordinates {(q, 9251)};
  \nextgroupplot[
    bar width=8pt,
    symbolic x coords={q},
    xticklabels={},
    xlabel={(d) Query $Q_{10}$},
    enlarge x limits=0.05,
  ]
  \addplot[barRCAS]         coordinates {(q, 1100)};
  \addplot[barLucene]       coordinates {(q, 1117)};
  \addplot[barVP]           coordinates {(q, 40143)};
  \addplot[barPV]           coordinates {(q, 36494)};
  \nextgroupplot[
    bar width=8pt,
    symbolic x coords={q},
    xticklabels={},
    xlabel={(e) Query $Q_{11}$},
    enlarge x limits=0.05,
  ]
  \addplot[barRCAS]         coordinates {(q, 1100)};
  \addplot[barLucene]       coordinates {(q, 10699)};
  \addplot[barVP]           coordinates {(q, 23114)};
  \addplot[barPV]           coordinates {(q, 23038)};
  \nextgroupplot[
    bar width=8pt,
    symbolic x coords={q},
    xticklabels={},
    xlabel={(f) Query $Q_{12}$},
    enlarge x limits=0.05,
  ]
  \addplot[barRCAS]         coordinates {(q, 1100)};
  \addplot[barLucene]       coordinates {(q, 1898)};
  \addplot[barVP]           coordinates {(q, 13456)};
  \addplot[barPV]           coordinates {(q, 13265)};
  \end{groupplot}
\end{tikzpicture}%
\caption{Runtime of queries $Q_7, \ldots, Q_{12}$ on cold caches}
\label{exp:q7-12-cold}
\end{figure}

In Figure \ref{exp:q7-12-cold} we show the runtime of queries
$Q_7, \ldots Q_{12}$ on the remaining three datasets (again on cold
caches).  The absolute runtimes are lower because the datasets are
considerably smaller than the GitLab dataset, see Table
\ref{tab:datasets}, but the relative differences between the
approaches are comparable to the previous set of queries, see Figure
\ref{exp:q1-6-cold}.

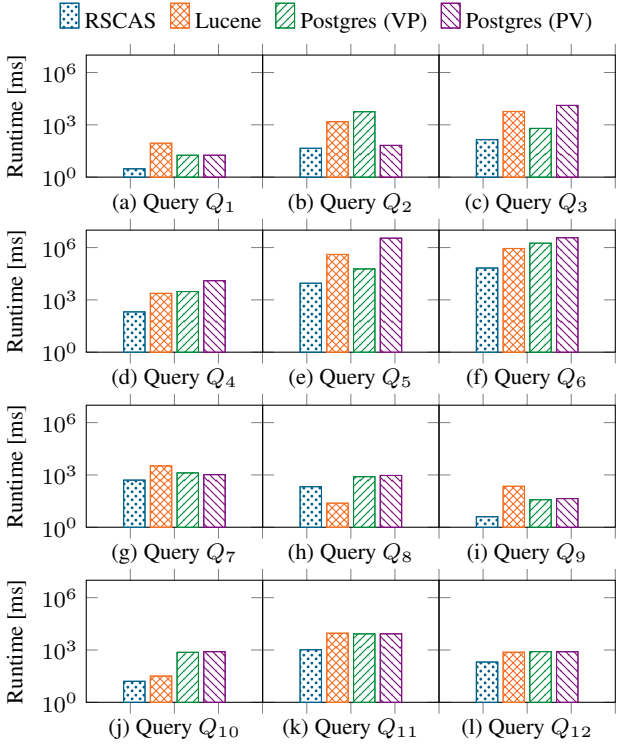
\begin{figure}[htb]
\centering
\begin{tikzpicture}
  \begin{groupplot}[
    col2,
    width=39mm,
    height=32mm,
    ytick scale label code/.code={},
    xticklabel shift={-7pt},
    group style={
      columns=3,
      rows=4,
      horizontal sep=0pt,
      vertical sep=20pt,
      yticklabels at=edge left,
      ylabels at=edge left,
    },
    ybar,
    ymode=log,
    ymin=1,
    ymax=10000000,
    ytick={1,1000,1000000},    
    ylabel={Runtime [ms]},
    legend style={
      at={(-0.2,1.10)},
      anchor=south west,
      legend columns=-1,
      draw=none,
      /tikz/every even column/.append style={column sep=3pt},
    },
  ]
  \nextgroupplot[
    bar width=8pt,
    symbolic x coords={q},
    xticklabels={},
    xlabel={(a) Query $Q_1$},
    enlarge x limits=0.05,
  ]
  \addplot[barRCAS]         coordinates {(q, 3)};
  \addplot[barLucene]       coordinates {(q, 88)};
  \addplot[barVP]           coordinates {(q, 18)};
  \addplot[barPV]           coordinates {(q, 18)};
  \legend{
    RSCAS,
    Lucene,
    Postgres (VP),
    Postgres (PV),
  }
  \nextgroupplot[
    bar width=8pt,
    symbolic x coords={q},
    xticklabels={},
    xlabel={(b) Query $Q_2$},
    enlarge x limits=0.05,
  ]
  \addplot[barRCAS]         coordinates {(q, 45)};
  \addplot[barLucene]       coordinates {(q, 1506)};
  \addplot[barVP]           coordinates {(q, 5639)};
  \addplot[barPV]           coordinates {(q, 66)};
  \nextgroupplot[
    bar width=8pt,
    symbolic x coords={q},
    xticklabels={},
    xlabel={(c) Query $Q_3$},
    enlarge x limits=0.05,
  ]
  \addplot[barRCAS]         coordinates {(q, 141)};
  \addplot[barLucene]       coordinates {(q, 5845)};
  \addplot[barVP]           coordinates {(q, 622)};
  \addplot[barPV]           coordinates {(q, 12961)};
  \nextgroupplot[
    bar width=8pt,
    symbolic x coords={q},
    xticklabels={},
    xlabel={(d) Query $Q_4$},
    enlarge x limits=0.05,
  ]
  \addplot[barRCAS]         coordinates {(q, 205)};
  \addplot[barLucene]       coordinates {(q, 2366)};
  \addplot[barVP]           coordinates {(q, 3005)};
  \addplot[barPV]           coordinates {(q, 12639)};
  \nextgroupplot[
    bar width=8pt,
    symbolic x coords={q},
    xticklabels={},
    xlabel={(e) Query $Q_5$},
    enlarge x limits=0.05,
  ]
  \addplot[barRCAS]         coordinates {(q, 8993)};
  \addplot[barLucene]       coordinates {(q, 403340)};
  \addplot[barVP]           coordinates {(q, 59464)};
  \addplot[barPV]           coordinates {(q, 3466071)};
  \nextgroupplot[
    bar width=8pt,
    symbolic x coords={q},
    xticklabels={},
    xlabel={(f) Query $Q_6$},
    enlarge x limits=0.05,
  ]
  \addplot[barRCAS]         coordinates {(q, 66900)};
  \addplot[barLucene]       coordinates {(q, 883617)};
  \addplot[barVP]           coordinates {(q, 1783140)};
  \addplot[barPV]           coordinates {(q, 3672385)};
  \nextgroupplot[
    bar width=8pt,
    symbolic x coords={q},
    xticklabels={},
    xlabel={(g) Query $Q_7$},
    enlarge x limits=0.05,
  ]
  \addplot[barRCAS]         coordinates {(q, 504)};
  \addplot[barLucene]       coordinates {(q, 3303)};
  \addplot[barVP]           coordinates {(q, 1308)};
  \addplot[barPV]           coordinates {(q, 1038)};
  \nextgroupplot[
    bar width=8pt,
    symbolic x coords={q},
    xticklabels={},
    xlabel={(h) Query $Q_8$},
    enlarge x limits=0.05,
  ]
  \addplot[barRCAS]         coordinates {(q, 212)};
  \addplot[barLucene]       coordinates {(q, 24)};
  \addplot[barVP]           coordinates {(q, 795)};
  \addplot[barPV]           coordinates {(q, 931)};
  \nextgroupplot[
    bar width=8pt,
    symbolic x coords={q},
    xticklabels={},
    xlabel={(i) Query $Q_9$},
    enlarge x limits=0.05,
  ]
  \addplot[barRCAS]         coordinates {(q, 4)};
  \addplot[barLucene]       coordinates {(q, 223)};
  \addplot[barVP]           coordinates {(q, 38)};
  \addplot[barPV]           coordinates {(q, 44)};
  \nextgroupplot[
    bar width=8pt,
    symbolic x coords={q},
    xticklabels={},
    xlabel={(j) Query $Q_{10}$},
    enlarge x limits=0.05,
  ]
  \addplot[barRCAS]         coordinates {(q, 16)};
  \addplot[barLucene]       coordinates {(q, 32)};
  \addplot[barVP]           coordinates {(q, 734)};
  \addplot[barPV]           coordinates {(q, 800)};
  \nextgroupplot[
    bar width=8pt,
    symbolic x coords={q},
    xticklabels={},
    xlabel={(k) Query $Q_{11}$},
    enlarge x limits=0.05,
  ]
  \addplot[barRCAS]         coordinates {(q, 1034)};
  \addplot[barLucene]       coordinates {(q, 9205)};
  \addplot[barVP]           coordinates {(q, 8392)};
  \addplot[barPV]           coordinates {(q, 8463)};
  \nextgroupplot[
    bar width=8pt,
    symbolic x coords={q},
    xticklabels={},
    xlabel={(l) Query $Q_{12}$},
    enlarge x limits=0.05,
  ]
  \addplot[barRCAS]         coordinates {(q, 203)};
  \addplot[barLucene]       coordinates {(q, 745)};
  \addplot[barVP]           coordinates {(q, 805)};
  \addplot[barPV]           coordinates {(q, 793)};
  \end{groupplot}
\end{tikzpicture}%
\caption{Runtime of queries $Q_1, \ldots, Q_{12}$ on warm caches.}
\label{exp:q1-12-warm}
\end{figure}

We repeat the same experiments on warm caches, see Figure
\ref{exp:q1-12-warm} (the y-axis shows the query runtime in
milliseconds).  Note that we did not implement a dedicated caching mechanism
and solely rely on the operating system's page cache. When the caches are
hot the CPU usage and memory access become the main bottlenecks.
Since RSCAS produces the smallest intermediate results, RSCAS requires
the least CPU time and memory accesses. As a result, RSCAS
consistently outperforms its competitors, see Figure
\ref{exp:q1-12-warm}.

\MREV{ \inRegularPaper{ An analysis of the query performance for an
    RSCAS index with different numbers of levels, i.e., tries, can be
    found in the accompanying technical report \cite{KW22}.
  }{
    To evaluate the impact of the number of levels on the query
    performance we ran an experiment for an RSCAS index with $10^9$
    keys from the GitLab dataset.  By varying the memory size to
    accommodate, respectively, $2^{20}$, $2^{26}$ and $2^{30}$ keys,
    we got an RSCAS index with 1, 4 and 7 levels (tries),
    respectively.  The total running time for running queries $Q_1$ to
    $Q_6$ is detailed in Figure~\ref{querfPerfDiffLevels}.
  \begin{figure}[htbp]\centering
    \begin{tikzpicture}
      \begin{axis}[
        xlabel = Number of levels (tries) in the RSCAS index,
        xtick={1,4,7},
        xticklabels={1,4,7},
        width=7cm,
        ylabel=Query time (sec),
        ymin=0,
        height=4cm,
        legend style={at={(0.0,.91)},anchor=west},
      ]
      \addplot[color=red,mark=x] coordinates { (1, 49) (4, 76.5) (7, 96.5) };
      \end{axis}
    \end{tikzpicture}
    \caption{Query performance for RSCAS index with different number of levels.}
    \label{querfPerfDiffLevels}
  \end{figure}
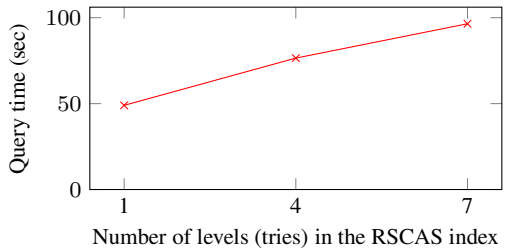
  }}

\subsection{Scalability}

RSCAS uses its LSM-based structure to gracefully and efficiently
handle large datasets that do not fit into main memory.  We discuss
how to choose threshold $\tau$, the performance of bulk-loading and
individual insertions, the accuracy of the cost model, and the index
size.

\subsubsection{Calibration}
\label{sec:calibration}

\begin{figure}[htb]
\begin{tikzpicture}
  \begin{groupplot}[
    col2,
    width=46mm,
    group style={
      columns=2,
      rows=2,
      horizontal sep=35pt,
      ylabels at=edge left,
    },
    legend style={
      at={(0.0,1.05)},
      anchor=south west,
      legend columns=-1,
      draw=none,
    },
    xmode=log,
  ]
  \nextgroupplot[
    xlabel={(a) Threshold $\tau$},
    ylabel={Bulk-loading [h]},
  ]
  \pgfplotstableread[col sep=tab]{experiments/02_threshold/data.csv}\plot
  \addplot[rscas]   table[x=threshold,y=runtime_h] {\plot};
  \draw[densely dotted] (rel axis cs:0.5,0) -- (rel axis cs:0.5,1);
  \node[anchor=south west] at (rel axis cs:0.5,0.5) {$\tau=100$};
  %
  \nextgroupplot[
    xlabel={(b) Threshold $\tau$},
    ylabel={Nr Nodes [B]},
    scaled y ticks=base 10:-9,
    ytick scale label code/.code={},
  ]
  \pgfplotstableread[col sep=tab]{experiments/02_threshold/data.csv}\plot
  \addplot[rscas]   table[x=threshold,y=partitions_created] {\plot};
  \draw[densely dotted] (rel axis cs:0.5,0) -- (rel axis cs:0.5,1);
  \node[anchor=south west] at (rel axis cs:0.5,0.5) {$\tau=100$};
  \nextgroupplot[
    xlabel={(c) Threshold $\tau$},
    ylabel={Query Runtime [s]},
    ymin=0,
  ]
  \pgfplotstableread[col sep=tab]{experiments/02_threshold/data.csv}\plot
  \addplot[rscas]   table[x=threshold,y=avg_query_s] {\plot};
  \draw[densely dotted] (rel axis cs:0.5,0) -- (rel axis cs:0.5,1);
  \node[anchor=south west] at (rel axis cs:0.5,0.4) {$\tau=100$};
  \nextgroupplot[
    xlabel={(d) Threshold $\tau$},
    ylabel={Index Size [GB]},
    ymin=0,
  ]
  \pgfplotstableread[col sep=tab]{experiments/02_threshold/data.csv}\plot
  \addplot[rscas]   table[x=threshold,y=index_bytes_written_gb] {\plot};
  \draw[densely dotted] (rel axis cs:0.5,0) -- (rel axis cs:0.5,1);
  \node[anchor=south west] at (rel axis cs:0.5,0.5) {$\tau=100$};
  %
  \end{groupplot}
\end{tikzpicture}
\caption{Calibrating partitioning threshold $\tau$}
\label{exp:tau}
\end{figure}
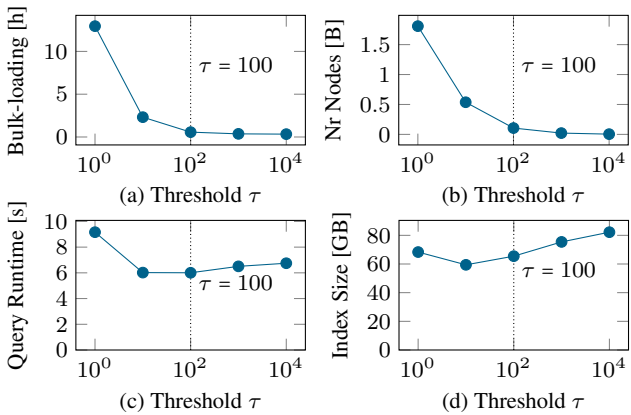

We start out by calibrating the partitioning threshold $\tau$, i.e.,
the maximum number of suffixes in a leaf node.  We calibrate $\tau$ in
Figure \ref{exp:tau} on a 100\,GB subset of the GitLab dataset.  Even
on the 100\,GB subset, bulk-loading RSCAS with $\tau=1$ takes more
than 12 hours, see Figure \ref{exp:tau}a.  When we increase $\tau$,
the recursive bulk-loading algorithm terminates earlier (see lines
15--21 in Algorithm \ref{alg:bulk}), hence fewer partitions are
created and the runtime improves. Since the bulk-loading algorithm
extracts from every partition its longest common prefixes and stores
them in a new node, the number of nodes in the index also decreases as
we increase $\tau$, see Figure \ref{exp:tau}b.  As a result, leaf
nodes get bigger and store more un-interleaved suffixes.  This
negatively affects the query performance and the index size, see
Figures \ref{exp:tau}c and \ref{exp:tau}d, respectively.  Figures
\ref{exp:tau}c shows the average runtime of the six queries
$Q_1, \ldots, Q_6$.  A query that reaches a leaf node must scan all
suffixes to find matches.  Making $\tau$ too small decreases query
performance because more nodes need to be traversed and making $\tau$
too big decreases query performance because a node must scan many
suffixes that do not match a query.  According to Figures
\ref{exp:tau}c, values $\tau \in [10,100]$ give the best query
performance. Threshold $\tau$ also affects the index size, see Figure
\ref{exp:tau}d. If $\tau$ is too small, many small nodes are created
and for each such node there is storage overhead in terms of node
headers, pointers, etc., see Figure \ref{fig:pagelayout}.  If $\tau$
is too big, leaf nodes contain long lists of suffixes for which we
could still extract common prefixes if we $\psi$-partitioned them
further. As a consequence, we choose medium values for $\tau$ to get a
good balance between bulk-loading runtime, query performance, and
index size. Based on Figure \ref{exp:tau} we choose $\tau = 100$ as
default value.
\MREV{
\inRegularPaper{
More details on a quantitative analysis on how $\tau$ affects certain
parameters can be found in the accompanying technical report
\cite{KW22}.
}{
More details on a quantitative analysis on how $\tau$ affects certain
parameters can be found in Appendix~\ref{sec:tuningtau}.
}}

\subsubsection{Bulk-Loading Performance}

Bulk-loading is a core operation that we use in two situations. First,
when we create RSCAS for an existing system with large amounts of data
we use bulk-loading to create RSCAS.  Second, our RSCAS index uses
bulk-loading to create a disk-based RSCAS trie whenever the in-memory
RSCAS trie $R^M_0$ overflows.  We compare our bulk-loading algorithm
with bulk-loading of composite B+ trees in Postgres (Lucene does not
support bulk-loading; as a reference point we include Lucene's
performance for the corresponding point insertions).

\begin{figure}[htb]
\begin{tikzpicture}
  \begin{groupplot}[
    col2,
    width=46mm,
    group style={
      columns=2,
      rows=2,
      horizontal sep=35pt,
      ylabels at=edge left,
    },
    legend style={
      at={(-0.3,1.05)},
      anchor=south west,
      legend columns=-1,
      draw=none,
    },
  ]
  \nextgroupplot[
    xlabel={(a) Nr Keys [$\times 10^9$]},
    ylabel={Time [h]},
    clip=true,
    ymax=6,
    scaled x ticks=base 10:-9,
    xtick scale label code/.code={},
  ]
  \addlegendimage{rscas}      \addlegendentry{RSCAS}
  \addlegendimage{postgresvp} \addlegendentry{Postgres (VP)}
  \addlegendimage{postgrespv} \addlegendentry{Postgres (PV)}
  \addlegendimage{lucene}     \addlegendentry{Lucene*}
  \pgfplotstableread[col sep=tab]{experiments/08_bulkloading/vp.csv}\plot
  \addplot[postgresvp] table[x=nr_keys,y=runtime_h] {\plot};
  \pgfplotstableread[col sep=tab]{experiments/08_bulkloading/pv.csv}\plot
  \addplot[postgrespv] table[x=nr_keys,y=runtime_h] {\plot};
  \pgfplotstableread[col sep=tab]{experiments/08_bulkloading/lucene.csv}\plot
  \addplot[lucene] table[x=nr_keys,y=runtime_h] {\plot};
  \pgfplotstableread[col sep=tab]{experiments/08_bulkloading/rcas.csv}\plot
  \addplot[rscas] table[x=nr_keys,y=runtime_h] {\plot};
  %
  %
  %
  \nextgroupplot[
    xlabel={(b) Nr Keys [$\times 10^9$]},
    ylabel={Disk I/O [TB]},
    clip=false,
    scaled x ticks=base 10:-9,
    xtick scale label code/.code={},
  ]
  \pgfplotstableread[col sep=tab]{experiments/08_bulkloading/vp.csv}\plot
  \addplot[postgresvp] table[x=nr_keys,y=disk_io_tb] {\plot};
  \pgfplotstableread[col sep=tab]{experiments/08_bulkloading/pv.csv}\plot
  \addplot[postgrespv] table[x=nr_keys,y=disk_io_tb] {\plot};
  \pgfplotstableread[col sep=tab]{experiments/08_bulkloading/lucene.csv}\plot
  \addplot[lucene] table[x=nr_keys,y=disk_io_tb] {\plot};
  \pgfplotstableread[col sep=tab]{experiments/08_bulkloading/rcas.csv}\plot
  \addplot[rscas] table[x=nr_keys,y=disk_io_tb] {\plot};
  \end{groupplot}
\end{tikzpicture}
\caption{Bulk-Loading performance}
\label{exp:bulkloading}
\end{figure}
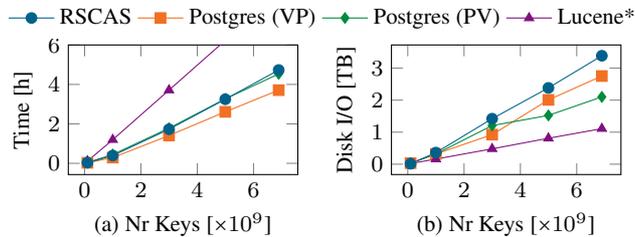

Figure~\ref{exp:bulkloading} evaluates the performance of the
bulk-loading algorithms for RSCAS and Postgres.  We give the
systems 8\,GB of main memory.  For a fair comparison, we set the fill
factor of the composite B+ trees in Postgres to 100\% to make them
read-optimized and as compact as possible since disk-based RSCAS tries
are read-only.  We compare the systems for our biggest dataset, the
GitLab dataset, in Figure~\ref{exp:bulkloading}.  The GitLab dataset
contains 6.9 billion keys and has a size of 550\,GB. Figure
\ref{exp:bulkloading}a confirms that bulk-loading RSCAS takes roughly
the same time as bulk-loading the PV and VP composite indexes in
Postgres (notice that RSCAS and the PV composite index have virtually
the same runtime, thus PV's curve is barely visible). The runtime and
disk I/O of all algorithms increase linearly in
Figure~\ref{exp:bulkloading}a, which means it is feasible to bulk-load
these indexes efficiently for very large datasets.  Postgres creates a
B+ tree by sorting the data and then building the index bottom up,
level by level. RSCAS partitions the data and builds the index top
down.  In practice, both paradigms perform similarly, both in terms of
runtime (Figure~\ref{exp:bulkloading}a) and disk I/O
(Figure~\ref{exp:bulkloading}b).

\subsubsection{Insertion Performance}

\begin{figure}[htb]
\begin{tikzpicture}
  \begin{groupplot}[
    col2,
    width=46mm,
    group style={
      columns=2,
      rows=2,
      horizontal sep=35pt,
      ylabels at=edge left,
    },
    legend style={
      at={(-0.3,1.05)},
      anchor=south west,
      legend columns=-1,
      draw=none,
    },
    ymin=0,
  ]
  \nextgroupplot[
    xlabel={(a) Nr Keys [$\times 10^6$]},
    ylabel={Time [s]},
    scaled x ticks=base 10:-6,
    xtick scale label code/.code={},
  ]
  \addlegendimage{rscas}  \addlegendentry{RSCAS}
  \addlegendimage{lucene} \addlegendentry{Lucene}
  \addlegendimage{postgresvp} \addlegendentry{Postgres* (VP)}
  \addlegendimage{postgrespv} \addlegendentry{Postgres* (PV)}
  \pgfplotstableread[col sep=tab]{experiments/09_mem_insertion/rcas.csv}\plot
  \addplot[rscas] table[x=nr_keys,y=runtime_s] {\plot};
  \nextgroupplot[
    xlabel={(b) Nr Keys [$\times 10^6$]},
    ylabel={Time [m]},
    scaled x ticks=base 10:-6,
    xtick scale label code/.code={},
    ymax=85,
  ]
  \pgfplotstableread[col sep=tab]{experiments/10_insertion/vp.csv}\plot
  \addplot[postgresvp] table[x=nr_keys,y=runtime_m] {\plot};
  \pgfplotstableread[col sep=tab]{experiments/10_insertion/pv.csv}\plot
  \addplot[postgrespv] table[x=nr_keys,y=runtime_m] {\plot};
  \pgfplotstableread[col sep=tab]{experiments/10_insertion/lucene.csv}\plot
  \addplot[lucene] table[x=nr_keys,y=runtime_m] {\plot};
  \pgfplotstableread[col sep=tab]{experiments/10_insertion/rcas.csv}\plot
  \addplot[rscas] table[x=nr_keys,y=runtime_m] {\plot};
  \end{groupplot}
\end{tikzpicture}
\caption{Insertion (a) in memory and (b) on disk}
\label{exp:mem_insertion}
\end{figure}
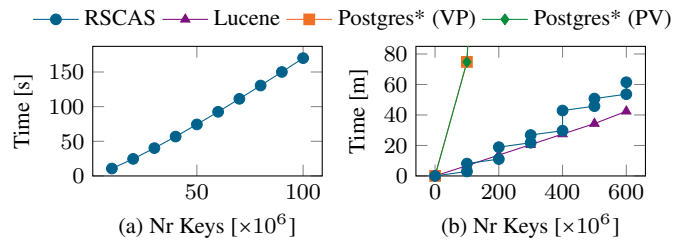

New keys are first inserted into the in-memory trie $R^M_0$ before
they are written to disk when $R^M_0$ overflows.  We evaluate
insertions into $R^M_0$ in Figure \ref{exp:mem_insertion}a and look at
the insertion speed when $R^M_0$ overflows in Figure
\ref{exp:mem_insertion}b.  For the latter case we compare RSCAS's
on-disk insertion performance to Lucene's and Postgres'.

Since $R^M_0$ is memory-based, insertions can be performed quickly,
see Figure \ref{exp:mem_insertion}a.  For example, inserting 100
million keys takes less than three minutes with one insertion taking
1.7$\mu$s, on average. In practice, the SWH archive crawls about one
million revisions per day and since a revision modifies on average
about 60 files in the GitLab dataset, there are 60 million insertions
into the RSCAS index per day, on average. Therefore, our RSCAS index
can easily keep up with the ingestion rate of the SWH archive. Every
two days, on average, $R^M_0$ overflows and a new disk-based RSCAS
trie $R_i$ is bulk-loaded.

In Figure \ref{exp:mem_insertion}b we show how the RSCAS index
performs when $R^M_0$ overflows.  In this experiment, we set the
maximum capacity of $R^M_0$ to $M = 100$ million keys and insert 600
million keys, thus $R^M_0$ overflows six times.  Typically when
$R^M_0$ overflows we bulk-load a disk-based trie in a background
process, but in this experiment we execute all insertions in the
foreground in one process to show all times.  As a result, we observe
a staircase runtime pattern, see Figure \ref{exp:mem_insertion}b.  A
flat part where insertions are performed efficiently in memory is
followed by a jump where a disk-based trie $R_i$ is bulk-loaded. Not
all jumps are equally high since their height depends on the size of
the trie $R_i$ that is bulk-loaded.  When $R^M_0$ overflows, the RSCAS
index looks for the smallest $i$ such that $R_i$ does not exist yet
and bulk-loads it from the keys in $R^M_0$ and all $R_j$, $j<i$.
Therefore, a trie $R_i$, containing $2^iM$ keys, is created for the
first time after $2^iM$ insertions. For example, after $M$ insertions
we bulk-load $R_0$ ($M$ keys); after $2M$ insertions we bulk-load
$R_1$ ($2M$ keys) and delete $R_0$; after $3M$ insertions we again
bulk-load $R_0$ ($M$ keys); after $4M$ insertions we bulk-load $R_2$
($4M$ keys) and delete $R_0$ and $R_1$, etc.  Lucene's insertion
performance is comparable to that of RSCAS, but insertion into
Postgres' B+ tree are expensive in comparison.\footnote{We measure
  insertion performance in Postgres by importing a dataset twice: once
  with index and once without index, and then we measure the
  difference in runtime} This is because insertions into Postgres' B+
trees are executed in-place, causing many random updates, while
insertions in RSCAS and Lucene are done out-of-place.

\subsubsection{Evaluating the Cost Model}
\label{sec:costmodeleval}

\MREV{We evaluate the cost model from Lemma \ref{lemma:bestcase} that measures
the I/O overhead of our bulk-loading algorithm for a uniform data distribution
and compare it to the I/O overhead of bulk-loading the real-world GitLab
dataset.} The I/O overhead is the number of page transfers to read/write
intermediate results during bulk-loading. We multiply the I/O overhead with
the page size to get the number of bytes that are transferred to and from
disk.  The cost model in Lemma~\ref{lemma:bestcase} has four parameters: $N$,
$M$, $B$, and $f$ (see Section~\ref{sec:costmodel}).  We set fanout $f=10$
since this is the average fanout of a node in RSCAS for the GitLab dataset,
see Figure \ref{exp:shape}a.  The cost model assumes that $M$ ($B$) keys fit
into memory (a page). Therefore, we set $B =
\ceil{\frac{16\,\text{KB}}{80\,\text{B}}} = 205$, where 16\,KB is the page
size and $80$ is the average key length (see Section~\ref{sec:expstructure}).
Similarly, if the memory size is 8\,GB we can store $M =
\ceil{\frac{8\,\text{GB}}{80\,\text{B}}} = 100$ million keys in memory.

\begin{figure}[htb]
\begin{tikzpicture}
  \begin{groupplot}[
    col2,
    width=48mm,
    group style={
      columns=2,
      rows=2,
      horizontal sep=20pt,
      ylabels at=edge left,
    },
    legend style={
      at={(0.0,1.05)},
      anchor=south west,
      legend columns=-1,
      draw=none,
    },
    ymin=0,
    ymax=2.75,
    ylabel={I/O Overhead [TB]},
  ]
  \nextgroupplot[
    xlabel={(a) Nr Keys $N$ [$\times 10^9$]},
    clip=false,
    scaled x ticks=base 10:-9,
    xtick scale label code/.code={},
  ]
  \addlegendimage{rscas}       \addlegendentry{Actual I/O overhead}
  \addlegendimage{costmodel}  \addlegendentry{Estimated I/O overhead}
  \pgfplotstableread[col sep=tab]{experiments/11_costmodel_bulk/real_bulk_keys.csv}\plot
  \addplot[rscas] table[x=nr_input_keys,y=disk_overhead_tb] {\plot};
  \pgfplotstableread[col sep=tab]{experiments/11_costmodel_bulk/model_bulk_keys.csv}\plot
  \addplot[costmodel] table[x=nr_input_keys,y=disk_io_tb] {\plot};
  \nextgroupplot[
    xlabel={(b) Memory Keys $M$ [$\times 10^6$]},
    xtick scale label code/.code={},
    xmode=log,
    log basis x={2},
    xtick=data,
    xtick={100,400,1600},
    xticklabels={100,400,1600},
  ]
  \pgfplotstableread[col sep=tab]{experiments/11_costmodel_bulk/real_bulk_mem.csv}\plot
  \addplot[rscas] table[x=mem_size_keys_m,y=disk_overhead_tb] {\plot};
  \pgfplotstableread[col sep=tab]{experiments/11_costmodel_bulk/model_bulk_mem.csv}\plot
  \addplot[costmodel] table[x=mem_size_keys_m,y=disk_overhead_tb] {\plot};
  \end{groupplot}
\end{tikzpicture}
\caption{Bulk-Loading cost model}
\label{exp:modelbulk}
\end{figure}
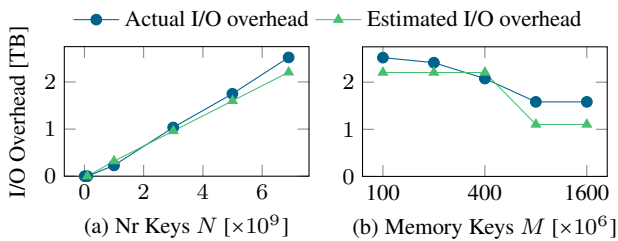

In Figure~\ref{exp:modelbulk}a we compare the actual and the estimated
I/O overhead to bulk-load RSCAS as we increase the number of keys $N$
in the dataset, keeping the memory size fixed at $M=100$ million
keys. The estimated and actual cost are close and within 15\% of each
other.  In Figure~\ref{exp:modelbulk}b we vary the memory size and fix
the full GitLab dataset as input.  The estimated cost is constant from
$M=100$ to $M=400$ million keys because of the ceiling operator in
$\log_f{\ceil{\frac{N}{M}}}$ to compute the number of levels of the
trie in Lemma~\ref{lemma:bestcase}.  If we increase $M$ to 800 million
keys, the trie in the cost model has one level less before partitions
fit entirely into memory and therefore the I/O overhead decreases and
remains constant thereafter since only the root partition does not fit
into main memory.

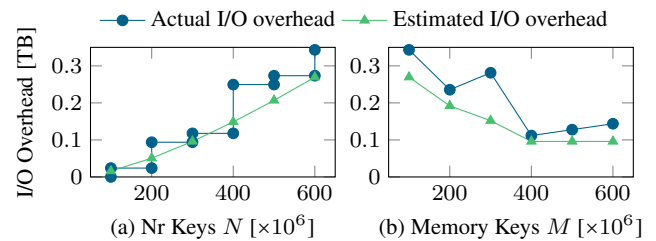
\begin{figure}[htb]
\begin{tikzpicture}
  \begin{groupplot}[
    col2,
    width=48mm,
    group style={
      columns=2,
      rows=2,
      horizontal sep=20pt,
      ylabels at=edge left,
    },
    legend style={
      at={(0.0,1.05)},
      anchor=south west,
      legend columns=-1,
      draw=none,
    },
    ymin=0,
    ymax=0.35,
    ylabel={I/O Overhead [TB]},
  ]
  \nextgroupplot[
    xlabel={(a) Nr Keys $N$ [$\times 10^6$]},
    clip=false,
    scaled x ticks=base 10:-6,
    xtick scale label code/.code={},
  ]
  \addlegendimage{rscas}       \addlegendentry{Actual I/O overhead}
  \addlegendimage{costmodel}  \addlegendentry{Estimated I/O overhead}
  \pgfplotstableread[col sep=tab]{experiments/12_costmodel_insertion/real_insertion_keys.csv}\plot
  \addplot[rscas] table[x=nr_input_keys,y=disk_overhead_tb] {\plot};
  \pgfplotstableread[col sep=tab]{experiments/12_costmodel_insertion/model_insertion_keys.csv}\plot
  \addplot[costmodel] table[x=nr_input_keys,y=disk_overhead_tb] {\plot};
  \nextgroupplot[
    xlabel={(b) Memory Keys $M$ [$\times 10^6$]},
    xtick scale label code/.code={},
    scaled x ticks=base 10:-6,
    xtick scale label code/.code={},
  ]
  \pgfplotstableread[col sep=tab]{experiments/12_costmodel_insertion/real_insertion_mem.csv}\plot
  \addplot[rscas] table[x=mem_size_keys,y=disk_overhead_tb] {\plot};
  \pgfplotstableread[col sep=tab]{experiments/12_costmodel_insertion/model_insertion_mem.csv}\plot
  \addplot[costmodel] table[x=mem_size_keys,y=disk_overhead_tb] {\plot};
  \end{groupplot}
\end{tikzpicture}
\caption{Insertion cost model}
\label{exp:modelinsertion}
\end{figure}

In Figure~\ref{exp:modelinsertion}a we compare the actual and the
estimated I/O overhead to insert $N$ keys one-by-one into RSCAS,
setting $M = 100 \times 10^6$.  We compute the estimated I/O overhead
by multiplying the amortized cost of one insertion according to Lemma
\ref{lemma:amortized} with the number of keys $N$. We observe a
staircase pattern for the actual I/O overhead because of the repeated
bulk-loading when the in-memory trie overflows after every $M$
insertions. Next we fix $N = 600$ million keys and increase $M$ in
Figure \ref{exp:modelinsertion}b. In general, increasing $M$ decreases
the actual and estimated overhead because less data must be
bulk-loaded. But this is not always the case. For example, the actual
I/O overhead increases from $M=200$ to $M=300$ million keys. To see
why, we have to look at the tries that need to be bulk-loaded. For
$M=200$ we create three tries: after $M$ insertions $R_0$ (200 mil.),
after $2M$ insertions $R_1$ (400 mil.), and after $3M$ insertions
again $R_0$ (200 mil.) for a total of 800 million bulk-loaded
keys. For $M=300$ we create only two tries: after $M$ insertions $R_0$
(300 mil.) and after $M$ insertions $R_1$ (600 mil.) for a total of
900 million bulk-loaded keys.

\subsubsection{Index Size}

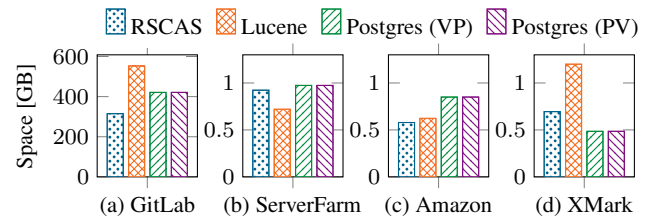
\begin{figure}[htb]
\begin{tikzpicture}
  \begin{groupplot}[
    col2,
    width=29mm,
    height=32mm,
    xtick=data,
    xticklabel style={
      align=center,
    },
    group style={
      columns=4,
      rows=1,
      horizontal sep=17pt,
      ylabels at=edge left,
    },
    ybar,
    legend style={
      at={(0.0,1.05)},
      anchor=south west,
      legend columns=6,
      draw=none,
      /tikz/every even column/.append style={column sep=3pt},
    },
    ylabel={Space [GB]},
    xticklabels={{{}}},
    ytick scale label code/.code={},
    xticklabel style = {yshift=+6pt},
    enlarge x limits=0.05,
    ymin = 0,
  ]
  \nextgroupplot[
    bar width=6pt,
    symbolic x coords={q},
    xticklabels={},
    xlabel={(a) GitLab},
    enlarge x limits=0.05,
  ]
  \addplot[barRCAS]   coordinates {(q, 315)};
  \addplot[barLucene] coordinates {(q, 553)};
  \addplot[barVP]     coordinates {(q, 421)};
  \addplot[barPV]     coordinates {(q, 421)};
  \legend{
    RSCAS,
    Lucene,
    Postgres (VP),
    Postgres (PV),
  }
  \nextgroupplot[
    bar width=6pt,
    symbolic x coords={q},
    xticklabels={},
    xlabel={(b) ServerFarm},
    enlarge x limits=0.05,
    ymax=1.3,
  ]
  \addplot[barRCAS]   coordinates {(q, 0.923)};
  \addplot[barLucene] coordinates {(q, 0.720)};
  \addplot[barVP]     coordinates {(q, 0.974)};
  \addplot[barPV]     coordinates {(q, 0.974)};
  \nextgroupplot[
    bar width=6pt,
    symbolic x coords={q},
    xticklabels={},
    xlabel={(c) Amazon},
    enlarge x limits=0.05,
    ymax=1.3,
  ]
  \addplot[barRCAS]   coordinates {(q, 0.579)};
  \addplot[barLucene] coordinates {(q, 0.623)};
  \addplot[barVP]     coordinates {(q, 0.850)};
  \addplot[barPV]     coordinates {(q, 0.850)};
  \nextgroupplot[
    bar width=6pt,
    symbolic x coords={q},
    xticklabels={},
    xlabel={(d) XMark},
    enlarge x limits=0.05,
    ymax=1.3,
  ]
  \addplot[barRCAS]   coordinates {(q, 0.695)};
  \addplot[barLucene] coordinates {(q, 1.200)};
  \addplot[barVP]     coordinates {(q, 0.485)};
  \addplot[barPV]     coordinates {(q, 0.485)};
  \end{groupplot}
\end{tikzpicture}
\caption{Space consumption}
\label{exp:space}
\end{figure}

Figure \ref{exp:space} shows the size of the RSCAS, Lucene, and
Postgres indexes for our four datasets. The RSCAS index is between
30\% to 80\% smaller than the input size (i.e., the size of the
indexed keys). The savings are highest for the XMark dataset because
it has only seven unique paths and therefore the RSCAS trie has fewer
nodes since there are fewer discriminative bytes.  But even for a
dataset with a large number of unique paths, e.g., the GitLab dataset,
RSCAS is 43\% smaller than the input. RSCAS's size is comparable to
that of the other indexes since all the indexes require space linear
in the number of keys in the input.

\section{Conclusion and Outlook}

We propose the RSCAS index, a robust and scalable index for
semi-structured hierarchical data. Its robustness is rooted in a
well-balanced integration of paths and values in a single index using
a new dynamic interleaving.  The dynamic interleaving does not
prioritize a particular dimension (paths or values), making the index
robust against queries with high individual selectivities that produce
large intermediate results and a small final result.  We use an
LSM-design to scale the RSCAS index to applications with a high
\MREV{insertion} rate.  We buffer \MREV{insertions} in a
memory-optimized RSCAS trie that we continuously flush to disk as a
series of read-only disk-optimized RSCAS tries.  We evaluate our index
analytically and experimentally.  We prove RSCAS's robustness by
showing that it has the smallest average query runtime over all
queries among interleaving-based approaches.  We evaluate RSCAS
experimentally on three real-world datasets and one synthetic data.
Our experiments show that the RSCAS index outperforms state-of-the-art
approaches by several orders of magnitude on real-world and synthetic
datasets.  We show-case RSCAS's scalability by indexing the revisions
(i.e., commits) of all public GitLab repositories archived by Software
Heritage, for a total of 6.9 billion modified files in 120 revisions.

In our future work we plan to support deletions.  In the in-memory
RSCAS trie we plan to delete the appropriate leaf node and efficiently
restructure the trie if necessary. To delete keys from the
disk-resident RSCAS trie we plan to flag the appropriate leaf nodes as
deleted to avoid expensive restructuring on disk. As a result, queries
need to filter flagged leaf nodes. Whenever a new disk-based trie is
bulk-loaded, we remove the elements previously flagged for deletion.
\MREV{It would also be interesting to implement RSCAS on top of a
  high-performance platform, such as an LSM-tree-based KV-store, the
  main challenge would be to adapt range filters to our complex
  interleaved queries.}

\inRegularPaper{
\MREVB{
  \section*{Acknowledgments}

  We thank the anonymous reviewers and the editor for their insightful
  and valuable comments and for insisting on precision.  }
}

\bibliographystyle{spmpsci}   
\bibliography{bibliography}

\inRegularPaper{
}{
\appendix

\section{Proofs}
\label{app:proofs}

\begin{proof}[Lemma \ref{lemma:op}]
  Consider the $\psi$-partitioning $\psi(K,D) = \{ K_1, \ldots,$ $K_m
  \}$.  Let $K_i \neq K_j$ be two different partitions of $K$ and let
  $k' \in K_i$ and $k'' \in K_j$ be two keys belonging to these
  partitions. Since the paths and values of our keys are
  binary-comparable \cite{VL13}, the most significant byte is the
  first byte and the least significant byte is the last byte.
  Therefore, $k'.D$ is smaller (greater) than $k''.D$ iff $k'.D$ is
  smaller (greater) than $k''.D$ at the first byte for which the two
  keys differ in dimension $D$.  All keys in $K$ have the same longest
  common prefix $s = \textsf{lcp}(K,D)$ in dimension $D$ and their
  discriminative byte is $g = |s| + 1 = \textsf{dsc}(K,D)$. By
  Definition \ref{def:partitioning}, keys $k'$ and $k''$ share the
  same longest common prefix $s$ in $D$, i.e., $k'.D[1,g-1] =
  k''.D[1,g-1] = s$ and they differ at the discriminative byte
  $k'.D[g] \neq k''.D[g]$. Therefore, if $k'.D[g] <  k''.D[g]$, we
  know that $k'.D < k''.D$ (and similarly for $>$).  By the
  correctness constraint in Definition \ref{def:partitioning}, all
  keys in $K_i$ have the same value at $k'.D[g]$ and are therefore all
  smaller or all greater than the keys in $K_j$, who all have the same
  value $k''.D[g]$. \hspace*{1cm} $\hfill\Box$
\end{proof}

\begin{proof}[Lemma \ref{lemma:pp}]
  By Definition \ref{def:partitioning}, all keys $k$ in a partition
  $K_i$ have the same value $k.D[\textsf{dsc}(K,D)]$ for the
  discriminative byte of $K$ in dimension $D$. Therefore,
  $\textsf{dsc}(K,D)$ is no longer a discriminative byte in $K_i$,
  instead $\textsf{dsc}(K_i,D) > \textsf{dsc}(K,D)$. Let $K_i \neq
  K_j$ be two different partitions. Again by Definition
  \ref{def:partitioning}, we know that $\textsf{dsc}(K_i \cup K_j,D) =
  \textsf{dsc}(K,D)$ since any two keys from these two partitions
  differ at byte $\textsf{dsc}(K,D)$.  Substituting
  $|\textsf{lcp}(K,D)| = \textsf{dsc}(K,D)-1$ concludes the proof that
  $\psi(K,D)$ is prefix-preserving in $D$. $\hfill\Box$
\end{proof}

\begin{proof}[Lemma \ref{lemma:gp}]
  Since not all keys in $K$ are equal in dimension $D$, we know there
  must be at least two keys $k_1$ and $k_2$ that differ in dimension
  $D$ at the discriminative byte $g = \textsf{dsc}(K,D)$, i.e.,
  $k_1.D[g] \neq k_2.D[g]$. According to the disjointness constraint
  of Definition~\ref{def:partitioning}, $k_1$ and $k_2$ must be in two
  different partitions of $\psi(K,D)$. Hence, $|\psi(K,D)|$ $\ge 2$.
  $\hfill\Box$
\end{proof}

\begin{proof}[Lemma \ref{lemma:monotonicity}]
  The first line states that $K_i \subset K$ is one of the partitions
  of $K$.  From Definition~\ref{def:partitioning} it follows that the
  value $k.D[\textsf{dsc}(K,D)]$ is the same for every key $k \in
  K_i$.  From Definition~\ref{def:dsc} it follows that
  $\textsf{dsc}(K_i,D) \neq \textsf{dsc}(K,D)$.  By removing one or
  more keys from $K$ to get $K_i$, the keys in $K_i$ will become more
  similar compared to those in $K$. That means, it is not possible for
  the keys in $K_i$ to differ in a position $g < \textsf{dsc}(K,D)$.
  Consequently, $\textsf{dsc}(K_i,D) \nless \textsf{dsc}(K,D)$ for any
  dimension $D$ (so this also holds for $\overline{D}$:
  $\textsf{dsc}(K_i,\overline{D}) \nless
  \textsf{dsc}(K,\overline{D})$).  Thus $\textsf{dsc}(K_i,D) >
  \textsf{dsc}(K,D)$ and $\textsf{dsc}(K_i,\overline{D}) \ge
  \textsf{dsc}(K,\overline{D})$.  $\hfill\Box$
\end{proof}

\begin{proof}[Theorem \ref{theo:avg}]
  We begin with a brief outline of the proof. We show for a level
  $l$ that the costs of query $Q$ and complementary query $Q'$ on
  level $l$ is smallest with the dynamic interleaving.  That is, for a
  level $l$ we show that $\prod_{i=1}^{l} (o \cdot \varsigma_{\phi_i}) +
  \prod_{i=1}^{l} (o \cdot \varsigma'_{\phi_i})$ is smallest with the
  vector $\phi_{\text{DY}} = (V,P,V,P,\ldots)$ of our dynamic
  interleaving.  Since this holds for any level $l$, it also holds for
  the sum of costs over all levels $l$, $1 \leq l \leq h$, and this
  proves the theorem.

  We only look at search trees with a height $h \geq 2$, as for $h=1$
  we do not actually have an interleaving (and the costs are all the
  same). W.l.o.g., we assume that the first level of the search tree
  always starts with a discriminative value byte, i.e., $\phi_1 = V$.
  Let us look at the cost for one specific level $l$ for query $Q$ and
  its complementary query $Q'$. We distinguish two cases: $l$ is even
  or $l$ is odd.

  \textbf{$l$ is even:} The cost for a perfectly
  alternating interleaving for $Q$ for level $l$ is equal to $o^l
  (\varsigma_V \cdot \varsigma_P \dots \varsigma_V \cdot
  \varsigma_P)$, while the cost for $Q'$ is equal to $o^l
  (\varsigma_V' \cdot \varsigma_P' \dots \varsigma_V' \cdot
  \varsigma_P')$, which is equal to $o^l (\varsigma_P \cdot
  \varsigma_V \dots \varsigma_P \cdot \varsigma_V)$. This is the same
  cost as for $Q$, so adding the two costs gives us $2 o^l
  \varsigma_V^{\sfrac{l}{2}} \varsigma_P^{\sfrac{l}{2}}$.

  For a non-perfectly alternating interleaving with the same
  number of $\varsigma_V$ and $\varsigma_P$ multiplicands up to level
  $l$ we have the same cost as for our dynamic interleaving, i.e., $2
  o^l \varsigma_V^{\sfrac{l}{2}} \varsigma_P^{\sfrac{l}{2}}$. Now let
  us assume that the number of $\varsigma_V$ and $\varsigma_P$
  multiplicands is different for level $l$ (there must be at least one
  such level $l$). Assume that for $Q$ we have $r$ multiplicands of
  type $\varsigma_V$ and $s$ multiplicands of type $\varsigma_P$, with
  $r+s = l$ and, w.l.o.g., $r>s$. This gives us $o^l \varsigma_V^s
  \varsigma_P^s \varsigma_V^{r-s} + o^l \varsigma_V^s \varsigma_P^s
  \varsigma_P^{r-s} = o^l \varsigma_V^s \varsigma_P^s
  (\varsigma_V^{r-s} + \varsigma_P^{r-s})$ for the cost.

  We have to show that $2 o^l \varsigma_V^{\sfrac{l}{2}}
  \varsigma_P^{\sfrac{l}{2}} \leq o^l \varsigma_V^s \varsigma_P^s
  (\varsigma_V^{r-s} + \varsigma_P^{r-s})$. As all values are greater
  than zero, this is equivalent to $2 \varsigma_V^{\sfrac{l}{2}-s}
  \varsigma_P^{\sfrac{l}{2}-s} \leq \varsigma_V^{r-s} +
  \varsigma_P^{r-s}$. The right-hand side can be reformulated:
  $\varsigma_V^{r-s} + \varsigma_P^{r-s} = \varsigma_V^{l-2s} +
  \varsigma_P^{l-2s} = \varsigma_V^{\sfrac{l}{2}-s}
  \varsigma_V^{\sfrac{l}{2}-s} + \varsigma_P^{\sfrac{l}{2}-s}
  \varsigma_P^{\sfrac{l}{2}-s}$.  Setting $a =
  \varsigma_V^{\sfrac{l}{2}-s}$ and $b =
  \varsigma_P^{\sfrac{l}{2}-s}$, this boils down to showing $2ab \leq
  a^2 + b^2 \Leftrightarrow 0 \leq (a - b)^2$, which is always true.

  \textbf{$l$ is odd:} W.l.o.g. we assume that for
  computing the cost for a perfectly alternating interleaving for $Q$,
  there are $\lceil \sfrac{l}{2} \rceil$ multiplicands of type
  $\varsigma_V$ and $\lfloor \sfrac{l}{2} \rfloor$ multiplicands of
  type $\varsigma_P$. This results in $o^l \varsigma_V^{\lfloor
  \sfrac{l}{2} \rfloor} \varsigma_P^{\lfloor \sfrac{l}{2} \rfloor}$
  $(\varsigma_V + \varsigma_P)$ for the sum of costs for $Q$ and $Q'$.

  For a non-perfectly alternating interleaving, we again have $o^l
  \varsigma_V^s \varsigma_P^s $ $(\varsigma_V^{r-s} + \varsigma_P^{r-s})$
  with $r+s = l$ and $r>s$, which can be reformulated to $o^l
  \varsigma_V^s \varsigma_P^s (\varsigma_V^{\lfloor \sfrac{l}{2}
  \rfloor - s} \varsigma_V^{\lfloor \sfrac{l}{2} \rfloor - s}
  \varsigma_V + \varsigma_P^{\lfloor \sfrac{l}{2} \rfloor - s}
  \varsigma_P^{\lfloor \sfrac{l}{2} \rfloor - s} \varsigma_P)$.

  What is left to prove is $o^l \varsigma_V^{\lfloor \sfrac{l}{2}
  \rfloor} \varsigma_P^{\lfloor \sfrac{l}{2} \rfloor} (\varsigma_V +
  \varsigma_P) \leq o^l \varsigma_V^s \varsigma_P^s
  (\varsigma_V^{\lfloor \sfrac{l}{2} \rfloor - s} $ $\varsigma_V^{\lfloor
  \sfrac{l}{2} \rfloor - s} \varsigma_V + \varsigma_P^{\lfloor
  \sfrac{l}{2} \rfloor - s} \varsigma_P^{\lfloor \sfrac{l}{2} \rfloor
  - s} \varsigma_P)$, which is equivalent to $\varsigma_V^{\lfloor
  \sfrac{l}{2} \rfloor - s} \varsigma_P^{\lfloor \sfrac{l}{2} \rfloor
  - s} $ $(\varsigma_V + \varsigma_P) \leq \varsigma_V^{\lfloor
  \sfrac{l}{2} \rfloor - s} \varsigma_V^{\lfloor \sfrac{l}{2} \rfloor
  - s} \varsigma_V + \varsigma_P^{\lfloor \sfrac{l}{2} \rfloor - s}
  \varsigma_P^{\lfloor \sfrac{l}{2} \rfloor - s} \varsigma_P$.
  Substituting $a = \varsigma_V$, $b = \varsigma_P$, and $x = {\lfloor
  \sfrac{l}{2} \rfloor - s}$, this means showing that $a^x b^x (a+b)
  \leq a^{2x + 1} + b^{2x + 1} \Leftrightarrow 0 \leq a^{2x + 1} +
  b^{2x + 1} - a^x b^x (a+b)$. Factorizing this polynomial gives us
  $(a^x - b^x)(a^{x+1} - b^{x+1})$ or $(b^x - a^x)(b^{x+1} -
  a^{x+1})$. We look at $(a^x - b^x)(a^{x+1} - b^{x+1})$, the argument
  for the other factorization follows along the same lines.  This term
  can only become negative if one factor is negative and the other is
  positive. Let us first look at the case $a < b$: since $0 \leq a,b
  \leq 1$, we can immediately follow that $a^x < b^x$ and $a^{x+1} <
  b^{x+1}$, i.e., both factors are negative. Analogously, from $a > b$
  (and $0 \leq a,b \leq 1$) immediately follows $a^x > b^x$ and
  $a^{x+1} > b^{x+1}$, i.e., both factors are positive.  $\hfill\Box$
\end{proof}

\begin{proof}[Theorem \ref{theo:robustness}]
  We assume a tree with fanout $o$ and height $h$. With the dynamic
  interleaving, the dimension on each level alternates, i.e.,
  $\phi_{\text{DY}} = (V,P,V,P,\ldots)$. We need to show that the cost
  for every query in a set of complementary queries $\mathbf{Q}$ is
  minimal with $\phi_{\text{DY}}$. Thus, we show that this inequality
  holds:
  \begin{align*}
    \forall \phi:
    & \sum_{(\varsigma_P,\varsigma_V) \in \mathbf{Q}}
      \widehat{C}(o,h,\phi_{\text{DY}},\varsigma_P,\varsigma_V)
      \leq \\
    & \sum_{(\varsigma_P,\varsigma_V) \in \mathbf{Q}}
      \widehat{C}(o,h,\phi,\varsigma_P,\varsigma_V)
  \end{align*}
  First, we double the cost on each side:
  \begin{align*}
    \forall \phi:~
    & 2 \times \sum_{(\varsigma_P,\varsigma_V) \in \mathbf{Q}}
      \widehat{C}(o,h,\phi_{\text{DY}},\varsigma_P,\varsigma_V)
      \leq \\
    & 2 \times \sum_{(\varsigma_P,\varsigma_V) \in \mathbf{Q}}
      \widehat{C}(o,h,\phi,\varsigma_P,\varsigma_V)
  \end{align*}
  This is the same as counting the cost of each query and its
  complementary query twice:
  {\small\begin{align*}
    \forall \phi:
  & \sum_{(\varsigma_P,\varsigma_V) \in \mathbf{Q}}
    \big(\widehat{C}(o,h,\phi_{\text{DY}},\varsigma_P,\varsigma_V) +
     \widehat{C}(o,h,\phi_{\text{DY}},\varsigma_V,\varsigma_P)\big)
    \leq \\
  & \sum_{(\varsigma_P,\varsigma_V) \in \mathbf{Q}}
    \big(\widehat{C}(o,h,\phi,\varsigma_P,\varsigma_V) +
     \widehat{C}(o,h,\phi,\varsigma_V,\varsigma_P)\big)
  \end{align*}}
  Since by Theorem \ref{theo:avg} each summand on the left side is
  smaller than or equal to its corresponding summand on the right
  side, the sum on the left side is is smaller than or equal to the
  sum on the right side. $\hfill\Box$
\end{proof}

\begin{proof}[Theorem \ref{theo:dev}]
  Similar to the proof of Theorem~\ref{theo:avg}, we show that for every level
  $l$, $|\prod_{i=1}^{l} (o \cdot \varsigma_{\phi_i}) - \prod_{i=1}^{l} (o
  \cdot \varsigma'_{\phi_i})|$ is smallest for the dynamic interleaving vector
  $\phi_{\text{DY}} = (V,P,V,P,\ldots)$.

  Again, we only look at search trees with a height $h \geq 2$ and, w.l.o.g.,
  we assume that the first level of the search tree always starts with a
  discriminative value byte, i.e., $\phi_1 = V$.  Let us look at the
  difference in costs for one specific level $l$ for query $Q$ and its
  complementary query $Q'$. We distinguish two cases: $l$ is even or $l$ is
  odd.

  \textbf{$l$ is even:} The cost for a perfectly
  alternating interleaving for $Q$ for level $l$ is equal to $o^l
  (\varsigma_V \cdot \varsigma_P \dots \varsigma_V \cdot
  \varsigma_P)$, while the cost for $Q'$ is equal to $o^l
  (\varsigma_V' \cdot \varsigma_P' \dots \varsigma_V' \cdot
  \varsigma_P')$, which is equal to $o^l (\varsigma_P \cdot
  \varsigma_V \dots \varsigma_P \cdot \varsigma_V)$. This is the same
  cost as for $Q$, so subtracting one cost from the other gives us 0.

  For a non-perfectly alternating interleaving with the same number of
  $\varsigma_V$ and $\varsigma_P$ multiplicands up to level $l$ we have the
  same difference in costs as for our dynamic interleaving, i.e., 0. Now let
  us assume that the number of $\varsigma_V$ and $\varsigma_P$ multiplicands
  is different for level $l$ (there must be at least one such level
  $l$). Assume that for $Q$ we have $r$ multiplicands of type $\varsigma_V$
  and $s$ multiplicands of type $\varsigma_P$, with $r+s = l$ and, w.l.o.g.,
  $r>s$. This gives us $|o^l \varsigma_V^s \varsigma_P^s \varsigma_V^{r-s} -
  o^l \varsigma_V^s \varsigma_P^s \varsigma_P^{r-s}|$ for the absolute value
  of the difference in costs, which is always greater than or equal to 0.

  \textbf{$l$ is odd:} W.l.o.g. we assume that for computing the cost for a
  perfectly alternating interleaving for $Q$, there are $\lceil \sfrac{l}{2}
  \rceil$ multiplicands of type $\varsigma_V$ and $\lfloor \sfrac{l}{2}
  \rfloor$ multiplicands of type $\varsigma_P$. This results in $|o^l
  \varsigma_V^{\lfloor \sfrac{l}{2} \rfloor} \varsigma_P^{\lfloor \sfrac{l}{2}
    \rfloor}$ $(\varsigma_V - \varsigma_P)|$ for the difference in costs
  between $Q$ and $Q'$.

  For a non-perfectly alternating interleaving, we again have
  $|o^l \varsigma_V^s \varsigma_P^s$ $\varsigma_V^{r-s} -
  o^l \varsigma_V^s \varsigma_P^s \varsigma_P^{r-s}| = |o^l
  \varsigma_V^s \varsigma_P^s $ $(\varsigma_V^{r-s} - \varsigma_P^{r-s})|$
  with $r+s = l$ and $r>s$, which can be reformulated to $|o^l
  \varsigma_V^s \varsigma_P^s (\varsigma_V^{\lfloor \sfrac{l}{2}
  \rfloor - s} \varsigma_V^{\lfloor \sfrac{l}{2} \rfloor - s}
  \varsigma_V - \varsigma_P^{\lfloor \sfrac{l}{2} \rfloor - s}
  \varsigma_P^{\lfloor \sfrac{l}{2} \rfloor - s} \varsigma_P)|$.

  What is left to prove is $|o^l \varsigma_V^{\lfloor \sfrac{l}{2} \rfloor}
  \varsigma_P^{\lfloor \sfrac{l}{2} \rfloor} (\varsigma_V - \varsigma_P)| \leq
  |o^l \varsigma_V^s \varsigma_P^s (\varsigma_V^{\lfloor \sfrac{l}{2} \rfloor
    - s} $ $\varsigma_V^{\lfloor \sfrac{l}{2} \rfloor - s} \varsigma_V -
  \varsigma_P^{\lfloor \sfrac{l}{2} \rfloor - s} \varsigma_P^{\lfloor
    \sfrac{l}{2} \rfloor - s} \varsigma_P)|$. W.l.o.g., assume that
  $\varsigma_V > \varsigma_P$ (if $\varsigma_V < \varsigma_P$, we just have to
  switch the minuend with the subtrahend in the subtractions and the roles of
  $\varsigma_V$ and $\varsigma_P$ in the following), then,
  as all numbers in the inequality
  are greater than or equal to 0, we have to prove
  $o^l \varsigma_V^{\lfloor \sfrac{l}{2} \rfloor} \varsigma_P^{\lfloor
    \sfrac{l}{2} \rfloor} (\varsigma_V - \varsigma_P) \leq o^l \varsigma_V^s
  \varsigma_P^s (\varsigma_V^{\lfloor \sfrac{l}{2} \rfloor - s} $
  $\varsigma_V^{\lfloor \sfrac{l}{2} \rfloor - s} \varsigma_V -
  \varsigma_P^{\lfloor \sfrac{l}{2} \rfloor - s} \varsigma_P^{\lfloor
    \sfrac{l}{2} \rfloor - s} \varsigma_P)$,
  which is equivalent to $\varsigma_V^{\lfloor
    \sfrac{l}{2} \rfloor - s} \varsigma_P^{\lfloor \sfrac{l}{2} \rfloor - s} $
  $(\varsigma_V - \varsigma_P) \leq \varsigma_V^{\lfloor \sfrac{l}{2} \rfloor
    - s} \varsigma_V^{\lfloor \sfrac{l}{2} \rfloor - s}$ $\varsigma_V -
  \varsigma_P^{\lfloor \sfrac{l}{2} \rfloor - s} \varsigma_P^{\lfloor
    \sfrac{l}{2} \rfloor - s} \varsigma_P$.  Substituting $a = \varsigma_V$,
  $b = \varsigma_P$, and $x = {\lfloor \sfrac{l}{2} \rfloor - s}$, this means
  showing that $a^x b^x (a-b) \leq a^{2x + 1} - b^{2x + 1} \Leftrightarrow
  a^{x+1} b^x - a^x b^{x+1} \leq a^{2x + 1} - b^{2x + 1} \Leftrightarrow b^{2x
    + 1} - a^x b^{x+1} \leq a^{2x + 1} - a^{x+1} b^x \Leftrightarrow
  b^{x+1}(b^x - a^x) \leq a^{x+1}(a^x - b^x)$. Since $a > b$, the left-hand
  side of the inequality is always less than 0, while the right-hand side is
  greater than 0. $\hfill\Box$
\end{proof}

\begin{proof}[Lemma \ref{lemma:bestcase}]
  There are $\ceil{\log_f{\ceil{\frac{N}{M}}}}$ levels before
  partitions fit completely into memory, at which point there is no
  further disk I/O except writing the final output (the index) to
  disk.  At each level we read and write $\ceil{\frac{N}{B}}$ pages.
  $\hfill\Box$
\end{proof}

\begin{proof}[Lemma \ref{lemma:worstcase}]
  We assume that $\psi(L,D)$ returns two partitions where the first
  contains one key and the second contains all other keys. Thus, on
  each level of the partitioning we have two partitions $L_{i,1}$ and
  $L_{i,2}$ such that $|L_{i,1}| = 1$ and $|L_{i,2}| = N-i$.
  Partition $L_{i,1}$ occupies one page on disk.  $L_{i,2}$ occupies
  $\ceil{\frac{N-i}{B}}$ pages on disk.  Setting the latter to
  $\ceil{\frac{M}{B}}$ (i.e., the number of pages that fit into
  memory) and solving for $i$ shows that $i = N-\ceil{\frac{M}{B}}B$
  is the smallest number of levels $i$ such that $L_{i,2}$ fits
  completely into memory.  $\hfill\Box$
\end{proof}

\begin{proof}[Theorem \ref{theorem:iooverhead}]
  The lower-bound follows from Lemma \ref{lemma:bestcase}.  In each
  level $O(\frac{N}{B})$ pages are transferred and there are
  $O(\log(\frac{N}{M}))$ levels in the partitioning.  The base of the
  logarithm is upper-bounded by $f=2^8$ since we $\psi$-partition at
  the granularity of bytes. The upper-bound follows from Lemma
  \ref{lemma:worstcase}.  In each level $O(\frac{N}{B})$ pages are
  transferred and there are $O(N-M)$ levels in the worst case.
  $\hfill\Box$
\end{proof}

\begin{proof}[Lemma \ref{lemma:amortized}]
  A key moves through a series of indexes in our indexing pipeline.
  First, it is stored in $R^M_0$ at no I/O cost and after that, it
  moves through a number of disk-based indexes $R_0, \ldots, R_k$.
  Importantly, when a key in an index $R_i$ is moved, it always moves
  to a larger index $R_j$, $j > i$. After inserting $N$ keys there
  exist at most $O(\log_2(\frac{N}{M}))$ RSCAS indexes, hence a key is
  moved at most $O(\log_2(\frac{N}{M}))$ times.  The amortized I/O
  overhead when all $N$ keys are bulk-loaded at once is $\frac{1}{N}
  \times \textsf{cost}(N,M,B)$. Hence, the amortized I/O overhead of a
  single insertion in a sequence of $N$ insertions is
  $O(\log_2(\frac{N}{M}) \times \frac{1}{N} \times
  \textsf{cost}(N,M,B))$.  $\hfill\Box$
\end{proof}

\section{Tuning $\tau$}
\label{sec:tuningtau}

The following provides more details on how to calibrate the partitioning
threshold $\tau$ discussed in Section~\ref{sec:calibration}. In particular, we
quantify the effects leading to the shape of the curves depicted in
Figure~\ref{exp:tau}. Although we have not been able to develop a closed
formula (so far), the results below are important steps towards such a
formula.

The diagrams in Figure~\ref{exp:tau}(a) and Figure~\ref{exp:tau}(b), showing
the impact of $\tau$ on the number of nodes and the overall time for
bulk-loading the index, are not particularly interesting for the calibration,
as there is a clear relationship: the larger the value of $\tau$, the faster
we can stop the partitioning and the smaller the number of created
nodes. Consequently, the execution time for bulk-loading goes down, as we can
skip more and more partitioning steps.

Figure~\ref{exp:tau}(c) and Figure~\ref{exp:tau}(d) are much more interesting,
as we have two effects counteracting each other in both figures. We start with
Figure~\ref{exp:tau}(d), showing the impact of $\tau$ on the index size. 
First of all, $\tau$ influences the total amount of metadata stored on each
disk page. Clearly, the fewer leaf nodes per page we have, the smaller the amount of this
metadata. Assuming that we actually fill every leaf node with exactly $\tau$
key suffixes and that we store $d$ bytes of metadata, the overhead is
$\frac{N_p \cdot d}{\tau}$ per disk page (with $N_p$ being the number of input
keys per disk page). This reciprocal function
flattens out quickly and explains why the curve in Figure~\ref{exp:tau}(d)
drops at the beginning. However, there is a second effect at play. The more
key suffixes share a leaf node (i.e., the larger the value of $\tau$), the
higher the probability that they share a common prefix that has not been
factored out, because we stopped the partitioning early. We found estimating
the expected value of overlapping prefixes in leaf nodes very hard to do, as
it depends on the data distribution. A simplified version can be computed as
follows. Assume we have $n$ different prefixes $x_1, x_2, \dots, x_n$
that appear in the key suffixes stored in leaf nodes and that all prefixes
have the same likelihood of appearing. Moreover, let $u$ be the number of
unique prefixes in a leaf node, then we have $\tau - u$ prefixes that are
stored multiple times (the first time a prefix shows up in a leaf node, it is
fine, but every subsequent appearance adds to the overhead). For a leaf node,
let $R_i$ be a random variable that is 1 if $x_i$ is in the node, and 0
otherwise. Then the number $X$ of different prefixes found in the node is
$\sum_{i=1}^{n} R_i$. Due to the linearity of expectation,
$\EX[X] = \sum_{i=1}^{n} \EX[R_i]$. The probability of at least one prefix
$x_i$ appearing in a leaf node is equal to one minus the probability that
there is none, which is equal to $1 - (\frac{n-1}{n})^\tau$. Thus,
$\EX[X] = n (1 - (\frac{n-1}{n})^\tau)$, which means that the expected
overhead is $\tau - \EX[X]$ per leaf node. This rises slowly for small values of $\tau$, but
ascends more quickly for larger values of $\tau$. Nevertheless, this is still
a simplification, as it counts the prefixes rather than summing their lengths.

We now turn to the impact of the threshold $\tau$ on the query runtime
(Figure~\ref{exp:tau}(c)). The threshold determines how many internal nodes we
have that distinguish subsets of keys. For $\tau = 1$, when evaluating a
query, we visit a path down the trie containing all discriminative bytes. When
increasing $\tau$, the path shortens, as we skip the final discriminative
bytes. Mapping $\tau$ to the path length is not straightforward, as this
depends on the distribution of the keys again. Assuming that every
discriminative byte splits a set of keys into $b$ subsets and that the full
length of a trie path for $\tau = 1$ is $l$, the number of internal nodes
visited by a query is equal to $l - \log_b \tau$. The curve of this function
drops at the beginning, but then quickly flattens out, explaining the
left-hand part of Figure~\ref{exp:tau}(c). The second effect of increasing
$\tau$ is that we are accessing more and more keys that are not relevant for
our query. The irrelevant keys just happen to be in the same leaf node,
because we no longer distinguish them from the relevant keys. There is at
least one key in the node that satisfies the query, for the other keys we
compute the probability that they are relevant. We cannot just use the
selectivity $\sigma_c$ of the complete query, since we need the selectivity
$\sigma_s$ of the suffix stored in the leaf node. Thus, the expected number of
keys in a leaf node not satisfying the query predicate is
$(1 - \sigma_s)(\tau - 1)$. Assuming uniform distribution and independence, we
can estimate $\sigma_s$ given $\sigma_c$: if the length of the suffix in the
leaf node is $\frac{1}{s}$ of the total length of a key, then
$\sigma_s = \sqrt[s]{\sigma_c}$. Since all selectivities are within the range
of $[0,1]$, $\sigma_s \geq \sigma_c$, which means that
$(1 - \sigma_s) \leq (1 - \sigma_c)$, so $(1 - \sigma_s)(\tau - 1)$ is usually
a relatively flat ascending line. This explains the shape of the curve on the
right-hand side of Figure~\ref{exp:tau}(c).

While the distribution of the keys has a direct impact on all of these
parameters, as far as we can see, the total number of input keys does not directly
influence them. Consequently, we can use a sample to calibrate $\tau$ (as we
have done in Section~\ref{sec:calibration}).

} 

\end{document}